\RequirePackage{etex}
\documentclass{aa} 
\let\linenumbers\nolinenumbers
\usepackage{graphicx}
\usepackage{amsmath}
\usepackage[varg]{txfonts}
\usepackage{enumitem}
\usepackage{hyperref}
\usepackage{xcolor}
\makeatletter\newcommand{\figurecaptype}{\def\@captype{figure}}\makeatother
\begin{document}

\title{Tracking the evolution of sub-Neptunes}
\subtitle{II. The radius valley on gyrochronological ages}

\author{N.~Greco\inst{\ref{inst:0001},\ref{inst:0002}}\corrauth{nino.greco@inaf.it}
        \and
        A.C.~Lanzafame\inst{\ref{inst:0001},\ref{inst:0002}}\email{a.lanzafame@unict.it}
        \and
        I.~Pagano\inst{\ref{inst:0003}}\email{isabella.pagano@inaf.it}
        \and      
        G.~Scandariato\inst{\ref{inst:0002}}\email{gaetano.scandariato@inaf.it}
        \and
        V.~Drago\inst{\ref{inst:0001}}\email{vincedrago89@gmail.com}
        \and
        F.~Spada\inst{\ref{inst:0001}}\email{federico.spada@dfa.unict.it}
        }

   \institute{
     Dipartimento di Fisica e Astronomia ``Ettore Majorana'', Universit\`{a} di Catania, Via S. Sofia 64, 95123 Catania, Italy\relax\label{inst:0001}
   \and 
     INAF - Osservatorio Astrofisico di Catania, Via S. Sofia 78, 95123 Catania, Italy\relax \label{inst:0002}
        \and
    INAF, Istituto Nazionale di Astrofisica, Viale del Parco Mellini 84, 00136 Roma, Italy\relax \label{inst:0003}
   }

\date{Received ; accepted }

\abstract
{The radius valley separating super-Earths from sub-Neptunes is understood as a signature of atmospheric mass loss, but the timescale over which it develops remains poorly constrained.}
{We investigate the age evolution of the radius valley and the super-Earth population around FGK stars using gyrochronological ages, and release the multi-source exoplanet database underlying this work.}
{We select $746$ planets around $510$ FGK hosts with reliable gyrochronological ages between $0.1$ and $4$\,Gyr. A subset of $581$ planets around $404$ hosts, for which detection and confirmation completeness can be characterised, forms the reference sample. We measure the valley morphology and its locus in the period--radius plane, and infer the intrinsic super-Earth fraction with a selection-conditioned mixture model that accounts for stellar-age uncertainties.}
{The intrinsic super-Earth fraction increases primarily during the first $\sim2$\,Gyr, with no significant subsequent evolution. An exponential approach-to-plateau model places $90\%$ of this increase within $1.9$\,Gyr at $90\%$ credibility.
The radius valley is consistently resolved from $\sim1.5$\,Gyr onwards, with no significant evolution in its depth or width. Its minimum may shift from $\sim1.5\,R_\oplus$ before $\sim2$\,Gyr to $\sim1.8\,R_\oplus$ at older ages, accompanied by a change in the shape of the super-Earth radius distribution, although these changes are only suggestive.
At fixed orbital period and metallicity, sub-Neptune radii increase with host-star mass, whereas no significant host-star mass dependence is detected for super-Earth radii. The super-Earth fraction shows no significant dependence on host metallicity.}
{The super-Earth population evolves predominantly during the first $\sim2$\,Gyr, over which the radius valley develops its mature morphology. This evolution is consistent with atmospheric loss continuing up to approximately $1$--$2$\,Gyr for a fraction of planets. The tentative shift of the valley minimum towards larger radii may indicate that planets with more massive cores complete envelope loss later, although the present data do not uniquely distinguish between atmospheric-loss mechanisms.}


\maketitle

\section{Introduction}\label{sec:intro}

The radii of close-in small planets are bimodally distributed. 
Using the precise stellar parameters of the California--Kepler Survey (CKS), \citet{Fulton2017} resolved a deficit of planets across $1.5$--$2.0\,R_\oplus$, the radius valley, separating two populations: super-Earths (SE; $\sim$$1.3\,R_\oplus$, generally interpreted as rocky, stripped cores) and sub-Neptunes (SN; $\sim$$2.4\,R_\oplus$, generally interpreted as retaining H/He envelopes of a few per cent by mass). 
This feature was subsequently confirmed, for instance with \textit{Gaia}-based stellar radii \citep{FultonPetigura2018} and asteroseismic samples \citep{VanEylen2018}, and is now established as a robust property of the small-planet population around FGK stars.

Two thermally driven atmospheric mass-loss mechanisms have been proposed to explain the radius valley.
In photoevaporation \citep{OwenWu2013,LopezFortney2013} stellar XUV irradiation drives a hydrodynamic wind that erodes the H/He envelope. 
The high-energy luminosity of the star remains saturated for only the first $\sim100$\,Myr, with a duration set by the initial rotation rate, and then declines steeply \citep{Tu2015,Johnstone2021}. 
Photoevaporation therefore acts most strongly at early ages, although the softer EUV component responsible for much of the later escape declines more gradually and can remain effective over Gyr timescales \citep{KingWheatley2021}.
In core-powered mass loss \citep{Ginzburg2018,GuptaSchlichting2019,GuptaSchlichting2020} the thermal energy of the planetary core is radiated through the blanketing envelope and drives the outflow. 
This process proceeds over Gyr timescales, so its cumulative effect grows with age.
Boil-off, a rapid loss of the outer envelope during the first $\sim$$1$\,Myr after disc dispersal, largely sets the initial envelope-mass-fraction distribution on which both mechanisms subsequently operate \citep{OwenWu2016,RogersOwenSchlichting2024}. 
A further possibility attributes the valley to a primordial compositional dichotomy between rocky and water-rich planets \citep{Zeng2019}; acting alone, such a mechanism would place the valley at formation and introduce no age dependence.

The two thermally driven mechanisms predict almost the same dependence of the valley radius on orbital period: the SE/SN boundary moves to smaller radii at longer periods, yielding a negative slope $\mathrm{d}\log R_p/\mathrm{d}\log P$, with $P$ the orbital period.
Core-powered mass loss gives an analytic slope of $-1/9\simeq-0.11$ \citep{GuptaSchlichting2019}, while energy-limited photoevaporation formally predicts a steeper value, $\simeq-0.25$ \citep{OwenWu2017}.
Full hydrodynamic escape calculations including boil-off, however, recover a slope close to $-0.11$ \citep{Affolter2023}. 
The measured values, $-0.09^{+0.02}_{-0.04}$ \citep{VanEylen2018} and $-0.11\pm0.02$ \citep{HoVanEylen2023}, are consistent with both mechanisms. 
The slope alone therefore does not distinguish between them. 
Their characteristic timescales differ more strongly in the standard pictures, although the persistence of EUV-driven escape blurs this distinction.

Testing these different time dependences requires reliable stellar ages for large planet samples, and existing age-resolved studies do not yet give a fully consistent picture.
Using isochrone ages, \citet{Berger2020} and \citet{Sandoval2021} find the SE-to-SN ratio increasing with age, as does the multi-survey sample of \citet{Kamulali2026}. 
In contrast, \citet{Chen2026} find the valley not fully established below $\sim$$3$\,Gyr and the ratio of planets outside to inside it to be non-monotonic at older ages. 
Considering instead the mean size of the SNs, \citet{Petigura2022} find no significant evolution over $1$--$10$\,Gyr and conclude that large-scale envelope contraction ends by $\sim$$1$\,Gyr. 
\citet{David2021}, combining isochrone ages with gyrochronology, report the valley to be most depleted at young ages ($\lesssim$2--3\,Gyr) and progressively filled at older ages.
Using galactic-kinematic ages, \citet{Chen2022} instead find the valley becoming more pronounced with age, with the accompanying rise in the SE-to-SN ratio largely complete within the first $1$--$2$\,Gyr.
Despite these differences, the emerging picture is that most of the evolution occurs at young ages, $\lesssim1$--$2$\,Gyr.

In \citet[hereafter Paper~I]{PaperI}, we derived homogeneous gyrochronological ages for exoplanet hosts by comparing their rotation periods and de-reddened \textit{Gaia} colours with rotational-evolution models. 
These ages are particularly informative at the young ages at which the radius valley is expected to evolve. 
Here we use them to resolve the valley as a function of age in a sample of FGK hosts. 
The sample contains $746$ planets around $510$ hosts with ages in the range $0.1$--$4$\,Gyr, drawn from a multi-source database of $6298$ confirmed planets that is also released with this work. 
For $581$ of these planets, around $404$ hosts, the selection function can be modelled and they form the reference sample.

This paper is organised as follows. Sect.~\ref{sec:data} presents the data and sample selection.
Sect.~\ref{sec:morph} describes the evolution of the radius valley and locates it in the period--radius plane. 
Sect.~\ref{sec:fse} measures the SE fraction as a function of age, while Sect.~\ref{sec:results} examines its dependence on the host metallicity and stellar mass. 
Sect.~\ref{sec:disc} discusses the results and their physical implications, and Sect.~\ref{sec:conclusions} summarises our conclusions. 


\section{Data}\label{sec:data}

\subsection{An integrated exoplanet database}\label{sec:db}

This work uses an integrated exoplanet database, publicly available online\footnote{[Database website link to be provided.]}, that combines approximately one hundred catalogues and individual studies into a provenance-tracked compilation of $6298$ confirmed planets around $4716$ host stars. The database provides both the planetary parameters used in this analysis and the stellar data from which the gyrochronological ages were derived in Paper~I. It also covers the wider exoplanet population beyond the small-planet regime considered here.

The NASA Exoplanet Archive \citep[NEA;][]{Christiansen2025}\footnote{\url{https://exoplanetarchive.ipac.caltech.edu/}} Planetary Systems table\footnote{doi:\href{https://doi.org/10.26133/NEA12}{10.26133/NEA12}}, retrieved on 2026 June~19 at 15:00~UTC, is the primary source of planetary identifications and published transit and orbital parameters. Stellar astrometry, photometry, and astrophysical parameters are provided by \textit{Gaia}~DR3 \citep{Gaia2016,GaiaDR3}. Rotation periods are compiled from approximately two dozen dedicated catalogues and numerous studies of individual systems, while chemical abundances are drawn from spectroscopic surveys \citep[including][]{Luo2015,Abdurrouf2022,Petigura2022} and individual studies.

The database also contains quantities not used here, including X-ray and ultraviolet measurements and independent stellar-age estimates from isochrones, asteroseismology, and kinematics. The complete source list accompanies the public release.

\subsection{Stellar and planetary parameters}
\label{sec:starpars}\label{sec:planetpars}

We adopt the stellar radii, effective temperatures, masses, metallicities, and de-reddened \textit{Gaia}~DR3 colours $(G_{\rm BP}-G_{\rm RP})_0$ selected in Paper~I\footnote{[Stellar catalogue link to be provided.]}. Each parameter has an associated reliability tier from 1 (highest reliability) to 4 (lowest reliability).

Stellar ages are taken from the gyrochronological analysis of Paper~I, based on rotational-evolution models covering $(G_{\rm BP}-G_{\rm RP})_0=0.65$--$2.99$ and masses of $0.40$--$1.25\,M_\odot$. Their uncertainties are represented by asymmetric two-sided Gaussian distributions with total uncertainties $\sigma_{\rm tot}^+$ and $\sigma_{\rm tot}^-$. The median fractional uncertainty is approximately 10\% at $1$--$4$\,Gyr and increases to 19\% at $0.1$--$1$\,Gyr. These individual age uncertainties are propagated into the population analysis of Sects.~\ref{sec:morph} and~\ref{sec:fse}.

Planetary parameters are selected from the multi-source database using criteria analogous to those applied to stellar parameters in Paper~I. In particular, we recompute planetary radii by combining the selected planet-to-star radius ratios with the stellar radii adopted in Paper~I, $R_{\star,\mathrm{I}}$:
\begin{equation}
R_p=(R_p/R_\star)\,R_{\star,\mathrm{I}},
\end{equation}
propagating the uncertainties of both quantities. This ensures consistency between the planetary radii and the stellar parameters underlying the gyrochronological ages.

The procedures for selecting the radius ratios, planetary masses, orbital periods, and other parameters, together with their reliability tiers, are described in Appendix~\ref{app:planetparams}. The resulting planetary catalogue is included in the public release\footnote{[Planetary catalogue link to be provided.]}. 

Throughout this work, uncertainties are quoted as 16th--84th percentile intervals unless otherwise stated, and significances expressed in units of $\sigma$ use half the width of these intervals.

\subsection{Sample selection}\label{sec:sample}

We construct three nested samples to address different aspects of the analysis. The first contains small, short-period planets with reliable measured parameters; the second restricts these planets to hosts with reliable gyrochronological ages; and the third retains systems for which the \textit{Kepler} detection efficiency can be modelled individually. The samples are summarised in Table~\ref{tab:samples}.

\begin{table}
\caption{Nested samples used in this work.}
\label{tab:samples}
\centering
\setlength{\tabcolsep}{4pt}
\begin{tabular}{lrrl}
\hline\hline
Sample & Planets & Hosts & Main purpose \\
\hline
$\mathcal{D}_1$ & 2619 & 1934 & Parent small-planet sample \\
$\mathcal{D}_2$ & 746 & 510 & Multi-survey age-resolved sample \\
$\mathcal{D}_3$ & 581 & 404 & Selection-corrected analysis \\
\hline
\end{tabular}
\tablefoot{The samples are nested:
$\mathcal{D}_3\subset\mathcal{D}_2\subset\mathcal{D}_1$.
The additional radius-restricted subsets used in individual
analyses are defined where introduced.}
\end{table}

We first exclude planets flagged as controversial in the NEA. The parent sample $\mathcal{D}_1$ is then defined by the following requirements:
\begin{enumerate}
\item[(a)] Planetary radius $R_p\in[0.5,4.0]\,R_\oplus$, recomputed using the stellar radius adopted in Paper~I;
\item[(b)] Orbital period $P_{\rm orb}\le100$\,d, restricting the analysis to a regime with numerous observable transits and well-characterised \textit{Kepler} detection efficiency \citep{Borucki2010};
\item[(c)] Relative uncertainties $\sigma(R_p)/R_p\le25\%$ and $\sigma(P_{\rm orb})/P_{\rm orb}\le10\%$, with reliability tiers $\le2$ for both parameters;
\item[(d)] De-reddened host-star colour $(G_{\rm BP}-G_{\rm RP})_0\in[0.65,1.90]$, restricting the sample to the FGK regime within the model domain of Paper~I;
\item[(e)] Non-grazing transit geometry, with impact parameter $\le0.9$;
\item[(f)] Non-circumbinary orbit, as flagged by the NEA.
\end{enumerate}
Here, relative uncertainties are computed using the mean of the upper and lower reported uncertainties. Applying these criteria, and requiring uncertainty estimates for planetary radius and orbital period, yields $\mathcal{D}_1$, containing $2619$ planets around $1934$ hosts.

The age-resolved sample $\mathcal{D}_2$ additionally requires: 
\begin{enumerate}
    \item[(g)] A gyrochronological age in $[0.1,4.0]$\,Gyr, given as a finite estimate rather than an upper or lower limit, with a reported uncertainty and reliability tier $\le2$.
\end{enumerate} 
$\mathcal{D}_2$ contains $746$ planets around $510$ hosts. It provides a wider sample, not restricted to a single transit survey, from which $\mathcal{D}_3$ is drawn.

The reference sample $\mathcal{D}_3$ is the subset of $\mathcal{D}_2$ consisting of \textit{Kepler} Objects of Interest (KOIs) from DR25 \citep{Thompson2018} whose hosts have the corresponding per-star detection metrics \citep{BurkeCatanzarite2017,BurkeCatanzarite2017b}. It contains $581$ planets around $404$ hosts. For these systems, the detection and confirmation probabilities can be estimated as functions of planetary radius, orbital period, and host properties. This makes $\mathcal{D}_3$ the reference sample for inferring intrinsic population properties after accounting for selection effects.

Because the database contains confirmed planets, inclusion in $\mathcal{D}_3$ requires both detection and confirmation. Twelve planets were classified as false positives by the DR25 vetting procedure but confirmed through independent analyses and are listed as confirmed in the NEA. We retain them and evaluate their selection probabilities as if they had passed DR25 candidate vetting (Appendix~\ref{app:compl:probit}).

\subsection{Detection and confirmation completeness}
\label{sec:compl}

The probability that a planet enters our sample decreases towards smaller radii. This selection effect is particularly important for SEs, whose shallower transits make them more difficult to detect and confirm than SNs. Consequently, the observed relative numbers of SEs and SNs cannot be interpreted directly as their intrinsic population fractions.

For each planet $i$ in $\mathcal{D}_3$, we define the selection function $S_i(x')$ as the probability that a transiting planet of true log-radius $x'=\log_{10}(R_p/R_\oplus)$, on the same orbit and around the same host, enters the sample. We model this probability in two stages: detection by the \textit{Kepler} pipeline and retention as a DR25 planet candidate, followed by confirmation and inclusion in our catalogue. The model accounts for the dependence of these processes on signal strength, orbital configuration, and host-star properties.

The first stage depends primarily on the multiple-event statistic (MES), which measures the signal-to-noise ratio of the transit signal. Following a form adapted from \citet{Burke2015}, we estimate the expected MES as
\begin{equation}
\mathrm{MES} =
\frac{10^6(R_p/R_\star)^2\sqrt{N_{\rm tr}}}
{\mathrm{CDPP}_{\rm dur}},
\end{equation}
where $R_p$ is the true planet radius, $N_{\rm tr}$ is the expected number of observed transits and $\mathrm{CDPP}_{\rm dur}$ is the combined differential photometric precision over the transit duration, expressed in parts per million. The transit search detects a signal only if its measured MES exceeds $7.1$ \citep{Twicken2016}. We calibrate the detection probability using DR25 transit-injection tests \citep{Christiansen2017}, accounting for differences between individual hosts and orbits.

The second stage accounts for incomplete confirmation of detected candidates, particularly at low MES. We estimate the confirmation probability from the DR25 candidate population, allowing it to depend on signal strength and on whether another planet of the same star is confirmed. At fixed MES, confirmation shows no dependence on whether a candidate is a SE or a SN. We also test for a possible dependence of confirmation on host age, through a coefficient $\mathcal{H}$. The calibration of both stages, their combination into $S_i(x')$, and the associated uncertainty tests are described in Appendix~\ref{app:compl:probit}.

At fixed host and orbit, factors independent of planetary radius cancel from the conditional likelihood of Sect.~\ref{sec:hier}. These include geometric transit probability, host-level selection criteria, rotation-period availability, and radius-independent efficiency ceilings. We therefore omit these factors from $S_i$, which approaches unity at large radii. The measured-radius restriction is incorporated through the likelihood normalisation. This approach corrects the radius-dependent selection of planets conditional on their observed host--orbit configurations, without modelling the selection of the parent stellar population.

The combined detection and confirmation selection function reaches half its efficiency ceiling at a median radius of $1.15\,R_\oplus$, with 16th--84th percentiles of $0.79$--$1.69\,R_\oplus$ across $\mathcal{D}_3$. Selection effects are therefore substantial in the SE radius range and must be accounted for when comparing planetary populations across age.

We assess the robustness of the inferred age dependence of the SE fraction against uncertainties in the selection model (Appendix~\ref{app:compl:probit}). 
Restricting the analysis to planets with $\mathrm{MES}\ge15$, where confirmation is close to its efficiency ceiling, gives an increase in the intrinsic SE fraction of $+0.32^{+0.09}_{-0.11}$ between $<1$ and $2$--$4$\,Gyr, consistent with $+0.24^{+0.10}_{-0.10}$ for the full sample (Sect.~\ref{sec:resglobal}).

We use the selection function in two complementary ways. In the mixture analysis of Sect.~\ref{sec:hier}, $S_i$ enters the likelihood for each planet, allowing the intrinsic SE fraction to be inferred. For the radius distributions and demographic comparisons of Sects.~\ref{sec:morph} and \ref{sec:results}, we instead apply inverse-completeness weights.

The population completeness $\bar S(R_p)$ is obtained by averaging the selection function over the host--orbit configurations of the relevant sample or age subsample, correcting for the over-representation of configurations in which planets are easier to detect. The completeness $\bar S_i$, with every configuration evaluated at the period and measured radius of planet $i$, provides the weight $1/\bar S_i$ for each planet. To prevent a few poorly detected planets from dominating the weighted distributions, planets with $\bar S_i<0.25$ are assigned zero weight. The definitions and normalisation of these quantities are given in Eqs.~\eqref{eq:sbar} and \eqref{eq:sbari} of Appendix~\ref{app:compl:probit}.

For $\mathcal{D}_3$, the population completeness is $0.26$ at $1\,R_\oplus$ and $0.86$ at $2\,R_\oplus$. At $1\,R_\oplus$, it is $0.31$, $0.27$, and $0.24$ for hosts aged $<1$, $1$--$2$, and $2$--$4$\,Gyr, respectively. These differences are accounted for when comparing the planetary populations across age.


\section{Age-resolved morphology of the radius valley}\label{sec:morph}

\subsection{The radius valley and its metrics}\label{sec:global}

The small-planet radius distribution is bimodal, with SEs and SNs separated by the radius valley. We first characterise its morphology using metrics that neither require classifying individual planets nor assume that a valley is resolved. By contrast, the support-vector method of Sect.~\ref{sec:prplane} assumes the presence of a valley, locates it in the period--radius plane, and defines the SE/SN boundary.

We estimate the SE and SN peak positions and the minimum between them from a fixed-bandwidth kernel-density estimate (KDE) of $\log R_p$, requiring exactly one peak in each of the SE and SN search intervals, $[1.0,1.7]$ and $[2.0,3.2]\,R_\oplus$, respectively. We define the \emph{valley depth} as $1-\rho_{\rm valley}/\overline{\rho}_{\rm peak}$, where $\rho_{\rm valley}$ is the density at the minimum and $\overline{\rho}_{\rm peak}$ is the mean density of the two peaks. The \emph{occupancy} is the fraction of planets lying within $\pm0.2\,R_\oplus$ of the valley locus of Sect.~\ref{sec:prplane} at their orbital periods (estimator settings in Appendix~\ref{app:morphtables:kde}).

All morphological statistics use planets with $R_p\in[1,4]\,R_\oplus$. We denote by $\tilde{\mathcal{D}}_i$ the corresponding restriction of sample $\mathcal{D}_i$ (Sect.~\ref{sec:sample}). The $0.5$--$1\,R_\oplus$ range is too incomplete for direct morphological analysis but is retained in the mixture likelihood of Sect.~\ref{sec:fse}, which uses the full samples. The KDE metrics are evaluated over Monte Carlo draws that resample the planets and, where ages enter, perturb stellar ages within their uncertainties; planetary radii are not perturbed. A draw \emph{resolves} the valley if it contains exactly one peak in each search interval. For each sample, we report the morphological measurements only when at least two thirds of its draws satisfy this condition (Appendix~\ref{app:morphtables:kde}).

For $\tilde{\mathcal{D}}_3$, the observed KDE places the SE peak at $1.42^{+0.05}_{-0.05}\,R_\oplus$, the SN peak at $2.34^{+0.06}_{-0.05}\,R_\oplus$, and the valley at $1.72^{+0.04}_{-0.05}\,R_\oplus$. The depth is $0.28^{+0.05}_{-0.05}$, the occupancy $0.122^{+0.015}_{-0.013}$, and $74\%$ of draws resolve the valley (Table~\ref{tab:gapmetrics}).

Because completeness decreases towards smaller radii, these observed counts under-represent SEs. We therefore also compute completeness-corrected metrics, weighting each planet by $1/\bar S_i$ (Sect.~\ref{sec:compl}). The weighted KDE places the SE peak at $1.35^{+0.06}_{-0.06}\,R_\oplus$, the SN peak at $2.30^{+0.05}_{-0.05}\,R_\oplus$, and the valley at $1.74^{+0.04}_{-0.05}\,R_\oplus$. The corresponding depth and occupancy are $0.26^{+0.06}_{-0.06}$ and $0.123^{+0.015}_{-0.015}$, with $98\%$ of draws resolving the valley. These uncertainties, for both the observed and the weighted KDE, do not include the effect of the smoothing choice or of planets sharing a host (Appendices~\ref{app:morphtables:lgcp} and~\ref{app:morphtables:tests}).

As a complementary description of the radius distribution in the $1$--$4\,R_\oplus$ range, we fit a log-Gaussian Cox process (LGCP) to the counts of the $N=541$ planets of $\tilde{\mathcal{D}}_3$ in bins of $\log R_p$. The counts in each radius bin follow a Poisson distribution with mean given by the intensity $\lambda(R_p)$, multiplied by the population completeness $\bar S(R_p)$ and the bin width. The log-intensity varies smoothly with $\log R_p$, with its curvature constrained by a smoothing penalty. We compute $\bar S(R_p)$ from $\mathcal{D}_3$; the model, uncertainty estimates, and extraction of peak and valley positions are detailed in Appendix~\ref{app:morphtables:lgcp}.

Applying the completeness correction shifts the LGCP SE peak from its uncorrected value of $1.41^{+0.09}_{-0.08}$ to $1.25^{+0.12}_{-0.11}\,R_\oplus$. The valley shifts from $1.75^{+0.05}_{-0.10}$ to $1.80^{+0.06}_{-0.11}\,R_\oplus$, while the SN peak moves from $2.36^{+0.07}_{-0.14}$ to $2.29^{+0.14}_{-0.07}\,R_\oplus$. The valley depth, defined analogously to the KDE as $1-\lambda_{\rm valley}/\overline{\lambda}_{\rm peak}$, increases from $0.40^{+0.09}_{-0.09}$ to $0.43^{+0.10}_{-0.11}$. 

The SE peaks measured by both methods are read within $1$--$4\,R_\oplus$ and need not coincide with the mode of the full SE population. The mixture model of Sect.~\ref{sec:fse}, which includes smaller planets, places that mode near $1.1\,R_\oplus$. Moreover, the weighted counts exclude $12$ planets with $\bar S_i<0.25$, all smaller than $1.4\,R_\oplus$. Finally, radius measurement errors, smoothing, and the superposition of period-dependent valley loci (Sect.~\ref{sec:prplane}) all reduce the apparent contrast. The measured valley depths should therefore be regarded as lower bounds on the intrinsic depth.

\subsection{The valley locus in the period--radius plane}\label{sec:prplane}

\begin{figure}\centering
\includegraphics[width=\linewidth]{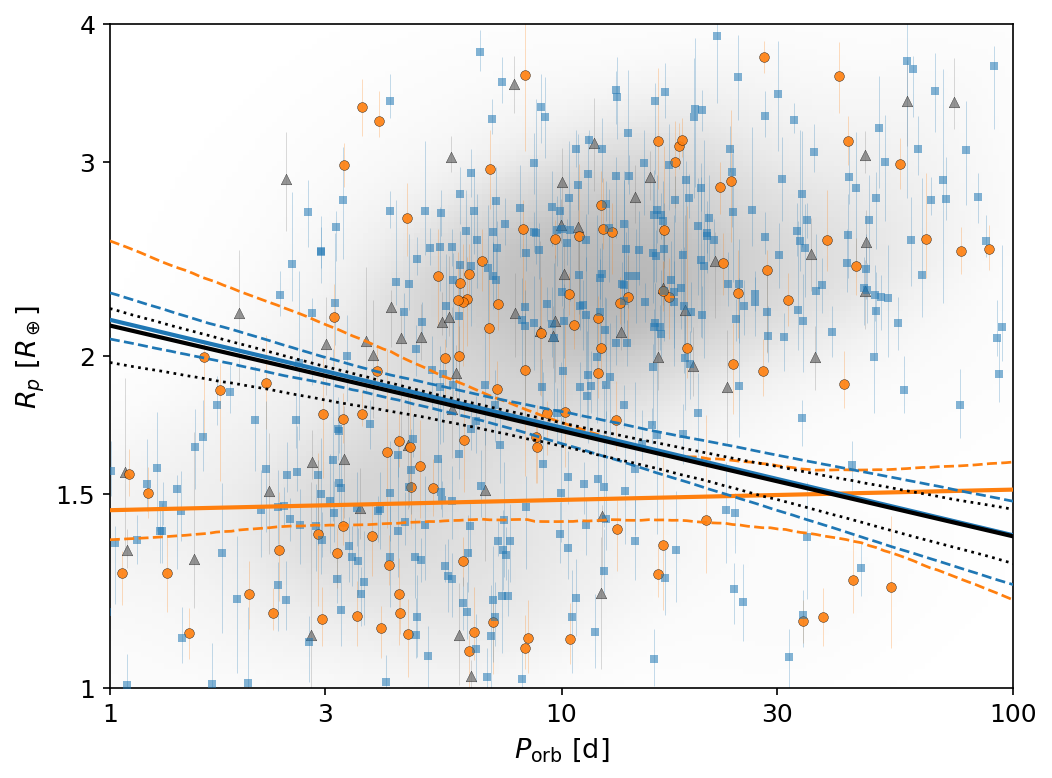}
\caption{Period--radius distribution of $\tilde{\mathcal{D}}_3$, with radius uncertainties, in three age groups: $0.1$--$1$\,Gyr (grey triangles), $1$--$2$\,Gyr (orange circles), and $2$--$4$\,Gyr (blue squares). The grey background shows the observed two-dimensional density without completeness weights. The black solid line is the reference valley locus (Eq.~\eqref{eq:slope}); coloured solid lines show the loci refitted on the two older age groups. Dotted (black) and dashed (coloured) lines delimit the $16$th--$84$th percentile bands of the corresponding loci. No locus is fitted below $1$\,Gyr because that group has fewer than the $N=60$ planets required for the fit.}
\label{fig:locibyage}
\end{figure}

We locate the radius valley in the $(\log P,\log R_p)$ plane of $\tilde{\mathcal{D}}_3$ using the observed counts (Fig.~\ref{fig:locibyage}). Following the gap-finder method of \citet{VanEylen2018}, we fit a linear support-vector machine (SVM) with soft-margin parameter $C_{\rm SVM}=10$. The SVM is trained on SE and SN labels assigned by a two-component Gaussian mixture, fitted after standardising $\log P$ and $\log R_p$ to zero mean and unit variance. The mixture and SVM are refitted in each of $1000$ bootstrap replicates, as in \citet{HoVanEylen2023}, resampling host stars rather than individual planets.

The resulting valley locus is
\begin{equation}\label{eq:slope}
\begin{aligned}
&\log_{10}[R_{\rm gap}(P)/R_\oplus]=m\log_{10}(P/\mathrm{d})+q,\\
&m=-0.096^{+0.027}_{-0.019},\qquad q=+0.329\pm0.025,
\end{aligned}
\end{equation}
corresponding to $R_{\rm gap}(10\,\mathrm{d})=1.71^{+0.04}_{-0.06}\,R_\oplus$. The fitted slope is robust to completeness weighting, the covariance structure adopted for the mixture classification, and the SVM regularisation. In particular, completeness weighting changes the slope by less than its statistical uncertainty and has a negligible effect on the SE/SN classification, the counted SE fractions and the valley offset across age. The quoted uncertainty includes the sensitivity to the mixture covariance model; the corresponding tests are detailed in Appendix~\ref{app:svmrobust}.

The slope is consistent with published measurements and with predictions from core-powered mass loss and hydrodynamic photoevaporation (Sect.~\ref{sec:intro}). We adopt the locus of Eq.~\eqref{eq:slope} as the SE/SN boundary, classifying planets with $R_p<R_{\rm gap}(P)$ at their own period as SEs.

Combining planets over a range of orbital periods superposes valleys at different radii, reducing their apparent depth. To correct for this effect, we also use the \emph{period-adjusted radius} 
\begin{equation}\label{eq:straight}
R_p'=R_p\,10^{-m\,[\log_{10}(P/\mathrm{d})-1]}=R_p\left(\frac{P}{10\,\mathrm{d}}\right)^{-m},
\end{equation} 
which rescales the radius of each planet to the reference period $P=10$\,d along the slope $m$ of Eq.~\eqref{eq:slope}. The transformation maps the valley onto $R_p'=R_{\rm gap}(10\,\mathrm{d})=1.71\,R_\oplus$ and preserves the SE/SN classification: $R_p<R_{\rm gap}(P)$ is equivalent to $R_p'<R_{\rm gap}(10\,\mathrm{d})$. Thus, $\log R_p'-\log R_{\rm gap}(10\,\mathrm{d})$ measures the vertical logarithmic distance from the valley locus at the planet's orbital period.

For $\tilde{\mathcal{D}}_3$ as a whole, using the observed counts, the valley lies at $1.72^{+0.04}_{-0.05}\,R_\oplus$ in $R_p$ and $1.71^{+0.04}_{-0.05}\,R_\oplus$ in $R_p'$. Its depth increases from $0.28^{+0.05}_{-0.05}$ to $0.39^{+0.06}_{-0.06}$ after period adjustment, as the superposition of period-dependent valley positions is removed.
 
The transformation also shifts the component peaks: the SE and SN peaks move from $1.42^{+0.05}_{-0.05}$ and $2.34^{+0.06}_{-0.05}\,R_\oplus$ in $R_p$ to $1.31^{+0.04}_{-0.04}$ and $2.45^{+0.06}_{-0.07}\,R_\oplus$ in $R_p'$, respectively. The opposite shifts reflect the tendency of SEs to occupy shorter orbital periods than SNs.

Finally, we measure the \emph{width at half depth} of the valley in $R_p'$ as the separation in $\log_{10}R_p'$ between the two radii where the KDE rises from its minimum to halfway towards the lower of the two peaks. For $\tilde{\mathcal{D}}_3$, this width is $0.103^{+0.014}_{-0.014}$\,dex in the observed counts and $0.124^{+0.023}_{-0.015}$\,dex after completeness weighting (Table~\ref{tab:gapmetrics}).

\subsection{The valley across age}\label{sec:ageshift}

\begin{figure}[t]\centering
\includegraphics[width=\linewidth]{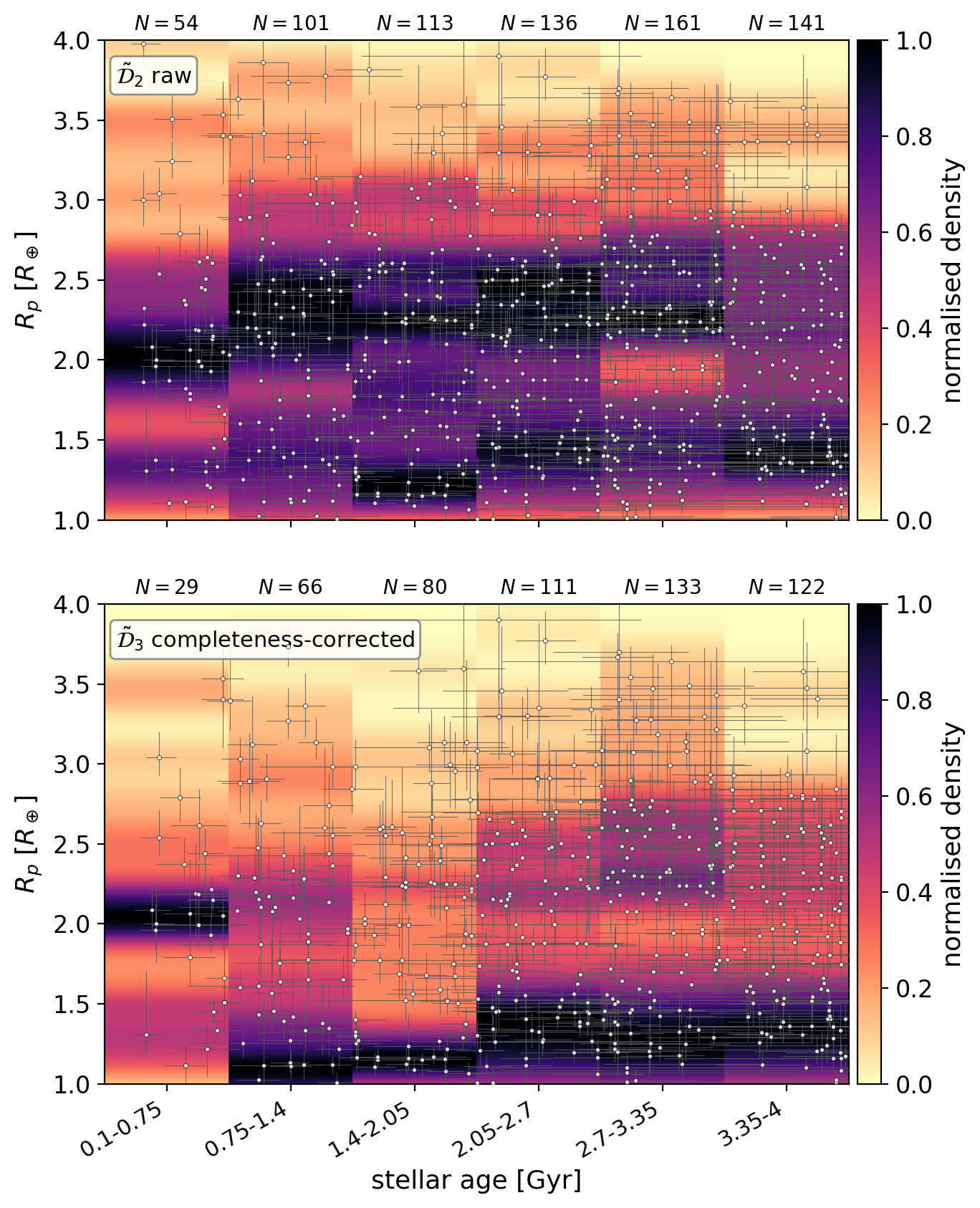}
\caption{Planet radius versus host age in six age bins (columns). The colour scale shows the radius density in each bin, normalised to its maximum; points show the planets with their age and radius uncertainties. \textit{Top:} uncorrected multi-survey sample $\tilde{\mathcal{D}}_2$. \textit{Bottom:} \textit{Kepler} sample $\tilde{\mathcal{D}}_3$, with the density of each column divided, at each radius, by the population completeness $\bar S$ of its bin (Sect.~\ref{sec:compl}). This correction is for display only; the mixture model of Sect.~\ref{sec:fse} includes selection in the likelihood.}\label{fig:heatmap6bin}
\end{figure}

\begin{figure*}[htbp]\centering
\includegraphics[width=\textwidth]{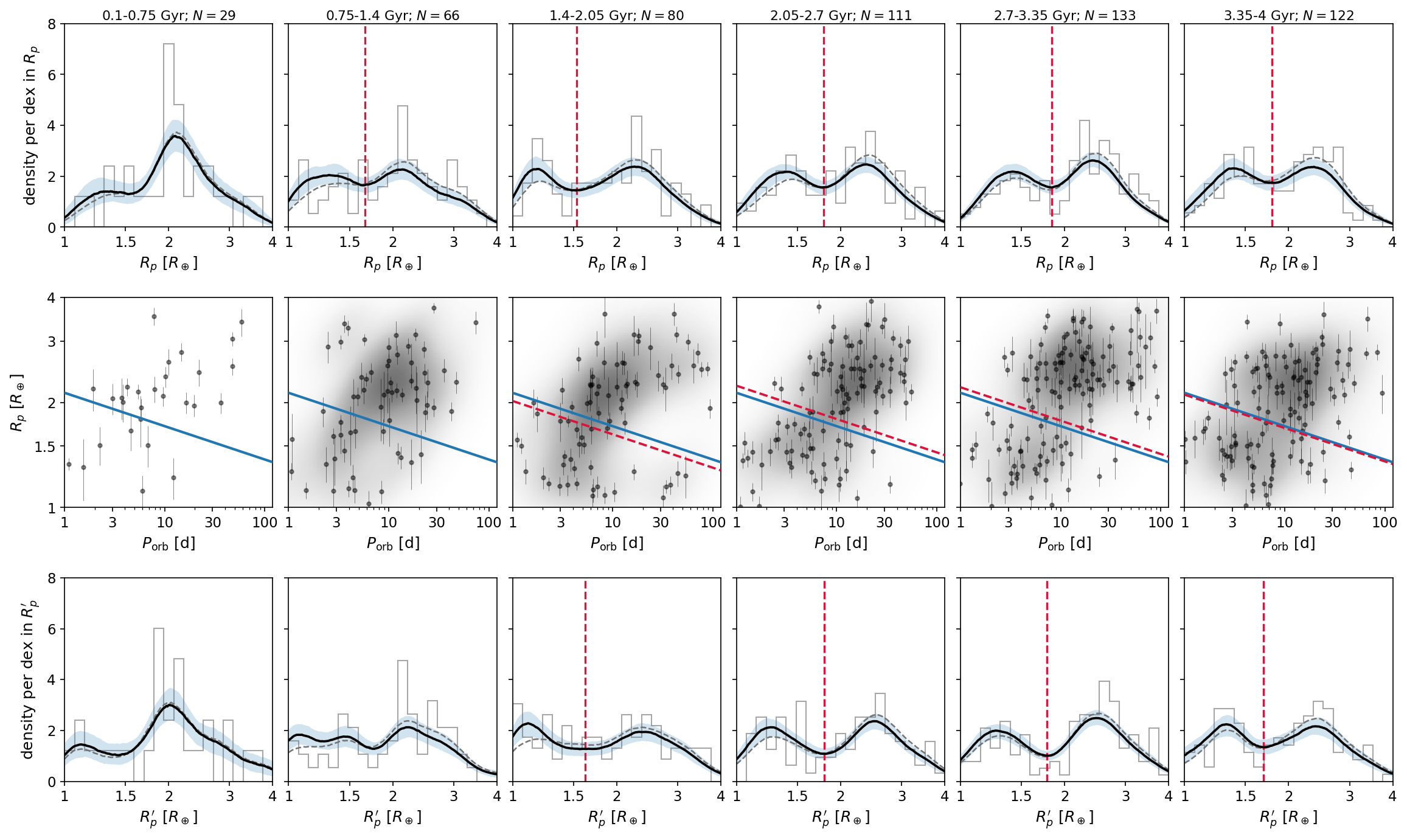}
\caption{Six age bins of $\tilde{\mathcal{D}}_3$ (columns). \textit{Top:} $R_p$ distributions. Observed counts: grey outline, histogram; grey dashed curve, median fixed-bandwidth KDE over Monte Carlo draws resampling planets and perturbing ages. Counts weighted by $1/\bar S_i$ (Sect.~\ref{sec:compl}): black curve and blue band, median and $16$th--$84$th percentile interval of the same KDE; crimson dashed line, the marginalised position of the weighted valley minimum, shown where at least two thirds of draws resolve the valley. \textit{Middle:} observed period--radius distribution. Background: two-dimensional density, shown for bins with more than $40$ planets. Points: the planets with their uncertainties. The blue solid line is the common valley locus of Eq.~\eqref{eq:slope}; the crimson dashed line shows the weighted valley minimum at the same slope, measured in $R_p'$ and plotted where at least two thirds of draws resolve the valley. \textit{Bottom:} distribution of the period-adjusted radius $R_p'$ (Eq.~\eqref{eq:straight}), with the conventions of the top row.}\label{fig:d3bins}
\end{figure*}

We examine the age dependence of the radius valley using $\tilde{\mathcal{D}}_3$, which avoids mixing different surveys (Sect.~\ref{sec:sample}). We divide the sample into six $0.65$\,Gyr-wide age bins (Figs.~\ref{fig:heatmap6bin} and~\ref{fig:d3bins}; Tables~\ref{tab:gapmetrics_agebins} and~\ref{tab:strbins}) and also analyse it continuously in sliding windows of width $0.5$, $0.8$ and $1.0$\,Gyr, read on $R_p$ (Appendix~\ref{app:morphtables:kde}). Since completeness at small radii varies with age (Sect.~\ref{sec:compl}), all age comparisons use counts weighted by $1/\bar S_i$.

\begin{figure}[t]\centering
\includegraphics[width=\linewidth]{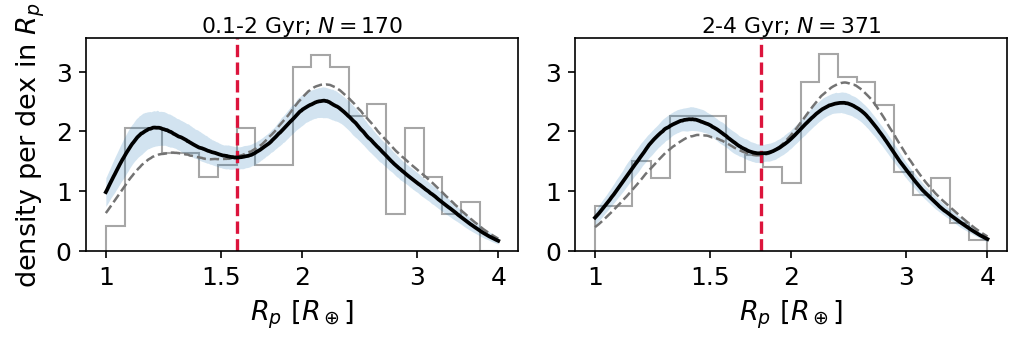}
\caption{$R_p$ distributions of $\tilde{\mathcal{D}}_3$ at $0.1$--$2$ and $2$--$4$\,Gyr, using the conventions of Fig.~\ref{fig:d3bins}.}
\label{fig:split2gyr}
\end{figure}

The youngest bin ($0.1$--$0.75$\,Gyr; $N=29$) resolves the valley in only $61\%$ of weighted draws in $R_p$ and $59\%$ in $R_p'$, below our two-thirds threshold. The $0.75$--$1.4$\,Gyr bin ($N=66$) passes this threshold in $R_p$ ($75\%$) but not in $R_p'$ ($54\%$). All four older bins resolve the valley in $85$--$99\%$ of weighted draws in both variables.

The valley minimum lies at smaller radii in the $1.4$--$2.05$\,Gyr bin than at older ages. In the adjacent $1.4$--$2.05$ and $2.05$--$2.7$\,Gyr bins, on either side of $2$\,Gyr, the weighted minima are $1.53^{+0.14}_{-0.10}$ and $1.79^{+0.08}_{-0.08}\,R_\oplus$, respectively ($1.50^{+0.16}_{-0.09}$ and $1.79^{+0.07}_{-0.07}\,R_\oplus$ on observed counts; Table~\ref{tab:gapmetrics_agebins}), and both bins resolve the valley in over $85\%$ of the draws. On the period-adjusted radius $R_p'$ (Eq.~\eqref{eq:straight}), the corresponding minima are $1.61^{+0.22}_{-0.17}$ and $1.79^{+0.08}_{-0.09}\,R_\oplus$, indicating a smaller shift within the uncertainties (Table~\ref{tab:strbins}). The two oldest bins have weighted minima on $R_p$ of $1.83^{+0.07}_{-0.08}$ and $1.78^{+0.13}_{-0.10}\,R_\oplus$, consistent with the $2.05$--$2.7$\,Gyr bin.

Sliding windows of width $0.5$, $0.8$ and $1.0$\,Gyr, advanced in steps of $0.05$\,Gyr, support this behaviour: their weighted minima in $R_p$ lie at $1.51$--$1.68\,R_\oplus$ for windows centred at $0.6$--$1.9$\,Gyr, compared with $1.73$--$1.86\,R_\oplus$ for those centred at $2.2$--$3.0$\,Gyr. The broader $0.1$--$2$ and $2$--$4$\,Gyr distributions of $R_p$ likewise show both the SE peak and valley minimum at smaller radii in the younger subset (Fig.~\ref{fig:split2gyr}).

We refit the period--radius valley locus (Eq.~\eqref{eq:slope}) on the planets of $\tilde{\mathcal{D}}_3$ with host ages of $1$--$2$\,Gyr ($N=115$) and $2$--$4$\,Gyr ($N=371$), using observed counts (Fig.~\ref{fig:locibyage}). At $P=10$\,d, the younger locus lies $0.24^{+0.07}_{-0.27}\,R_\oplus$ below the older one, a $1.4\sigma$ difference.
Restricting the bootstrap replicates to the $88\%$ in which the younger subset favours two mixture components over one ($\Delta\mathrm{BIC}>0$) increases the significance only marginally, to $1.6\sigma$.
The younger subset poorly constrains the slope ($+0.01\pm0.09$), whereas the older subset gives $-0.10\pm0.03$, consistent with the slope of the full sample (Eq.~\eqref{eq:slope}). We therefore retain the fixed classifier of Sect.~\ref{sec:prplane} at all ages. Lowering the boundary by the measured offset below $2$\,Gyr would reclassify $11$ planets in the $1$--$2$\,Gyr subset from SEs to SNs, increasing the difference in counted SE fraction across $2$\,Gyr by $+0.08^{+0.05}_{-0.04}$.

The completeness-corrected intensity model of Sect.~\ref{sec:global} provides a further comparison. Scaling the smoothing penalty to the number of planets in each subset, so that each has the same smoothing per planet as the two subsets together (Appendix~\ref{app:morphtables:lgcp}), the model resolves the valley in $71\%$ of draws at $1$--$2$\,Gyr and places it at $1.60^{+0.26}_{-0.09}\,R_\oplus$, compared with $1.85^{+0.12}_{-0.06}\,R_\oplus$ at $2$--$4$\,Gyr. The difference is $-0.24^{+0.24}_{-0.13}\,R_\oplus$, computed draw by draw over the draws that resolve the valley in both subsets. With a common smoothing penalty, however, the younger subset, which holds fewer planets, is smoothed more strongly and resolves the valley in only $47\%$ of draws, below the two-thirds threshold. The identification of its valley therefore depends on the smoothing choice.

We tested whether differences in host properties could account for the apparent shift. Older bins contain bluer, more massive hosts: the fraction of planets around F and G stars increases from $0.21$ in the youngest to $0.41$ in the oldest bin. Matching the host colours of the $1$--$2$ and $2$--$4$\,Gyr subsets leaves the locus offset unchanged (Appendix~\ref{app:colourmatch}). Conversely, metal-rich hosts are more common at younger ages: $70\%$ of planets at $1$--$2$\,Gyr orbit stars with $[\mathrm{Fe/H}]\geq0$, compared with $59\%$ at $2$--$4$\,Gyr.

Splitting the $1.5$--$2.5$ and $2.5$--$4$\,Gyr bins at solar metallicity, where the valley is resolved (Sect.~\ref{sec:composition}), 
the largest difference between the median valley positions of metal-rich and metal-poor hosts, over both bins and both radius variables, is $+0.14^{+0.18}_{-0.23}\,R_\oplus$, measured in $R_p$ at $1.5$--$2.5$\,Gyr. Approximating each subset's valley position by the mean of the metal-rich and metal-poor valleys weighted by their fractions, the difference of $0.11\pm0.05$ in metal-rich fractions between the $1$--$2$ and $2$--$4$\,Gyr subsets could produce an offset of at most $0.11\times0.14\simeq0.02^{+0.02}_{-0.03}\,R_\oplus$, where $0.14\,R_\oplus$ is the largest difference above. This is less than one tenth of the $0.26\,R_\oplus$ shift between the weighted valleys of the adjacent bins around $2$\,Gyr. Neither host colour nor metallicity therefore explains the observed shift. \citet{David2021} similarly found the valley at smaller radii below $3$\,Gyr, using both isochrone and gyrochronological ages, mainly because SEs of $1.5$--$1.8\,R_\oplus$ were scarce at younger ages.
 
The SE peak also appears to evolve. In the completeness-corrected intensity model of Sect.~\ref{sec:global}, it lies at $1.15^{+0.07}_{-0.03}\,R_\oplus$ at $1$--$2$\,Gyr and $1.37^{+0.04}_{-0.12}\,R_\oplus$ at $2$--$4$\,Gyr, a difference of $-0.19^{+0.12}_{-0.09}\,R_\oplus$ computed draw by draw ($1.8\sigma$). To test these differences against a random assignment of ages, we permute host ages, retain the $2$\,Gyr division, and refit both subsets with the most probable scaled smoothing penalty. At the measured ages, this fit gives differences of $-0.26\,R_\oplus$ for the valley and $-0.23\,R_\oplus$ for the SE peak. Among the $72\%$ of permutations in which both subsets resolve the valley, differences at least as large in absolute value occur in $4\%$ and $7\%$ of cases, respectively.
 
These shifts are not driven by differences in completeness. At $1\,R_\oplus$, $\bar S$ is $0.27$ at $1$--$2$\,Gyr and $0.24$ at $2$--$4$\,Gyr. Refitting each subset with the completeness curve of the other, or with their combined curve, leaves the valley minimum unchanged and moves the SE peak by at most $0.03\,R_\oplus$.

Unlike the SE peak, the median radius $R_{\rm SE}$ of planets below $1.71\,R_\oplus$ (the valley position at $10$\,d) shows no significant age dependence. On weighted counts, it is $1.32^{+0.05}_{-0.05}$, $1.31^{+0.05}_{-0.04}$ and $1.38^{+0.03}_{-0.03}\,R_\oplus$ in the $0.1$--$1.5$, $1.5$--$2.5$ and $2.5$--$4$\,Gyr bins, respectively ($1.33^{+0.05}_{-0.05}$, $1.34^{+0.05}_{-0.05}$ and $1.38^{+0.03}_{-0.03}\,R_\oplus$ on observed counts). The fitted slope is $\mathrm{d}\log R_{\rm SE}/\mathrm{d}\log t =+0.018\pm0.022$ ($+0.016\pm0.020$ unweighted; Appendix~\ref{app:compl:widths}). Thus, the SE size distribution appears to change in shape and relative weight rather than shifting uniformly in radius.

Consistently, the SE radius distributions at $1$--$2$ and $2$--$4$\,Gyr differ in quantile skewness, although the weighted difference is not significant. Their skewnesses $\mathcal{A}_{\rm skew}$ (Appendix~\ref{app:skewtest}) are $+0.25^{+0.31}_{-0.20}$ and $-0.07^{+0.11}_{-0.12}$, respectively, giving a difference of $+0.34^{+0.31}_{-0.24}$ computed draw by draw ($+0.34^{+0.17}_{-0.18}$ unweighted).
The positive skewness of the younger distribution indicates a more extended tail towards the SE/SN boundary. Mock catalogues generated from the fixed-shape mixture model of Sect.~\ref{sec:fse} produce an absolute difference at least as large as that at the measured ages, $+0.28$ ($+0.39$ unweighted), in $19\%$ of cases ($5\%$ unweighted; Appendix~\ref{app:skewtest}).

The completeness-weighted fraction $f_{\rm SE}^{{\rm c},w}$ of planets below the fixed classifier of Sect.~\ref{sec:prplane} rises from $0.34^{+0.11}_{-0.11}$ in the $0.1$--$0.75$\,Gyr bin to $0.47^{+0.06}_{-0.06}$ at $0.75$--$1.4$\,Gyr. Between the $0.1$--$1$ and $1$--$2$\,Gyr subsets, the increase is $+0.16^{+0.08}_{-0.09}$ ($1.8\sigma$ computed resample by resample), compared with $+0.14^{+0.08}_{-0.08}$ on observed counts.

Sliding windows show the same rise at young ages. For widths of $0.5$, $0.8$ and $1.0$\,Gyr, respectively, $f_{\rm SE}^{{\rm c},w}$ increases from $0.35^{+0.11}_{-0.12}$, $0.37^{+0.09}_{-0.09}$ and $0.40^{+0.08}_{-0.07}$ in the youngest windows (centred at $0.5$, $0.5$ and $0.6$\,Gyr; Appendix~\ref{app:morphtables:kde}) to $0.51^{+0.08}_{-0.09}$, $0.50^{+0.06}_{-0.07}$ and $0.50^{+0.06}_{-0.06}$ in windows centred at $1.25$, $1.4$ and $1.4$\,Gyr.

Above $1.4$\,Gyr, the four bins show no systematic ordering in $f_{\rm SE}^{{\rm c},w}$ (Table~\ref{tab:gapmetrics_agebins}). Their fitted linear trend is $-0.026^{+0.049}_{-0.046}$ per Gyr, and permuting their hosts among the age bins produces a spread at least as large as observed in $42\%$ of cases. The $2.7$--$3.35$\,Gyr bin has a weighted SE fraction $0.10\pm0.06$ below the inverse-variance mean of the other three bins at the measured ages ($0.08\pm0.05$ unweighted). Its width of $0.65$\,Gyr is $2.4$ times the median age error of its hosts, $0.27$\,Gyr. Its difference from the inverse-variance mean of all four bins is $-0.07$ at the measured ages and $-0.03^{+0.05}_{-0.05}$ with the ages redrawn within their errors.

Shifting the bin boundaries by up to $\pm0.3$\,Gyr consistently places the low-fraction bin at starting ages of $2.65$--$3.25$\,Gyr. However, host-age permutations produce a deficit at least as deep, at any position and bin shift, in $12\%$ of cases. Differences in orbital-period distributions may also contribute to the observed fluctuations. The fraction of planets with $P<10$\,d is $48\%$ in the $2.7$--$3.35$\,Gyr bin, compared with $56\%$ in both the $1.4$--$2.05$ and $3.35$--$4$\,Gyr bins, and $47\%$ at $2.05$--$2.7$\,Gyr. Since the SE weight of the mixture model decreases with orbital period ($\beta_P=-2.55^{+0.43}_{-0.47}$ per dex; Eq.~\eqref{eq:wlogit}; Appendix~\ref{app:compl:posterior}), differences in period distributions can affect the counted SE fractions. At the measured ages, the unweighted fractions $f_{\rm SE}^{\rm c}$ in these four bins, in age order, are $0.44^{+0.05}_{-0.06}$, $0.40^{+0.05}_{-0.05}$, $0.33^{+0.05}_{-0.05}$ and $0.40^{+0.05}_{-0.05}$.

These counted fractions are descriptive (Appendix~\ref{app:morphtables:kde}). Our reference population SE fractions are instead obtained from the mixture model of Sect.~\ref{sec:fse}, which incorporates selection directly into the likelihood (Appendix~\ref{app:compl}). 
The restriction to the completeness-characterised \textit{Kepler} sample also avoids the age-dependent survey composition of the multi-survey sample $\mathcal{D}_2$, for which an equivalent selection correction is unavailable (Appendix~\ref{app:surveymix}).

Finally, we find no significant age trend in the remaining valley metrics over the four bins above $1.4$\,Gyr. The two younger bins are excluded because the valley fails the two-thirds resolution criterion in $R_p$, $R_p'$, or both (Tables~\ref{tab:gapmetrics_agebins} and~\ref{tab:strbins}; Appendix~\ref{app:morphtables:tests}).

The weighted valley occupancy (Sect.~\ref{sec:global}) is $0.13^{+0.04}_{-0.04}$, $0.15^{+0.04}_{-0.03}$, $0.08^{+0.02}_{-0.02}$ and $0.13^{+0.04}_{-0.03}$ in the $1.4$--$2.05$, $2.05$--$2.7$, $2.7$--$3.35$ and $3.35$--$4$\,Gyr bins, respectively.
A $\chi^2$ test of a common occupancy gives $\chi^2=3.2$ ($p=0.35$) for weighted counts and $\chi^2=2.8$ ($p=0.42$) for observed counts, each with three degrees of freedom. The smallest occupancy difference between the first and last of these bins detectable at $2\sigma$ with $80\%$ probability is $0.14$ in either case.

The width at half depth (Sect.~\ref{sec:prplane}), measured in $R_p'$ using each bin's slope on observed counts, is $0.10^{+0.04}_{-0.02}$, $0.15^{+0.03}_{-0.05}$, $0.13^{+0.02}_{-0.02}$ and $0.12^{+0.01}_{-0.03}$\,dex in the same four bins.
We use observed counts for these measurements because completeness-weighted widths are unreliable in individual age bins (Appendix~\ref{app:morphtables:kde}). 
In the same four bins, the weighted valley depths (Sect.~\ref{sec:global}) are $0.44^{+0.12}_{-0.11}$, $0.37^{+0.11}_{-0.11}$, $0.38^{+0.12}_{-0.11}$ and $0.34^{+0.11}_{-0.13}$.
The intervals overlap across all four bins for both metrics.

\subsection{The valley across host metallicity and colour}\label{sec:composition}

We investigate the dependence of the valley locus on host colour and metallicity by splitting $\tilde{\mathcal{D}}_3$ at $(G_{\rm BP}-G_{\rm RP})_0=0.95$ (F and G versus K and late-G hosts) and $[\mathrm{Fe/H}]=0$ (metal-poor versus metal-rich hosts). As in Sect.~\ref{sec:prplane}, the SVM fits use observed counts. We assess differences between subsets through a host-level bootstrap, refitting both subsets on each common resample and re-deriving their class labels.

Neither the slope nor the position of the valley locus shows a significant dependence on host colour or metallicity. In the full $\tilde{\mathcal{D}}_3$, the F and G hosts have a steeper locus than the K and late-G hosts in $84\%$ of bootstrap replicates, while the metal-poor hosts have a steeper locus than the metal-rich hosts in $56\%$. For comparison, the $2$--$4$\,Gyr hosts have a steeper locus than the $0.1$--$2$\,Gyr hosts in $81\%$ of replicates. Restricted to the $2$--$4$\,Gyr hosts, the corresponding fractions for colour and metallicity are $70\%$ and $36\%$, respectively. In this rotation-selected sample colour is correlated with age (Sect.~\ref{sec:ageshift}).

At $P=10$\,d, the fitted valley lies at $1.65^{+0.15}_{-0.03}\,R_\oplus$ for F and G hosts and $1.71^{+0.04}_{-0.06}\,R_\oplus$ for K and late-G hosts. The corresponding values are $1.72^{+0.09}_{-0.08}\,R_\oplus$ for metal-poor hosts and $1.71^{+0.04}_{-0.07}\,R_\oplus$ for metal-rich hosts. The offsets between the two subsets are $-0.05\pm0.10\,R_\oplus$ (F and G minus K and late G) and $-0.01\pm0.10\,R_\oplus$ (metal-rich minus metal-poor), both consistent with zero. For comparison, the offset between the $1$--$2$ and $2$--$4$\,Gyr subsets is $0.24^{+0.07}_{-0.27}\,R_\oplus$ (Sect.~\ref{sec:ageshift}).

To examine the valley morphology at fixed host colour or metallicity, we further divide each colour and metallicity subset into three age bins: $0.1$--$1.5$, $1.5$--$2.5$ and $2.5$--$4$\,Gyr. These wider bins retain sufficient planets in each cell for the KDE analysis of Sect.~\ref{sec:global} (Table~\ref{tab:strbins}; Figs.~\ref{fig:splitmetstr} and~\ref{fig:splitcolstr}). In the metallicity cells, the Monte Carlo draws also perturb [Fe/H], allowing metallicity-bin membership to vary along with age. Since median orbital periods range from $6.5$ to $14.5$\,d across cells, with no ordering with age common to the colour and metallicity subsets, we measure the valley minimum in both $R_p$ and the period-adjusted radius $R_p'$.

In the $1.5$--$2.5$ and $2.5$--$4$\,Gyr bins, the completeness-weighted valley of F and G hosts lies above that of K and late-G hosts by $+0.26^{+0.15}_{-0.20}$ and $+0.10^{+0.12}_{-0.12}\,R_\oplus$ in $R_p$, respectively. The differences decrease to $+0.11^{+0.23}_{-0.20}$ and $+0.03^{+0.10}_{-0.12}\,R_\oplus$ in $R_p'$. 
The metal-rich minus metal-poor offsets are $+0.13^{+0.19}_{-0.22}$ and $+0.09^{+0.10}_{-0.10}\,R_\oplus$ in $R_p$, but become $-0.04^{+0.25}_{-0.19}$ and $-0.05^{+0.16}_{-0.14}\,R_\oplus$ in $R_p'$.
All offsets are computed draw by draw, with the individual valley positions on $R_p'$ reported in Table~\ref{tab:strbins}. The reduction in the metallicity offsets after period adjustment reflects the shorter median periods of metal-rich hosts ($8.5$ and $8.2$\,d, compared with $11.2$ and $11.5$\,d for metal-poor hosts).

The youngest age bin ($0.1$--$1.5$\,Gyr) is less well constrained. In $R_p'$, the weighted counts resolve the valley in $47\%$ of draws for F and G hosts, $61\%$ for K and late-G hosts, $55\%$ for metal-poor hosts and $74\%$ for metal-rich hosts. Thus, only the metal-rich cell exceeds the two-thirds resolution threshold. On observed counts, only the K and late-G cell passes this threshold ($66.8\%$, compared with $66.3\%$ for the metal-rich cell). By contrast, all cells in the two older age bins resolve the valley on $R_p'$ in more than $80\%$ of draws, both weighted and unweighted (Table~\ref{tab:strbins}).

Within the metal-rich population, where the youngest valley is resolved, its position in $R_p'$ increases by $0.27^{+0.18}_{-0.20}\,R_\oplus$ between $0.1$--$1.5$ and $2.5$--$4$\,Gyr. The tendency for the valley to lie at smaller radii at younger ages therefore persists at fixed metallicity. The same ordering is found in $R_p$ in all four colour and metallicity subsets, although the individual differences are not significant. The corresponding offsets in $R_p$ are reported in Appendix~\ref{app:morphtables:kde}. These results are consistent with the age-dependent shift identified in Sect.~\ref{sec:ageshift}, while providing no compelling evidence for an independent dependence of the valley position on host colour or metallicity.

\section{The super-Earth fraction across age}\label{sec:fse}

\subsection{Model and likelihood}\label{sec:hier}

\paragraph{The population being modelled.}
We infer the age dependence of the intrinsic SE population using a two-component mixture model of planetary radii. It is important to distinguish the mixture weight from the counted SE fraction $f_{\rm SE}^{\rm c}$ of Sect.~\ref{sec:ageshift}. The latter is the fraction of \emph{observed} planets with radii in $[1,4]\,R_\oplus$ that lie below the period-dependent valley locus. By contrast, the intrinsic mixture weight $f_{\rm SE}^{\rm m}$ describes the relative contribution of the SE component before radius-dependent detection and confirmation losses, in the broader $[0.5,4]\,R_\oplus$ analysis range. The two quantities need not agree: the mixture components overlap across the valley, extend outside the counted range, and are selected with different efficiencies. In an evolutionary interpretation, an increase in the intrinsic SE weight is consistent with SNs losing their envelopes, but the mixture alone does not establish the evolutionary history of individual planets.

We model $x=\log_{10}(R_p/R_\oplus)$ as a mixture of two Gaussian components, with means and intrinsic widths $\{\mu_{\rm SE},\sigma_{\rm SE},\mu_{\rm SN},\sigma_{\rm SN}\}$. These are held fixed with age: the evolution is described by changing component weights rather than by shifting the component distributions. This model estimates relative population abundances; the detailed valley morphology is measured independently in Sect.~\ref{sec:morph}. Tests of age-dependent component means and widths are given in Appendix~\ref{app:compl:widths}.

For planet $i$, the intrinsic probability of belonging to the SE component, before using its measured radius, is $w_i$. We allow this weight to depend on age, orbital period, and confirmed-sibling status:
\begin{equation}\label{eq:wlogit}
{\rm logit}\,w_i={\rm logit}\,f_{\rm SE}^{\rm m}(t)
+\beta_P\left(\log_{10}P_i-\log_{10}P_{\rm ref}\right)+\gamma g_i,
\end{equation}
where ${\rm logit}\,w=\ln[w/(1-w)]$, $P_i$ is the orbital period of planet $i$, $P_{\rm ref}=8.7$\,d is the median period of $\mathcal{D}_3$, and $g_i=1$ when another KOI around the same host is confirmed ($0$ otherwise). Thus $f_{\rm SE}^{\rm m}(t)$ is the intrinsic SE weight at age $t$ of a planet at the reference period without a confirmed sibling; $\beta_P$ and $\gamma$ describe period and confirmed-multiplicity dependence. To compare ages without changes in the period and confirmed-sibling distributions, we report, unless otherwise stated, the \emph{standardised SE weight}
\begin{equation}\label{eq:wstd}
\bar f_{\rm SE}^{\rm m}(t)=\frac{1}{N}\sum_{i=1}^{N}w_i(t),
\end{equation}
where all $N$ planets of $\mathcal{D}_3$ are assigned the same age $t$ while retaining their observed periods and sibling statuses.

\paragraph{Representing age dependence.}
We use two complementary forms for $f_{\rm SE}^{\rm m}(t)$. The \emph{three-interval model} assigns independent constant weights $f_{<1}$, $f_{1\text{--}2}$, and $f_{2\text{--}4}$ to the intervals $0.1$--$1$, $1$--$2$, and $2$--$4$\,Gyr. Its primary measure of evolution is the difference between the corresponding standardised weights, $\bar f_{2\text{--}4}-\bar f_{<1}$.

The second form estimates the timescale of an approach to a plateau:
\begin{equation}\label{eq:satexp}
f_{\rm SE}^{\rm m}(t)=f_\infty-(f_\infty-f_0)e^{-t/\tau_{\rm E}},
\end{equation}
where $f_0$ is the extrapolated initial weight, $f_\infty$ the asymptotic weight, and $\tau_{\rm E}$ the e-folding timescale. For a rising solution ($f_0<f_\infty$), $63\%$ of the increase from $f_0$ to $f_\infty$ is complete at $t=\tau_{\rm E}$. We constrain $f_\infty$ with a Gaussian prior based on the $2$--$4$\,Gyr interval of the first fit. This assumes that the plateau has been reached within that interval and reuses its data in the prior and likelihood; mock-catalogue tests show that the resulting timescale bound retains approximately its nominal coverage (Appendix~\ref{app:compl:injrec}). We define $t_{90}$ as the age at which the \emph{standardised} weight reaches $90\%$ of its asymptotic value, and compute its bounds only for posterior draws starting below that threshold. The corresponding quantity with a flat prior on $f_\infty$ is denoted $t_{90}^{f}$ (Appendix~\ref{app:compl:posterior}). Since the youngest host is approximately $0.3$\,Gyr old, $f_0$ is an extrapolation and cannot distinguish primordial SEs from planets stripped before the sampled ages.

\paragraph{Selection and the observed population.}
The smaller radii of SEs make them less likely to enter the observed sample. We therefore condition the likelihood on detection and confirmation rather than correcting the observed counts by inverse-completeness weights. The selection function $S_i(x')$ of Sect.~\ref{sec:compl} gives the probability of selecting a planet of true log-radius $x'$ at the host and orbital configuration of planet $i$. For component ${\rm k}\in\{{\rm SE},{\rm SN}\}$, its selection factor at the measured radius $x_i$ is
\begin{equation}\label{eq:seldef}
\epsilon_{{\rm k},i}=\int S_i(x')\,p(x'\mid x_i,{\rm k})\,\mathrm{d}x',
\end{equation}
where $p(x'\mid x_i,{\rm k})$ combines the intrinsic Gaussian component with the measurement likelihood. The fraction of each component admitted by the selection and measured-radius limits is
\begin{equation}\label{eq:zk}
Z_{{\rm k},i}=\int S_i(x')\,W_i(x')\,
\mathcal{N}(x';\mu_{\rm k},\sigma_{\rm k}^2)\,\mathrm{d}x',
\end{equation}
where $W_i(x')$ is the probability that a planet with true radius $x'$ is measured in $[0.5,4]\,R_\oplus$ (Appendix~\ref{app:compl:probit}). At an age $t_i^{(r)}$ drawn from the age uncertainty of planet $i$, the total selected fraction is
\begin{equation}\label{eq:znorm}
Z_i^{(r)}=w_i(t_i^{(r)})Z_{{\rm SE},i}
+\big[1-w_i(t_i^{(r)})\big]Z_{{\rm SN},i}.
\end{equation}
This normalisation removes factors that multiply selection probability independently of radius at fixed host and orbit, but not radius-dependent selection, which is modelled by $S_i$, or age-dependent changes in the host population. These changes are assessed in Appendix~\ref{app:resgate}; a possible radius-dependent detection ceiling is tested in Appendix~\ref{app:compl:probit}.

\paragraph{Propagating uncertainties.}
We use an unbinned likelihood that convolves each intrinsic component with the measured radius uncertainty and marginalises over the uncertainty in stellar age. Writing $x_i=\log_{10}(R_{p,i}/R_\oplus)$ and approximating its error by a Gaussian of width $\sigma_{x,i}$, the contribution of planet $i$ is
\begin{equation}\label{eq:like}
\begin{split}
\mathcal{L}_i=\frac{1}{N_{\rm d}}\sum_{r=1}^{N_{\rm d}}\frac{1}{Z_i^{(r)}}\Big[&w_i(t_i^{(r)})\,\epsilon_{{\rm SE},i}\,
\mathcal{N}(x_i;\mu_{\rm SE},\sigma_{\rm SE}^2+\sigma_{x,i}^2)\\
&+\big(1-w_i(t_i^{(r)})\big)\,\epsilon_{{\rm SN},i}\,
\mathcal{N}(x_i;\mu_{\rm SN},\sigma_{\rm SN}^2+\sigma_{x,i}^2)\Big],
\end{split}
\end{equation}
where $\mathcal{N}(x;\mu,\sigma^2)$ is a normal density. We use $N_{\rm d}=200$ draws $t_i^{(r)}$ from the asymmetric age distributions of Paper~I; for the three-interval model, each draw contributes to its corresponding interval. The intrinsic widths $\sigma_{\rm k}$ thus exclude the measurement scatter represented by $\sigma_{x,i}$. The age-dependent confirmation threshold in $S_i$ is evaluated at the Paper~I point age rather than at each draw, so the component selection factors remain fixed for each planet at fixed model parameters. The precise error approximation, integration, and numerical checks are documented in Appendix~\ref{app:compl:probit}.

We fit the model to $\mathcal{D}_3$ using \texttt{emcee} \citep{ForemanMackey2013}. The confirmation-age coefficient $\mathcal{H}$ is marginalised using its likelihood profile from the DR25 candidate calibration as a prior (Appendices~\ref{app:compl:probit} and \ref{app:compl:posterior}). Tests on mock catalogues are reported in Appendix~\ref{app:compl:injrec}.

\subsection{The global age trend}\label{sec:resglobal}

In the three-interval model, the standardised SE weights are $\bar f_{<1}=0.43^{+0.10}_{-0.09}$, $\bar f_{1\text{--}2}=0.75\pm0.06$, and $\bar f_{2\text{--}4}=0.67\pm0.04$, constrained by $59$, $128$, and $394$ planets whose estimated host ages fall within the respective intervals. The age contrast, $\bar f_{2\text{--}4}-\bar f_{<1} =+0.24^{+0.10}_{-0.10}$, is positive in $99.5\%$ of the posterior draws, indicating that the intrinsic SE fraction increases with age. Some of the inferred SE population lies below $1.15\,R_\oplus$, the median radius at which the selection efficiency reaches half its ceiling. In this low-completeness regime, the inferred weight depends partly on the assumed Gaussian shape of the SE component. Nevertheless, restricting the intrinsic population to true radii of $1.2$--$4\,R_\oplus$, where selection is more efficient, still yields an increase in the standardised SE fraction from $0.27^{+0.08}_{-0.07}$ below $1$\,Gyr to $0.51\pm0.05$ at $2$--$4$\,Gyr. There is no evidence for a further increase beyond $1$--$2$\,Gyr: $\bar f_{1\text{--}2}$ exceeds $\bar f_{2\text{--}4}$ by $0.07\pm0.07$.

We estimate the timescale of this increase using the exponential model of Eq.~\eqref{eq:satexp}, with the plateau constrained by the $2$--$4$\,Gyr population. The inferred standardised plateau is $\bar f_\infty=0.69\pm0.04$. Among the $88\%$ of posterior draws in which the initial SE weight lies below $0.9\,\bar f_\infty$, the age $t_{90}$ at which the standardised weight reaches $90\%$ of its plateau satisfies $t_{90}<1.9$\,Gyr at $90\%$ probability and $t_{90}<2.6$\,Gyr at $95\%$ probability. Allowing the plateau to vary freely weakens the constraint to $t_{90}^{f}<3.8$\,Gyr (Appendix~\ref{app:compl:posterior}). These limits constrain how late the increase can be completed, rather than establishing a minimum timescale: with only seven planets younger than $0.5$\,Gyr, a substantially earlier rise cannot be excluded.

The initial weight $f_0$ is less directly constrained. The youngest host is $0.3$\,Gyr old, so $f_0$ is obtained entirely by extrapolating the assumed exponential form to $t=0$. By contrast, $t_{90}$ lies within the sampled age range and is only weakly correlated with $f_0$ in the posterior (Pearson coefficient $+0.08$). The exponential model should be regarded as an empirical parametrisation of the observed rise, rather than a physical law of atmospheric loss. More flexible age dependences, or age-dependent means and widths of the two Gaussian components, are not justified by the current sample, particularly by the limited number of planets younger than $1$\,Gyr. Our analysis demonstrates the constraints already achievable with an age-resolved, per-planet likelihood, while larger samples from future PLATO releases should enable more detailed and physically motivated models \citep{Rauer2024}.

Finally, the mixture weights $f_{\rm SE}^{\rm m}$ and $\bar f_{\rm SE}^{\rm m}$ quantify membership of the SE component, not the fraction of planets that have completely lost their envelopes. Some SEs may have formed without substantial envelopes, so $\bar f_\infty$ provides only an upper bound on the fraction that underwent envelope loss by the time the SE fraction reaches its plateau. Moreover, a planet undergoing atmospheric loss can become predominantly associated with the SE component once its radius falls below the crossing point of the weighted component distributions, even while retaining a residual envelope. Consequently, the SE fraction may reach its plateau before the SE radius distribution has completed its evolution (Sect.~\ref{sec:ageshift}).

\section{Demographic trends across host properties}
\label{sec:results}

We examine whether the relative abundance and characteristic radii of SEs and SNs depend on stellar mass and metallicity. These are distinct questions: host properties may affect the probability that a planet retains an envelope without changing the typical radius of either population, or may alter planet sizes without appreciably changing their relative numbers. We address both possibilities using the selection-conditioned mixture model of Sect.~\ref{sec:hier}, supplemented by a completeness-weighted comparison of planets classified relative to the radius valley.

For each component ${\rm k}\in\{{\rm SE},{\rm SN}\}$, we allow the mean log-radius to depend on host mass $M_{\star,i}$, metallicity $[\mathrm{Fe/H}]_i$, and orbital period $P_i$:
\begin{equation}\label{eq:mumass}
\begin{split}
\mu_{{\rm k},i}={}&\mu_{\rm k}
+\kappa^{M}_{\rm k}\,(\log_{10}M_{\star,i}-\log_{10}M_0)\\
&+\kappa^{\rm Fe}_{\rm k}\,([\mathrm{Fe/H}]_i-[\mathrm{Fe/H}]_0)\\
&+\kappa^{P}_{\rm k}\,(\log_{10}P_i-\log_{10}P_{\rm ref}),
\end{split}
\end{equation}
where $M_0$, $[\mathrm{Fe/H}]_0$ and $P_{\rm ref}$ are the medians over $\mathcal{D}_3$. We also allow the SE mixture weight of Eq.~\eqref{eq:wlogit} to depend on mass and metallicity, through the terms $\beta_M(\log_{10}M_{\star,i}-\log_{10}M_0)$ and $\beta_{\rm Fe}([\mathrm{Fe/H}]_i-[\mathrm{Fe/H}]_0)$ added to ${\rm logit}\,w_i$. The model is fitted to all $581$ planets of $\mathcal{D}_3$, using three age intervals and component widths independent of host properties. Host-parameter and planetary-radius uncertainties are propagated jointly; the treatment of correlated errors, priors, and alternative model specifications is described in Appendix~\ref{app:hostdemog}.

\subsection{Host metallicity and the super-Earth fraction}\label{sec:resfeh}

We find no significant dependence of the SE fraction on host metallicity. This result is relevant because more metal-rich envelopes may cool or lose mass differently, potentially affecting the survival of SN atmospheres \citep{OwenMurrayClay2018,GuptaSchlichting2020}.

We first compare the completeness-weighted fractions of planets classified as SEs and SNs relative to the period-dependent valley locus (Eq.~\eqref{eq:slope}). We use $\tilde{\mathcal{D}}_3$ and exclude $13$ planets whose completeness falls below $0.25$ in either metallicity subset, leaving $198$ planets around metal-poor hosts and $330$ around metal-rich hosts, split at $[\mathrm{Fe/H}]=0$. Because the subsets differ in stellar mass, colour, apparent magnitude, and orbital-period distribution, their raw SE fractions cannot be compared directly. We adjust for these variables and age using an augmented inverse-probability weighting (AIPW) estimator, with completeness weights and host-level bootstrap uncertainties (Appendix~\ref{app:hostdemog}).

The adjusted difference between metal-rich and metal-poor hosts is
$\hat\delta_{\rm Fe}=+0.03^{+0.12}_{-0.11}$, consistent with zero ($0.3\sigma$). Alternative classification boundaries, completeness weights, control-variable choices, and metallicity splits yield differences between $-0.02$ and $+0.10$, within the quoted uncertainty. Thus, the data do not reveal a metallicity dependence of the fraction classified below the valley.

The mixture model provides a test that does not assign planets to classes with a fixed valley boundary. Its metallicity coefficient for the intrinsic SE weight is $\beta_{\rm Fe}=-1.79^{+1.96}_{-1.92}$ per dex in ${\rm logit}\,w_i$, at fixed host mass and period, with only a marginal improvement relative to zero ($2\Delta\ln\mathcal{L}=0.8$). Across the $0.24$\,dex difference in mean metallicity between metal-rich and metal-poor hosts of $\mathcal{D}_3$, this corresponds to $-0.07^{+0.08}_{-0.07}$ in the standardised intrinsic SE weight. Although this quantity differs from the counted fraction, both analyses are consistent with no metallicity effect.

\subsection{Planet size across host mass and metallicity}\label{sec:resmass}

The characteristic SN radius increases with host-star mass, whereas no significant mass dependence is detected for SEs. Metal-rich hosts may also harbour larger SNs, because their planets may have more massive cores, which accrete more substantial envelopes \citep{OwenMurrayClay2018}, while higher envelope opacity can slow their contraction \citep{GuptaSchlichting2020}. Stellar mass and metallicity are correlated in our sample (Spearman $\rho=+0.35$ for the $375$ hosts retained in the completeness-weighted comparison), so we fit their contributions jointly rather than attributing the radius trend to either property in isolation.

For SNs, the joint mass--metallicity dependence is supported relative to a model with both slopes fixed to zero ($2\Delta\ln\mathcal{L}=29.5$; $\Delta\mathrm{BIC}=16.8$ in favour of the more complex model). We obtain $\kappa^{M}_{\rm SN}=+0.35\pm0.16$ and $\kappa^{\rm Fe}_{\rm SN}=+0.15\pm0.08$ per dex. 
Their posterior correlation of $-0.72$ indicates substantial degeneracy: fixing only the mass slope to zero worsens the fit by $2\Delta\ln\mathcal{L}=5.5$, and fixing only the metallicity slope by $4.2$, against $29.5$ for both. The individual slopes should therefore be interpreted with more caution than their joint significance.
For SEs, neither dependence is significant: $\kappa^{M}_{\rm SE}=-0.33^{+0.43}_{-0.40}$ and $\kappa^{\rm Fe}_{\rm SE}=+0.15^{+0.21}_{-0.24}$, with $2\Delta\ln\mathcal{L}=0.8$ for the two slopes jointly.

\begin{figure}
\centering
\includegraphics[width=\linewidth]{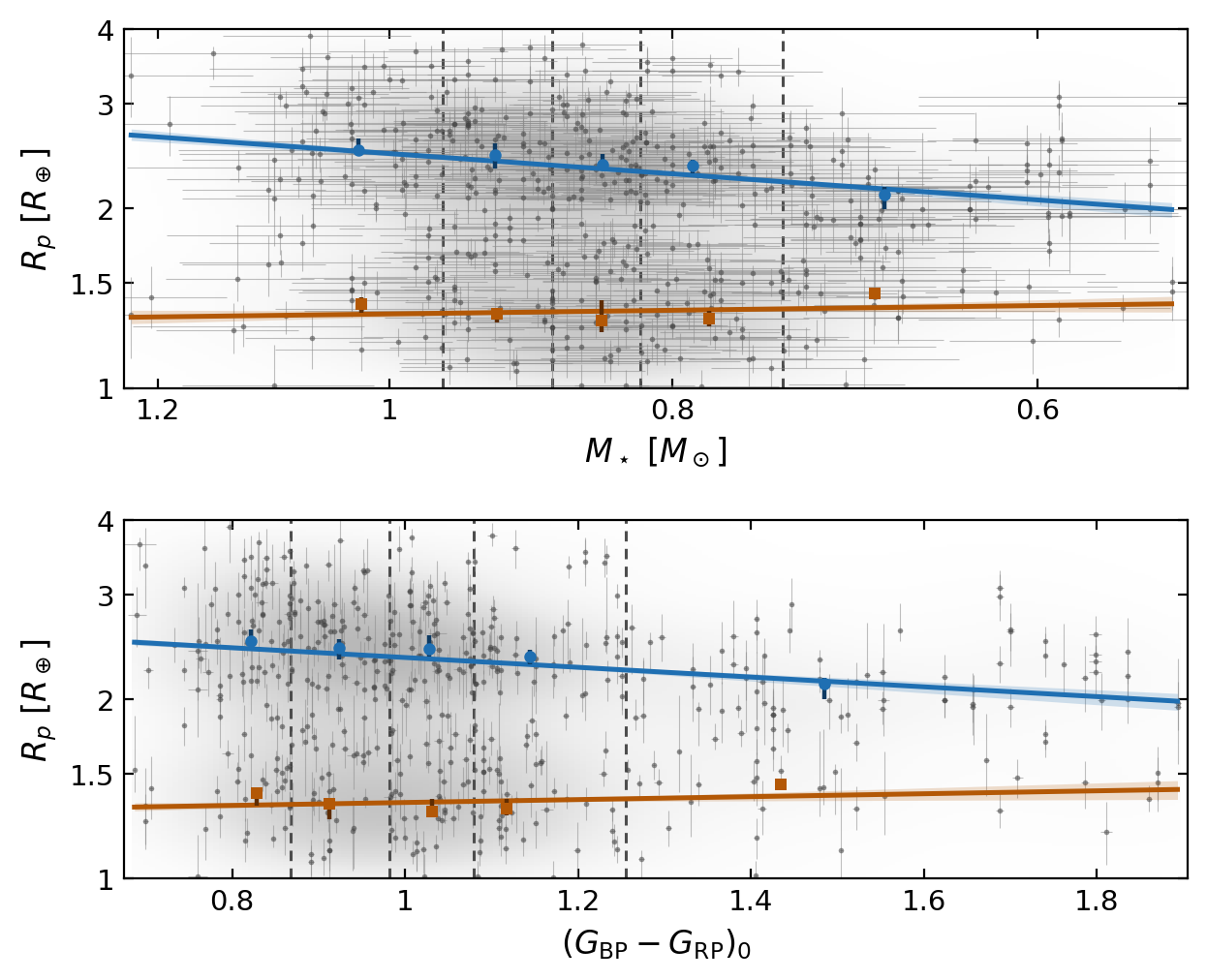}
\caption{Planet radii versus host mass (\textit{top}) and de-reddened \textit{Gaia} colour (\textit{bottom}) in $\tilde{\mathcal{D}}_3$. The mass axis decreases to the right, so both panels progress from bluer to redder hosts. Blue circles and orange squares show the completeness-weighted median radii of SNs ($1.71$--$4\,R_\oplus$) and SEs ($1$--$1.71\,R_\oplus$), respectively, in five equal-count bins shared by the two classes. Markers are placed at the median abscissa of each bin, with $16$--$84\%$ bootstrap intervals; dashed vertical lines indicate bin boundaries. Solid lines and shaded bands show the within-class regressions of Sect.~\ref{sec:resmass} and their $16$--$84\%$ intervals. Grey points represent individual planets and their uncertainties, with colour errors generally smaller than the symbols (median $0.003$\,mag). The grey background shows the two-dimensional sample density.}
\label{fig:massdecomp}
\end{figure}

Replacing host mass with de-reddened \textit{Gaia} colour gives $\kappa^{\rm col}_{\rm SN}=-0.12\pm0.02$ and $\kappa^{\rm col}_{\rm SE}=+0.05\pm0.08$ per mag: SNs are smaller around redder, generally lower-mass hosts, while SE sizes show no significant colour dependence. The completeness-weighted medians and direct within-class regressions in Fig.~\ref{fig:massdecomp} support the same qualitative result. These regressions yield host-mass slopes of $+0.35\pm0.06$ for SNs and $-0.06\pm0.07$ for SEs. 
 Their amplitudes should not be compared directly with the intrinsic mixture slopes: the regressions omit metallicity (the mixture fit without metallicity terms gives $\kappa^{M}_{\rm SN}=+0.58\pm0.11$), and their coefficients can be attenuated by classification at a fixed radius and by measurement errors in host mass (Appendix~\ref{app:hostdemog}).

The increasing SN radius with host mass is consistent with more massive SN cores around more massive hosts, since larger cores can accrete and retain more substantial envelopes \citep{Ginzburg2016,OwenMurrayClay2018}. Its implications for atmospheric loss depend on the initial core population. Under photoevaporation, XUV-driven loss varies comparatively weakly with host mass at fixed period, so a host-dependent core-mass distribution could produce larger SNs while leaving the characteristic SE radius approximately unchanged \citep{Petigura2022}. Core-powered mass loss instead depends on bolometric irradiation, which increases with host mass at fixed period; this allows more massive cores to be stripped and shifts the SE population towards larger radii, even without changes to the underlying planet population \citep{GuptaSchlichting2020}.

The SE mass slope is positive in only $22\%$ of posterior draws, but its uncertainty is too large to discriminate decisively between these interpretations. A nearly constant SE radius together with increasing SN radius would be compatible with photoevaporation acting on a host-dependent core-mass distribution. Larger SEs around more massive hosts, by contrast, would not single out core-powered mass loss, since both mechanisms produce them if the entire core-mass distribution shifts with host mass \citep{Wu2019,GuptaSchlichting2020}. Stronger constraints on the SE radius--host-mass relation are needed.

\section{Discussion}\label{sec:disc}

\subsection{The emergence and evolution of the radius valley}

Gyrochronological ages allow us to trace the development of the radius valley over the first $4$\,Gyr of planetary evolution. In the completeness-characterised sample of $541$ planets around $381$ hosts ($\tilde{\mathcal{D}}_3$; Sect.~\ref{sec:global}), the valley becomes identifiable at $0.75$--$1.4$\,Gyr and is consistently resolved from $1.4$\,Gyr onwards. Its depth, width, and occupancy show no significant subsequent evolution (Sect.~\ref{sec:ageshift}). The data therefore indicate that the valley has acquired its mature morphology by approximately $2$\,Gyr, although the limited number of younger planets prevents us from establishing when it first appeared.

Previous studies have also investigated the age dependence of the radius valley, although with partly differing results. \citet{David2021} found the valley nearly empty among systems younger than $\sim2$--$3$\,Gyr and progressively filled at older ages, with its location shifted towards smaller radii compared with older populations. \citet{Chen2022} instead reported an increasing valley prominence with age, while \citet{Kamulali2026} found a shallower valley at older ages. More recently, \citet{Chen2026} suggested that the valley is not fully established below $\sim3$\,Gyr and that its subsequent evolution is non-monotonic, partly reflecting the changing mixture of Galactic stellar populations. Our results show a clearly resolved valley from $1.4$\,Gyr onwards, with no statistically significant evolution in its depth, width, or occupancy up to $4$\,Gyr. Differences in stellar-age determinations, sample selection, age binning, and the metrics used to characterise the valley may contribute to these apparently discrepant findings.

There is tentative evidence that the valley minimum shifts from approximately $1.5\,R_\oplus$ before $\sim2$\,Gyr to $1.8\,R_\oplus$ at older ages. Although the displacement is of modest statistical significance, its direction is consistent across colour and metallicity subsets. Differences in host colour and metallicity, as well as dilution by unresolved companions, do not account for the observed shift (Sects.~\ref{sec:ageshift} and~\ref{sec:composition}; Appendices~\ref{app:colourmatch} and~\ref{app:dilution}). The younger SE radius distribution also peaks at smaller radii and is more strongly skewed towards the class boundary, whereas the older distribution is approximately symmetric. The median SE radius, however, does not change significantly.

These tentative morphological differences may reflect the final stages of atmospheric loss. In both photoevaporation and core-powered mass-loss models, planets with more massive cores generally retain their envelopes longer at fixed orbital period \citep{OwenWu2017,GuptaSchlichting2020}. If such planets cross the valley later, they would replenish the upper-radius part of the SE population, while planets still retaining envelopes would contribute to the distribution above the valley. This could produce a smaller valley radius and a more asymmetric SE distribution at younger ages, followed by a shift towards larger radii as the transitions are completed.

This interpretation is consistent with the scarcity of SEs of $1.5$--$1.8\,R_\oplus$ at ages below $\sim2$--$3$\,Gyr reported by \citet{David2021}. However, neither the displacement of the valley minimum nor the change in the SE radius distribution is individually significant. They should therefore be regarded as possible signatures of late atmospheric loss rather than direct evidence of planets crossing the valley.

\subsection{The timescale of super-Earth population evolution}

The strongest evidence for population evolution comes from the intrinsic SE fraction. The standardised mixture weight $\bar f_{\rm SE}^{\rm m}$ (Eq.~\eqref{eq:wstd}) increases by $0.24^{+0.10}_{-0.10}$ between ages below $1$\,Gyr and $2$--$4$\,Gyr (Sect.~\ref{sec:resglobal}). The increase is concentrated within the first $\sim2$\,Gyr, with no significant subsequent trend. In the exponential approach-to-plateau model, the age $t_{90}$ at which the SE weight reaches $90\%$ of its asymptotic value is below $1.9$\,Gyr with $90\%$ posterior probability. Alternative sample selections give compatible results (Appendix~\ref{app:altsamples}).

The limited number of planets younger than $0.5$\,Gyr prevents us from resolving the earliest evolution. In particular, the SEs already present around the youngest hosts, at approximately $0.3$\,Gyr, may have formed without substantial envelopes or lost them before the ages accessible to this study. The inferred timescale therefore constrains the evolution of the population rather than the envelope-loss history of individual planets.

Nevertheless, the increase in the SE fraction extending beyond the first Gyr is difficult to explain solely through processes confined to the earliest evolutionary stages, such as photoevaporation restricted to the saturated phase of stellar XUV emission \citep{OwenWu2017,GuptaSchlichting2020}. Interpreted as the conversion of SNs into SEs, the observed trend requires that some planets are subject to mass loss beyond that very early phase. Both core-powered mass loss \citep{GuptaSchlichting2020} and photoevaporation sustained by long-lived stellar EUV emission \citep{KingWheatley2021} can produce extended evolution.

The approach of the SE fraction to a plateau by approximately $2$\,Gyr, together with the absence of significant subsequent changes in valley morphology, suggests that any further evolution affects a progressively smaller fraction of planets. The tentative shift of the valley minimum towards larger radii may trace the last planets to lose their envelopes, potentially those with the most massive cores that can still be stripped.

\subsection{Constraints from host-star properties}

Host-star properties provide complementary constraints on the origin of the two planetary populations. We find no significant dependence of the SE fraction on host metallicity (Sect.~\ref{sec:resfeh}). At fixed orbital period and metallicity, SN radii increase with stellar mass, whereas SE radii show no significant mass dependence. SN radii also tend to increase with host metallicity at fixed stellar mass, although the two effects are partially degenerate. No significant dependence of SE radius on metallicity or colour is detected (Sect.~\ref{sec:resmass}).

The different mass dependences of the two classes may help distinguish atmospheric-loss mechanisms. In photoevaporation models, the relatively weak dependence of integrated XUV irradiation on host mass at fixed period allows an approximately constant SE size, while larger SNs around more massive stars may arise from a primordial core-mass distribution extending to higher masses \citep{Petigura2022,OwenMurrayClay2018}. Core-powered mass loss, by contrast, responds to the stronger bolometric irradiation of planets around more massive stars and also shifts the upper boundary of the SE population towards larger radii (\citealt{GuptaSchlichting2020}; Sect.~\ref{sec:resmass}).

The observed combination of increasing SN size and approximately constant SE size is therefore compatible with photoevaporation acting on a host-mass-dependent core population. However, the SE mass slope remains insufficiently constrained, and the predictions of the two mechanisms depend on the underlying core-mass distribution. The present data consequently do not distinguish photoevaporation from core-powered mass loss decisively.

\section{Conclusions}\label{sec:conclusions}

Using gyrochronological ages for FGK exoplanet hosts, we have investigated the evolution of the radius valley and the relative populations of SEs and SNs over $0.1$--$4$\,Gyr. Our principal findings are:

\begin{enumerate}
\item The intrinsic SE fraction increases predominantly during the first $\sim2$\,Gyr, with no significant subsequent trend. Its increase between ages below $1$\,Gyr and $2$--$4$\,Gyr is $0.24^{+0.10}_{-0.10}$. An exponential model places $90\%$ of the approach to its asymptotic fraction within $1.9$\,Gyr at $90\%$ credibility.

\item The radius valley is consistently resolved from approximately $1.4$\,Gyr onwards, with no significant subsequent change in depth, width, or occupancy. A possible shift of its minimum from $\sim1.5$ to $\sim1.8\,R_\oplus$, accompanied by a change in the shape of the SE radius distribution, remains tentative.

\item SN radii increase with host-star mass at fixed orbital period and metallicity, whereas SE radii show no significant corresponding dependence. We find no significant dependence of the SE fraction on host metallicity.
\end{enumerate}

The continuing increase in the intrinsic SE fraction is consistent with atmospheric loss remaining effective for part of the planetary population up to approximately $1$--$2$\,Gyr. Both core-powered mass loss and long-lived EUV-driven photoevaporation remain viable explanations, potentially acting together. The tentative morphological evolution of the valley may reflect the final envelope-loss transitions of planets with the most massive cores that can still be stripped, but requires confirmation.

The principal observational limitation is the small number of planets younger than $1$\,Gyr, which restricts our ability to determine when the valley first becomes established and how rapidly the SE population evolves at early ages. Larger age-resolved samples from PLATO \citep{Rauer2024} should improve these constraints. The dependence of the evolution identified here on planetary irradiation will be investigated in the next paper of this series \citep[][Paper~III]{PaperIII}.


\begin{acknowledgements}
This work was partially supported by the European Union -- NextGenerationEU, under the Italian Ministry of University and Research (MUR) National Recovery and Resilience Plan (PNRR), Mission 4, Component 2, Investment 1.1, PRIN 2022, project No. 2022HY2NSX, ``CHRONOS: adjusting the clock(s) to unveil the CHRONO-chemo-dynamical Structure of the Galaxy'' and the European Union's Horizon 2020 research and innovation programme under Grant Agreement No. 824135 (SOLARNET -- Integrating High Resolution Solar Physics).
\end{acknowledgements}

\bibliographystyle{aa}
\bibliography{references}

\appendix
\section{Planetary parameter selection}\label{app:planetparams}

This appendix describes the selection of planetary parameters, complementing the stellar-parameter procedure of Paper~I. We derive the primary quantities---radius, mass, orbital period, and eccentricity---from the multi-source measurements compiled in our database. Dependent quantities, including semi-major axis, insolation, density, and equilibrium temperature, are then computed consistently from the adopted planetary parameters and the stellar parameters of Paper~I. Each selected or derived quantity is assigned an integer reliability tier, with tier~1 indicating the highest reliability.

\subsection{Radius}\label{app:planetparams:radius}

For planets whose hosts are included in Paper~I, we recompute the planetary radius from the published planet-to-star radius ratio and the adopted stellar radius, $R_{\star,\mathrm{I}}$, as
$R_p=(R_p/R_\star)R_{\star,\mathrm{I}}$, propagating the uncertainties of both quantities.

We first consolidate the published radius ratios to one measurement per source, removing duplicate quotations and superseded catalogue editions. Only measurements with a quoted relative uncertainty of at most $60\%$ enter the selection, unless none satisfies this condition. The admitted measurements are then grouped into clusters, with values differing by less than $15\%$ from a neighbouring measurement assigned to the same cluster. Each cluster is scored according to
\[
Q=\sum_s\frac{1}{\sigma_{{\rm rel},s}},
\]
where the sum runs over the independent sources contributing to the cluster and $\sigma_{{\rm rel},s}$ is the relative uncertainty of the most precise measurement contributed by source $s$, subject to a floor of $3\%$. The linear dependence on inverse uncertainty, rather than inverse variance, limits the influence of potentially overoptimistic precision estimates. We adopt the median of the highest-scoring cluster, with an uncertainty equal to the larger of the median quoted uncertainty and the scatter within that cluster.

The selected radius ratio receives reliability tier~1 when at least two independent sources in the winning cluster agree within $5\%$. Tier~2 is assigned when the agreement is less stringent or when a single source provides a measurement with relative uncertainty below $15\%$ ($25\%$ if it is the only available measurement). Tier~3 identifies less secure determinations, including a single imprecise measurement, a winning cluster spanning more than $30\%$, a competing cluster with a score at least $60\%$ of the winning score, or a single-source winning cluster contradicted by two or more independent sources. In the latter two cases, the adopted uncertainty is increased to at least half the separation between the competing clusters.

The reliability tier of the derived $R_p$ is the worse of those assigned to $R_p/R_\star$ and $R_{\star,\mathrm{I}}$. The latter may reach tier~4 when Paper~I identifies an unreliable photometric stellar radius. Grazing transits ($b>0.9$) further limit the planetary-radius reliability to tier~3 or worse. We adopt the recomputed radii for all $4651$ planets with both a selected radius ratio and an available $R_{\star,\mathrm{I}}$.

For completeness, the catalogue also provides radii for $91$ planets lacking either of these quantities. Their radii are selected directly from published $R_p$ measurements using the same clustering procedure. Because these values are not computed from a selected radius ratio and the Paper~I stellar radius, they are assigned tier~3 or worse, explicitly flagged, and excluded from all samples analysed in this work.

\subsection{Mass, period and other parameters}\label{app:planetparams:other}

Planetary masses are selected using the same clustering procedure as the radius ratios, but with thresholds adapted to their greater observational dispersion. Specifically, the clustering tolerance is increased from $15\%$ to $20\%$, the agreement threshold from $5\%$ to $12\%$, and the maximum admitted relative uncertainty from $60\%$ to $80\%$. For non-transiting planets, measurements of $M\sin i$ are treated as lower limits and superseded whenever a true-mass determination is available. Masses reported only as limits are assigned tier~4, as are any corresponding density limits. Planets whose only mass estimates come from empirical mass--radius relations also receive tier~4, but no density is computed to avoid a circular inference.

Orbital periods are selected using the radius-ratio procedure, with the agreement threshold tightened to $1\%$. Owing to their generally high precision, most published periods form a single cluster, whose median is adopted. Discrepant clusters typically arise from harmonic aliases or erroneous measurements accompanied by unrealistically small quoted uncertainties. The $3\%$ uncertainty floor in the cluster score prevents such measurements from dominating the selection. A competing cluster supported by only one source does not increase the adopted uncertainty; instead, the winning cluster is retained with a reliability no better than tier~2. Unlike for the radius ratios, a single-source winning cluster opposed by two or more sources is not demoted to tier~3, since such rival clusters are usually aliases. Ties are resolved in favour of the more recent source.

Eccentricity requires a different treatment. Approximately $74\%$ of the published eccentricities in the database are exactly zero, usually without uncertainties, indicating that circular orbits were assumed rather than measured. Such entries are excluded from comparisons with measured eccentricities and adopted at tier~3 only when no measurement is available.
Among genuine eccentricity measurements, we initially select the most precise. It is retained if a majority of the other measurements agree with it within
$2(\sigma_c^2+\sigma_j^2)^{1/2}$, or within $0.02$ if this is larger, where $\sigma_c$ and $\sigma_j$ are the uncertainties of the candidate and comparison measurements, respectively. Otherwise, the candidate is replaced by the next most precise measurement only if at least two discordant measurements agree with one another and thus support a coherent alternative. If neither a majority nor a coherent alternative exists, we adopt the most recent measurement.
Following \citet{LucySweeney1971}, an adopted eccentricity satisfying $e\leq2.45\sigma_e$ is not regarded as a significant detection. It is replaced by the upper limit $e+2.45\sigma_e$, capped at unity, unless a tighter published limit is available. Eccentricity limits are assigned tier~4.

The impact parameter is selected as the median of the measurements reported by the sources contributing to the adopted radius-ratio cluster. If none of these sources provides an impact parameter, we use the median of the remaining published values and flag the result.

Finally, we recompute the semi-major axis, insolation, planetary density, and equilibrium temperature (assuming zero Bond albedo) from the adopted planetary quantities and the stellar parameters of Paper~I. Insolation and equilibrium temperature are calculated assuming circular orbits. Implausible results, namely densities exceeding $100$\,g\,cm$^{-3}$ or equilibrium temperatures outside $10$--$10^4$\,K, are left undefined. Each derived quantity inherits the least reliable tier among its input parameters.


\section{Valley metrics and their estimators}\label{app:morphtables}

\begin{table*}[!t]\centering
\caption{Valley metrics of the full samples in the $[1,4]\,R_\oplus$ range.
$\tilde{\mathcal{D}}_3$ is given on the observed counts (obs) and with each planet weighted by $1/\bar S_i$ ($w$; Sect.~\ref{sec:compl}), and in the row $w$ the column $f_{\rm SE}^{\rm c}$ gives the weighted fraction $f_{\rm SE}^{{\rm c},w}$. The multi-survey samples $\tilde{\mathcal{D}}_1$ and $\tilde{\mathcal{D}}_2$ have no per-star completeness and are given on the observed counts only. The first block of columns is measured on the radius distribution, and gives the KDE peak positions and valley minimum, the valley depth ($1-\rho_{\rm valley}/\overline{\rho}_{\rm peak}$), the valley width at half depth on $R_p'$ (Sect.~\ref{sec:prplane}), the counted SE fraction (Sect.~\ref{sec:ageshift}), and the occupancy of the fixed $\pm0.2\,R_\oplus$ band around the adopted locus (Sect.~\ref{sec:global}). 
The second block is measured in the period--radius plane with the SVM, and gives the valley slope and the valley position at $P=10$\,d (Sect.~\ref{sec:prplane}). The age-resolved metrics are in Table~\ref{tab:gapmetrics_agebins}. All uncertainties are the $16$th and $84$th percentiles (Appendix~\ref{app:morphtables:kde}).}
\label{tab:gapmetrics}
\footnotesize
\setlength{\tabcolsep}{3pt}\renewcommand{\arraystretch}{1.3}
\begin{tabular}{lr|cccccccc|cc}
\hline\hline
& & \multicolumn{8}{c|}{radius distribution} & \multicolumn{2}{c}{period--radius plane}\\
Sample & $N$ & $R_{\rm SE}^{\rm pk}$ & $R_{\rm SN}^{\rm pk}$ & valley & depth & width & $f_{\rm SE}^{\rm c}$\tablefootmark{a} & occ.\tablefootmark{b} & res.\tablefootmark{c} & slope & valley ($10$\,d)\\
 & & [$R_\oplus$] & [$R_\oplus$] & [$R_\oplus$] & & [dex] & & & [\%] & & [$R_\oplus$]\\
\hline
$\tilde{\mathcal{D}}_1$ (obs) & 2489 & $1.52^{+0.02}_{-0.02}$ & $2.45^{+0.03}_{-0.02}$ & $1.78^{+0.02}_{-0.02}$ & $0.31^{+0.02}_{-0.02}$ & $0.075^{+0.008}_{-0.005}$ & $0.33^{+0.01}_{-0.01}$ & $0.113^{+0.006}_{-0.006}$ & $88$ & $-0.055^{+0.005}_{-0.010}$\tablefootmark{d} & $1.81^{+0.02}_{-0.02}$\\
$\tilde{\mathcal{D}}_2$ (obs) & 706 & $1.45^{+0.05}_{-0.04}$ & $2.38^{+0.03}_{-0.05}$ & $1.72^{+0.05}_{-0.05}$ & $0.27^{+0.04}_{-0.04}$ & $0.097^{+0.017}_{-0.017}$ & $0.36^{+0.02}_{-0.02}$ & $0.127^{+0.013}_{-0.013}$ & $36$ & $-0.096^{+0.020}_{-0.017}$ & $1.70^{+0.03}_{-0.06}$\\
$\tilde{\mathcal{D}}_3$ (obs) & 541 & $1.42^{+0.05}_{-0.05}$ & $2.34^{+0.06}_{-0.05}$ & $1.72^{+0.04}_{-0.05}$ & $0.28^{+0.05}_{-0.05}$\tablefootmark{e} & $0.103^{+0.014}_{-0.014}$ & $0.38^{+0.02}_{-0.02}$ & $0.122^{+0.015}_{-0.013}$ & $74$ & $-0.096^{+0.027}_{-0.019}$ & $1.71^{+0.04}_{-0.06}$\\
$\tilde{\mathcal{D}}_3$ ($w$) & 541 & $1.35^{+0.06}_{-0.06}$ & $2.30^{+0.05}_{-0.05}$ & $1.74^{+0.04}_{-0.05}$ & $0.26^{+0.06}_{-0.06}$ & $0.124^{+0.023}_{-0.015}$ & $0.44^{+0.02}_{-0.02}$ & $0.123^{+0.015}_{-0.015}$ & $98$ & $-0.080^{+0.021}_{-0.037}$ & $1.73^{+0.03}_{-0.07}$\\
\hline
\end{tabular}
\tablefoot{\tablefoottext{a}{The intervals of $f_{\rm SE}^{\rm c}$ and $f_{\rm SE}^{{\rm c},w}$ come from resampling the hosts.}
\tablefoottext{b}{Occupancy, the fraction of planets within the band, with intervals from resampling the planets.}
\tablefoottext{c}{Resolved draws, the percentage of draws with one peak in each search interval (Appendix~\ref{app:morphtables:kde}). The peak positions, valley and depth are medians over the resolved draws.}
\tablefoottext{d}{The slope of $\tilde{\mathcal{D}}_1$ depends on the radius range through the class labels on which the classifier is trained, and over $[0.5,4]\,R_\oplus$ the $1922$ planets of $\mathcal{D}_1$ discovered by the \textit{Kepler} mission, excluding K2, give $-0.088\pm0.017$.}
\tablefoottext{e}{The depth given by the intensity model of Sect.~\ref{sec:global} is $0.43^{+0.10}_{-0.11}$ on $\tilde{\mathcal{D}}_3$ corrected for the selection and $0.40^{+0.09}_{-0.09}$ on the observed counts. The difference between the latter and the $0.28$ tabulated here comes from the estimator, the intensity model against the KDE.}}
\end{table*}

\begin{table*}[!t]\centering
\caption{The six age bins of $\tilde{\mathcal{D}}_3$ (Sect.~\ref{sec:ageshift}), on the observed counts (obs) and with each planet weighted by $1/\bar S_i$ ($w$; Sect.~\ref{sec:compl}). Under $f_{\rm SE}^{\rm c}$, the column $w$ gives the weighted fraction $f_{\rm SE}^{{\rm c},w}$. The columns headed ``res.''\ (resolved draws) give the percentage of draws in which the valley is resolved (Appendix~\ref{app:morphtables:kde}). The peak positions and the valley are given only where at least two thirds of the draws resolve the valley, whereas $f_{\rm SE}^{\rm c}$ does not require a resolved valley and is always given. All uncertainties are the $16$th and $84$th percentiles (Appendix~\ref{app:morphtables:kde}).}
\label{tab:gapmetrics_agebins}
\footnotesize
\setlength{\tabcolsep}{4pt}\renewcommand{\arraystretch}{1.3}
\begin{tabular}{lr|cc|cc|cc|cc|cc}
\hline\hline
& & \multicolumn{2}{c|}{$R_{\rm SE}^{\rm pk}$ [$R_\oplus$]} & \multicolumn{2}{c|}{$R_{\rm SN}^{\rm pk}$ [$R_\oplus$]} & \multicolumn{2}{c|}{valley [$R_\oplus$]} & \multicolumn{2}{c|}{$f_{\rm SE}^{\rm c}$} & \multicolumn{2}{c}{res.~[\%]}\\
Age [Gyr] & $N$ & obs & $w$ & obs & $w$ & obs & $w$ & obs & $w$ & obs & $w$\\
\hline
$0.1$--$0.75$ & 29 & -- & -- & -- & -- & -- & -- & $0.31^{+0.11}_{-0.10}$ & $0.34^{+0.11}_{-0.11}$ & $50$ & $61$\\
$0.75$--$1.4$ & 66 & -- & $1.29^{+0.11}_{-0.14}$ & -- & $2.15^{+0.11}_{-0.08}$ & -- & $1.67^{+0.10}_{-0.12}$ & $0.38^{+0.05}_{-0.05}$ & $0.47^{+0.06}_{-0.06}$ & $65$ & $75$\\
$1.4$--$2.05$ & 80 & $1.20^{+0.05}_{-0.03}$ & $1.17^{+0.05}_{-0.03}$ & $2.28^{+0.12}_{-0.12}$ & $2.26^{+0.12}_{-0.12}$ & $1.50^{+0.16}_{-0.09}$ & $1.53^{+0.14}_{-0.10}$ & $0.44^{+0.06}_{-0.05}$ & $0.48^{+0.07}_{-0.06}$ & $86$ & $88$\\
$2.05$--$2.7$ & 111 & $1.45^{+0.06}_{-0.05}$ & $1.40^{+0.07}_{-0.10}$ & $2.42^{+0.09}_{-0.09}$ & $2.37^{+0.09}_{-0.09}$ & $1.79^{+0.07}_{-0.07}$ & $1.79^{+0.08}_{-0.08}$ & $0.40^{+0.04}_{-0.05}$ & $0.47^{+0.05}_{-0.06}$ & $87$ & $95$\\
$2.7$--$3.35$ & 133 & $1.43^{+0.07}_{-0.07}$ & $1.41^{+0.08}_{-0.08}$ & $2.44^{+0.09}_{-0.08}$ & $2.41^{+0.09}_{-0.08}$ & $1.81^{+0.06}_{-0.07}$ & $1.83^{+0.07}_{-0.08}$ & $0.33^{+0.04}_{-0.04}$ & $0.38^{+0.05}_{-0.05}$ & $93$ & $96$\\
$3.35$--$4$ & 122 & $1.43^{+0.07}_{-0.06}$ & $1.39^{+0.07}_{-0.06}$ & $2.40^{+0.12}_{-0.12}$ & $2.36^{+0.13}_{-0.13}$ & $1.76^{+0.13}_{-0.10}$ & $1.78^{+0.13}_{-0.10}$ & $0.40^{+0.05}_{-0.05}$ & $0.48^{+0.05}_{-0.06}$ & $76$ & $88$\\
\hline
\end{tabular}
\end{table*}

\begin{figure}[!t]\centering
\includegraphics[width=\linewidth]{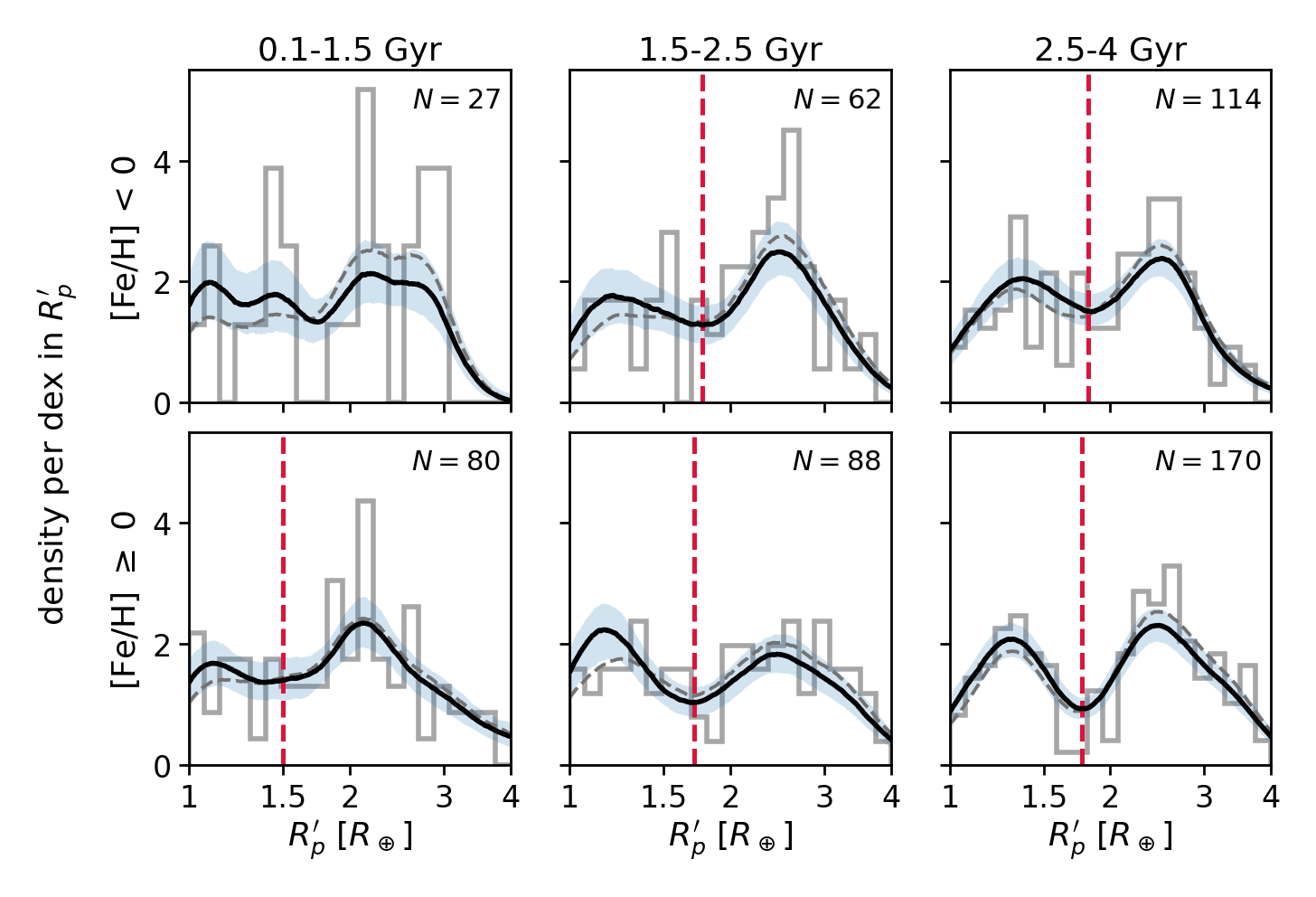}
\caption{Distribution of $R_p'$ (Eq.~\eqref{eq:straight}) in $\tilde{\mathcal{D}}_3$, with the hosts split by metallicity into metal-poor, $[\mathrm{Fe/H}]<0$ (top row), and metal-rich, $[\mathrm{Fe/H}]\geq0$ (bottom row), and by age into $0.1$--$1.5$, $1.5$--$2.5$ and $2.5$--$4$\,Gyr (columns, from left to right). $R_p'$ removes the different period mixing of the cells, whose median periods range from $6.5$ to $11.5$\,d. Each panel is drawn as the bottom row of Fig.~\ref{fig:d3bins}, and the draws perturb [Fe/H] as well as the ages, so that a planet can move between cells. The line of the valley minimum is drawn where the weighted counts resolve the valley in at least two thirds of the draws, which among the $0.1$--$1.5$\,Gyr cells happens only for the metal-rich one ($74\%$ of the draws). On the observed counts neither $0.1$--$1.5$\,Gyr cell reaches two thirds, the metal-rich one stopping at $66.3\%$ (Table~\ref{tab:strbins}).}\label{fig:splitmetstr}
\vspace{4mm}
\includegraphics[width=\linewidth]{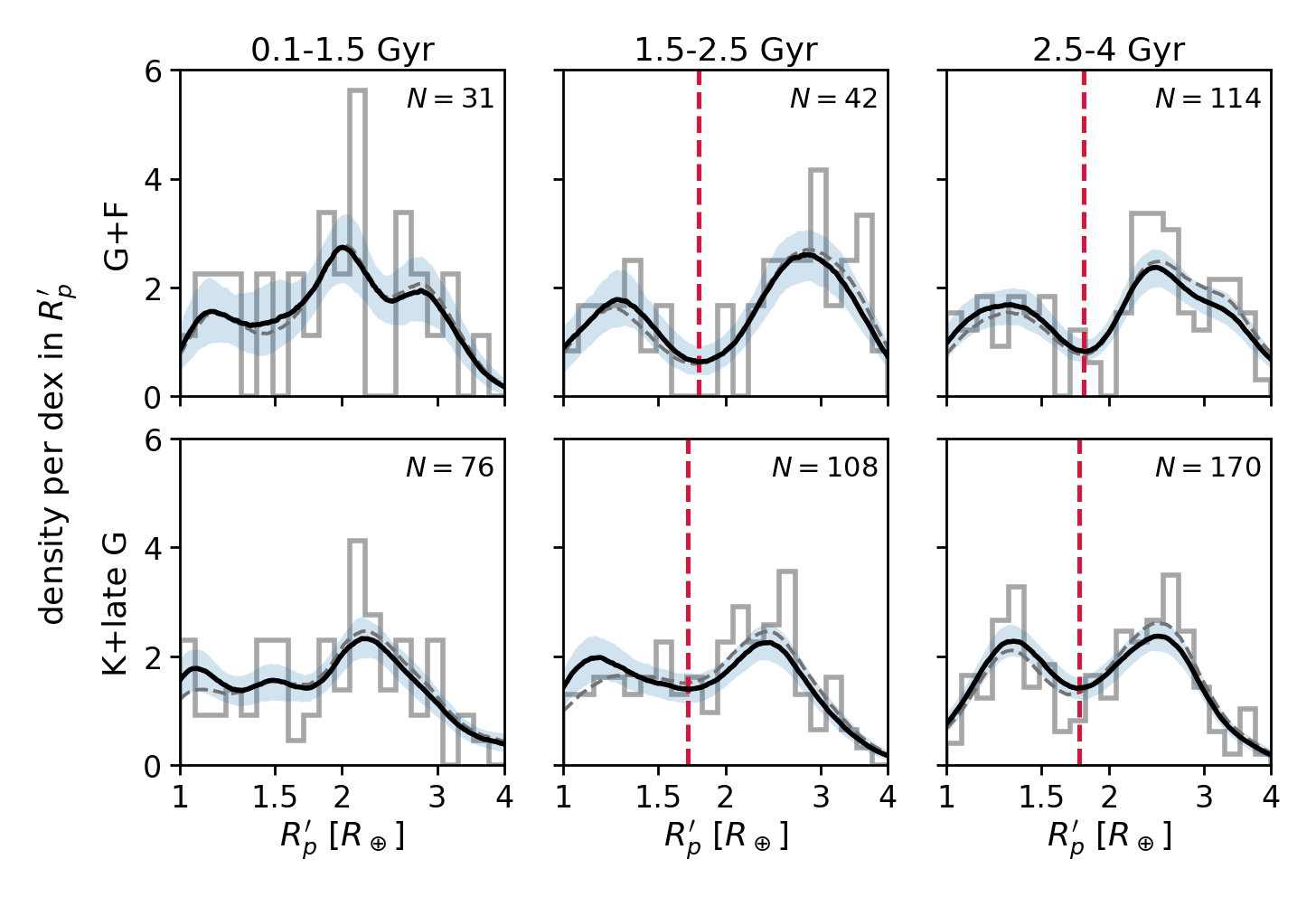}
\caption{As Fig.~\ref{fig:splitmetstr}, with the hosts split by colour into F and G, $(G_{\rm BP}-G_{\rm RP})_0\leq0.95$ (top row), and K and late G, $(G_{\rm BP}-G_{\rm RP})_0>0.95$ (bottom row), with median periods from $6.5$ to $14.5$\,d. Here a planet moves between cells only through its age. Neither $0.1$--$1.5$\,Gyr cell resolves the valley in two thirds of the weighted draws ($47\%$ and $61\%$), so no line of the valley minimum is drawn. On the observed counts the $0.1$--$1.5$\,Gyr cell of the K and late-G hosts reaches $66.8\%$, just above two thirds (Table~\ref{tab:strbins}).}\label{fig:splitcolstr}
\end{figure}

\begin{table}[!t]\centering
\caption{The six age bins and the composition cells of $\tilde{\mathcal{D}}_3$ (Sects.~\ref{sec:ageshift} and~\ref{sec:composition}; Fig.~\ref{fig:d3bins}), measured on $R_p'$ (Eq.~\eqref{eq:straight}), on the observed counts (obs) and with each planet weighted by $1/\bar S_i$ ($w$; Sect.~\ref{sec:compl}). $\tilde{P}$ is the median orbital period. The columns headed ``resolved'' give the percentage of draws in which the valley is resolved (Appendix~\ref{app:morphtables:kde}). The draws resample the planets and perturb their ages, and in the metallicity cells also their [Fe/H], so that a planet can move between cells.
The colour classes, split at $(G_{\rm BP}-G_{\rm RP})_0=0.95$, are held fixed, since the colour errors are negligible against the split. The valley is the median over the resolved draws and is given only where at least two thirds of the draws resolve the valley. The age bins on $R_p$ are in Table~\ref{tab:gapmetrics_agebins}. All uncertainties are the $16$th and $84$th percentiles over the draws.}
\label{tab:strbins}
\footnotesize
\setlength{\tabcolsep}{3.5pt}\renewcommand{\arraystretch}{1.3}
\begin{tabular}{lrr|cc|cc}
\hline\hline
& & & \multicolumn{2}{c|}{resolved [\%]} & \multicolumn{2}{c}{valley [$R_\oplus$]}\\
Age [Gyr] & $N$ & $\tilde{P}$ [d] & obs & $w$ & obs & $w$\\
\hline
$0.1$--$0.75$ & 29 & 6.8 & $53$ & $59$ & -- & --\\
$0.75$--$1.4$ & 66 & 8.9 & $65.9$ & $54$ & -- & --\\
$1.4$--$2.05$ & 80 & 8.2 & $78$ & $85$ & $1.68^{+0.17}_{-0.23}$ & $1.61^{+0.22}_{-0.17}$\\
$2.05$--$2.7$ & 111 & 10.8 & $98$ & $99$ & $1.75^{+0.08}_{-0.08}$ & $1.79^{+0.08}_{-0.09}$\\
$2.7$--$3.35$ & 133 & 10.3 & $100$ & $99$ & $1.76^{+0.05}_{-0.06}$ & $1.77^{+0.06}_{-0.07}$\\
$3.35$--$4$ & 122 & 8.2 & $95$ & $93$ & $1.67^{+0.08}_{-0.06}$ & $1.71^{+0.13}_{-0.08}$\\
\hline
\hline
\multicolumn{7}{l}{\textit{metal-poor} ([Fe/H]$<0$)}\\[2pt]
$0.1$--$1.5$ & 27 & 10.0 & $60$ & $55$ & -- & --\\
$1.5$--$2.5$ & 62 & 11.2 & $83$ & $89$ & $1.74^{+0.13}_{-0.23}$ & $1.77^{+0.12}_{-0.22}$\\
$2.5$--$4$ & 114 & 11.5 & $93$ & $93$ & $1.71^{+0.12}_{-0.10}$ & $1.82^{+0.12}_{-0.15}$\\
\hline
\multicolumn{7}{l}{\textit{metal-rich} ([Fe/H]$\geq0$)}\\[2pt]
$0.1$--$1.5$ & 80 & 6.5 & $66.3$ & $74$ & -- & $1.50^{+0.19}_{-0.18}$\\
$1.5$--$2.5$ & 88 & 8.5 & $90$ & $93$ & $1.72^{+0.11}_{-0.12}$ & $1.71^{+0.12}_{-0.11}$\\
$2.5$--$4$ & 170 & 8.2 & $100$ & $99$ & $1.74^{+0.05}_{-0.05}$ & $1.77^{+0.05}_{-0.06}$\\
\hline
\hline
\multicolumn{7}{l}{\textit{F and G} ($(G_{\rm BP}-G_{\rm RP})_0\leq0.95$)}\\[2pt]
$0.1$--$1.5$ & 31 & 9.0 & $48$ & $47$ & -- & --\\
$1.5$--$2.5$ & 42 & 14.5 & $87$ & $89$ & $1.75^{+0.11}_{-0.11}$ & $1.79^{+0.11}_{-0.11}$\\
$2.5$--$4$ & 114 & 10.0 & $94$ & $92$ & $1.79^{+0.06}_{-0.06}$ & $1.80^{+0.06}_{-0.07}$\\
\hline
\multicolumn{7}{l}{\textit{K and late G} ($(G_{\rm BP}-G_{\rm RP})_0>0.95$)}\\[2pt]
$0.1$--$1.5$ & 76 & 6.5 & $66.8$ & $61$ & $1.62^{+0.12}_{-0.35}$ & --\\
$1.5$--$2.5$ & 108 & 9.0 & $81$ & $88$ & $1.72^{+0.12}_{-0.18}$ & $1.70^{+0.14}_{-0.22}$\\
$2.5$--$4$ & 170 & 9.2 & $99$ & $99$ & $1.70^{+0.07}_{-0.06}$ & $1.76^{+0.09}_{-0.08}$\\
\hline
\end{tabular}
\end{table}

\begin{table*}[!t]\centering
\caption{Counted SE fractions of $\tilde{\mathcal{D}}_3$ and their counterparts from the three-interval form of the mixture fit of Sect.~\ref{sec:hier}, $\widehat{f}_{\rm SE}^{\rm c}$ and $\widehat{f}_{\rm SE}^{{\rm c},w}$ (Appendix~\ref{app:morphtables:kde}). $\bar f_{\rm SE}^{\rm m}$ is the standardised SE weight of each age interval of the fit (Sect.~\ref{sec:resglobal}). $f_{\rm SE}^{{\rm c},w}$ excludes the planets with $\bar S_i<0.25$, which $f_{\rm SE}^{{\rm c},w_+}$ counts with the weight $1/0.25$. $\widehat{f}_{\rm SE}^{\rm c}$ and $\widehat{f}_{\rm SE}^{{\rm c},w}$ are medians over $2000$ draws of the posterior, and their uncertainties come from the posterior alone, without the scatter of the counts. All uncertainties are the $16$th and $84$th percentiles.}
\label{tab:fsepred}
\footnotesize
\setlength{\tabcolsep}{6pt}\renewcommand{\arraystretch}{1.3}
\begin{tabular}{lr|c|cc|ccc}
\hline\hline\noalign{\vskip 2pt}
Age [Gyr] & $N$ & $\bar f_{\rm SE}^{\rm m}$ & $f_{\rm SE}^{\rm c}$ & $\widehat{f}_{\rm SE}^{\rm c}$ & $f_{\rm SE}^{{\rm c},w}$ & $f_{\rm SE}^{{\rm c},w_+}$ & $\widehat{f}_{\rm SE}^{{\rm c},w}$\\
\noalign{\vskip 2pt}\hline
$0.1$--$4$ & 541 & -- & $0.38\pm0.02$ & $0.37\pm0.01$ & $0.44\pm0.02$ & $0.48\pm0.03$ & $0.47\pm0.02$\\
\hline
$0.1$--$1$ & 55 & $0.43^{+0.10}_{-0.09}$ & $0.31\pm0.07$ & $0.33\pm0.04$ & $0.37\pm0.08$ & $0.37\pm0.08$ & $0.37\pm0.06$\\
$1$--$2$ & 115 & $0.75\pm0.06$ & $0.44\pm0.05$ & $0.43\pm0.03$ & $0.52\pm0.05$ & $0.57^{+0.05}_{-0.06}$ & $0.54\pm0.04$\\
$2$--$4$ & 371 & $0.67\pm0.04$ & $0.37\pm0.03$ & $0.36\pm0.02$ & $0.43\pm0.03$ & $0.47\pm0.03$ & $0.47\pm0.03$\\
\hline
\end{tabular}
\end{table*}

\subsection{Kernel density estimates}\label{app:morphtables:kde}
 
We estimate the radius distribution using a Gaussian KDE with a fixed bandwidth of $0.045$\,dex in $\log_{10}R_p$ over $1$--$4\,R_\oplus$. The KDE is evaluated over Monte Carlo draws that resample planets and perturb their host ages according to their uncertainties. 
Planetary radii are not perturbed: their measurement errors are already represented in the observed distribution, and adding further perturbations would introduce additional smoothing and systematically alter the inferred morphology. 
For the period-adjusted radii $R_p'$, we evaluate the KDE over $0.9$--$4.5\,R_\oplus$ to accommodate planets displaced across the nominal boundaries by the period correction while retaining the sample selected on $R_p$.

The valley occupancy is the fraction of planets within a fixed $\pm0.2\,R_\oplus$ band around the adopted valley locus. Unlike the other tabulated quantities, it places each planet in the age bin of its measured age, so its draws resample the planets without perturbing the ages, as in the test of Sect.~\ref{sec:ageshift}.

For the four age bins from $1.4$ to $4$\,Gyr, where the valley is resolved (Sect.~\ref{sec:ageshift}), the observed-count valley occupancies, in increasing age order, are $0.15^{+0.04}_{-0.04}$, $0.13^{+0.03}_{-0.03}$, $0.08^{+0.02}_{-0.02}$, and $0.13^{+0.03}_{-0.03}$. The corresponding observed-count valley depths are $0.42^{+0.12}_{-0.12}$, $0.39^{+0.11}_{-0.09}$, $0.43^{+0.10}_{-0.10}$, and $0.33^{+0.11}_{-0.12}$. Completeness-weighted widths are not reported for individual age bins because they are unreliable: in two of the four bins, the width measured from the full bin lies at or beyond the 84th percentile of its host-resampling distribution. 

We identify a KDE maximum as a peak only if its height exceeds the higher of its two flanking minima by at least $3\%$ of the global KDE maximum. Each flanking minimum is the lowest point before the KDE rises above the candidate maximum again or reaches the boundary of the evaluation range. A draw has a \emph{resolved valley} if exactly one qualifying peak lies in each of the search intervals: $[1.0,1.7]\,R_\oplus$ for SEs and $[2.0,3.2]\,R_\oplus$ for SNs. Otherwise, the draw is classified as unresolved, and its peak positions, valley minimum, and depth are undefined. We report these quantities as measurements only when at least two thirds of the draws resolve the valley; otherwise they are omitted from Tables~\ref{tab:gapmetrics_agebins} and~\ref{tab:strbins} and from the figures. Values for the full samples are retained in Table~\ref{tab:gapmetrics} for completeness.
Draws containing fewer than $20$ planets, or fewer than $20$ planets contributing to the weighted counts, are excluded from the age-bin analysis. This occurs only for $14$ of the $1000$ draws in the $0.1$--$0.75$\,Gyr bin of $\tilde{\mathcal{D}}_3$. 

The youngest two bins of $\tilde{\mathcal{D}}_3$, containing $29$ and $66$ planets, resolve the valley less frequently than older bins (Sect.~\ref{sec:ageshift}). To determine whether this difference is due solely to sample size, we randomly subsample each of the four older bins to $29$ and $66$ planets. On the observed counts, these subsamples still resolve the valley in at least $73\%$ and $76\%$ of draws on $R_p$, respectively, and in at least $66.3\%$ and $78\%$ on $R_p'$. The corresponding fractions for the actual $0.1$--$0.75$ and $0.75$--$1.4$\,Gyr bins are only $53\%$ and $65.9\%$ on $R_p'$. Thus, the weaker evidence for a valley at young ages cannot be attributed to their smaller samples alone.

We also examine the age-dependent valley positions within the host-colour and metallicity subsets of Sect.~\ref{sec:composition}. Comparing the youngest ($0.1$--$1.5$\,Gyr) and oldest ($2.5$--$4$\,Gyr) bins, the completeness-weighted young-minus-old offsets in $R_p$, computed draw by draw over realisations in which both cells resolve the valley, are $-0.29^{+0.14}_{-0.12}\,R_\oplus$ for metal-rich hosts, $-0.11^{+0.11}_{-0.14}\,R_\oplus$ for metal-poor hosts, $-0.26^{+0.16}_{-0.13}\,R_\oplus$ for F and G hosts, and $-0.19^{+0.12}_{-0.20}\,R_\oplus$ for K and late-G hosts. All four offsets have the same sign, although the individual differences are not significant.

The radius-distribution entries in Tables~\ref{tab:gapmetrics}, \ref{tab:gapmetrics_agebins}, and~\ref{tab:strbins} are medians over resolved draws, with age-bin membership recalculated after each age perturbation. By contrast, the period--radius slopes and positions in Table~\ref{tab:gapmetrics} are fitted freely to the full set of planets of each sample, with uncertainties obtained by resampling hosts and reassigning planetary classes in each replicate (Sect.~\ref{sec:prplane}).

For the sliding-window analysis of Sect.~\ref{sec:ageshift}, we use widths of $0.5$, $0.8$, and $1.0$\,Gyr, advancing in steps of $0.05$\,Gyr. 
Windows containing fewer than $25$ planets at the measured ages are excluded; consequently, the youngest retained $0.5$\,Gyr window is centred at $0.5$ rather than $0.35$\,Gyr. In each Monte Carlo draw, planets are assigned to all windows containing their perturbed ages, and windows containing fewer than $25$ planets in that draw are classified as unresolved. The same KDE and valley-resolution criteria are then applied to the remaining windows. The sliding-window analysis also provides estimates of $f_{\rm SE}^{\rm c}$ and $f_{\rm SE}^{{\rm c},w}$ as continuous functions of age. 
Their uncertainties are obtained by resampling planets within each window, whereas the age-bin fractions use host-level resampling.

For completeness-corrected counts, each planet is weighted by $1/\bar S_i$, where $\bar S_i$ is calculated over the host--orbit configurations of the group to which the planet belongs in that draw: the full sample, an age bin or subset, or a composition cell (Sect.~\ref{sec:compl}). Planets with $\bar S_i<0.25$ are excluded from these counts. In the $0.1$--$1$, $1$--$2$, and $2$--$4$\,Gyr subsets, respectively, this removes $0$, $4$, and $8$ planets, all smaller than $1.4\,R_\oplus$. The completeness weights are held fixed across resamples, so the reported intervals do not propagate uncertainty in $\bar S_i$. As a sensitivity test, we instead exclude from every age bin the $18$ planets whose $\bar S_i$ would fall below $0.25$ in any of the six bins or three broader age subsets. This changes the valley minima by at most $0.01\,R_\oplus$ and the counted SE fractions by at most $0.04$; no bin crosses the two-thirds resolution threshold.

Three SE-fraction estimators are used throughout this work. The counted fraction $f_{\rm SE}^{\rm c}$ is the proportion of observed planets with radii of $1$--$4\,R_\oplus$ below the adopted class boundary. The completeness-weighted fraction $f_{\rm SE}^{{\rm c},w}$ applies $1/\bar S_i$ weights to the same classification. The standardised mixture weight $\bar f_{\rm SE}^{\rm m}$ (Sect.~\ref{sec:hier}) instead measures membership in the intrinsic SE component before selection, including radii outside the counted interval. 
The mixture weight is therefore higher than the counted fractions, since the fitted SE Gaussian places $10\%$ of its probability above the class boundary at $10$\,d, where planets are counted as SNs, and $38\%$ below $1\,R_\oplus$, outside the counted range.

To compare these quantities on a common basis, we use the three-interval mixture fit to predict the counted fractions for the observed periods and radius uncertainties of $\tilde{\mathcal{D}}_3$. The prediction $\widehat{f}_{\rm SE}^{\rm c}$ includes the selection function of Sect.~\ref{sec:compl}, whereas $\widehat{f}_{\rm SE}^{{\rm c},w}$ is calculated without selection, as the counterpart of the completeness-weighted counts (Table~\ref{tab:fsepred}). Over the full sample, the predictions are $\widehat{f}_{\rm SE}^{\rm c}=0.37\pm0.01$ and $\widehat{f}_{\rm SE}^{{\rm c},w}=0.47\pm0.02$, compared with observed estimates of $f_{\rm SE}^{\rm c}=0.38\pm0.02$ and $f_{\rm SE}^{{\rm c},w}=0.44\pm0.02$.

The increase in the mixture weight from $\bar f_{<1}=0.43^{+0.10}_{-0.09}$ to $\bar f_{2\text{--}4}=0.67\pm0.04$ is consequently much less apparent in the counted fractions: $f_{\rm SE}^{\rm c}$ increases from $0.31\pm0.07$ to $0.37\pm0.03$, while its mixture prediction changes from $0.33\pm0.04$ to $0.36\pm0.02$. The $12$ small SEs with $\bar S_i<0.25$ are omitted from $f_{\rm SE}^{{\rm c},w}$. Retaining them with capped inverse-completeness weights $1/\max(\bar S_i,0.25)$ gives $f_{\rm SE}^{{\rm c},w_+}=0.48\pm0.03$ over the full sample. In every row of Table~\ref{tab:fsepred}, $\widehat{f}_{\rm SE}^{{\rm c},w}$ lies between $f_{\rm SE}^{{\rm c},w}$ and $f_{\rm SE}^{{\rm c},w_+}$.

The three-interval mixture assumes a constant SE weight throughout $2$--$4$\,Gyr and therefore does not intrinsically predict the dip in the counted fraction at $2.7$--$3.35$\,Gyr. Nevertheless, simulations that draw each planet's measured-radius classification, including outcomes outside $1$--$4\,R_\oplus$, yield counted fractions of $0.37\pm0.04$, $0.34\pm0.04$, and $0.38\pm0.04$ in the $2.05$--$2.7$, $2.7$--$3.35$, and $3.35$--$4$\,Gyr bins. These agree with the observed $0.40\pm0.05$, $0.33\pm0.05$, and $0.40\pm0.05$. The lowest simulated value falls in the $2.7$--$3.35$\,Gyr bin, which holds fewer planets within $10$\,d, and a fraction at least as low as observed occurs there in $47\%$ of realisations. Thus, the dip does not require an additional age dependence of the intrinsic SE weight.

\subsection{Intensity model}\label{app:morphtables:lgcp}

The log-Gaussian Cox process used in Sect.~\ref{sec:global} models the radius distribution through a penalised-Poisson smoother. The log-intensity is fitted in $46$ equal-width bins of $\log_{10}R_p$ over $1$--$4\,R_\oplus$, with a penalty on second differences between adjacent bins to suppress excessive curvature.

We marginalise over $25$ logarithmically spaced penalty strengths between $1$ and $10^4$. Each is weighted by the marginal likelihood under a Laplace approximation, which represents the posterior log-intensity locally as a Gaussian around its maximum. We obtain uncertainties from $4000$ posterior draws: first selecting a penalty strength according to its weight, then drawing a log-intensity curve from the corresponding Gaussian approximation. In each draw, the SE and SN peaks are the maxima within the search intervals of Appendix~\ref{app:morphtables:kde}, and the valley is the minimum between them. A draw is unresolved if either maximum occurs at an interval boundary, indicating that no interior peak is identified.

For the $1$--$2$ and $2$--$4$\,Gyr subsets, penalty strengths are weighted by the marginal likelihood of the two subsets fitted together, and each subset uses the strength multiplied by its fraction of the planets. The valley is resolved in $85\%$ of draws for the full completeness-corrected $\tilde{\mathcal{D}}_3$ ($95\%$ for observed counts), and in $87\%$ and $71\%$ for the $2$--$4$ and $1$--$2$\,Gyr subsets, respectively. For the $464$ planets whose period-adjusted radii $R_p'$ remain within $1$--$4\,R_\oplus$, the $1$--$2$\,Gyr subset resolves the valley in only $12\%$ of draws.

These posterior draws treat planets as independent, although multiple planets may share a host. We therefore repeat the fit on $400$ host-level bootstrap samples, using the most probable penalty for each. Among the $345$ samples with a resolved valley, the $16$--$84\%$ interval of valley depth is $1.19$ times as wide as that from posterior draws ($1.14$ for observed counts). We inflate the depth uncertainties reported in Sect.~\ref{sec:global} accordingly. The corresponding intervals for peak and valley positions are no wider than the posterior-draw intervals, which are retained.

\subsection{Sensitivity to the criterion for a resolved valley}\label{app:morphtables:tests}

Our definition of a resolved valley depends on three choices: the minimum fraction of resolved draws, the peak search intervals, and the KDE bandwidth. We test each using $\tilde{\mathcal{D}}_3$.

First, we vary the required fraction of resolved draws. On $R_p'$ and using observed counts, the six age bins resolve the valley in $53$, $65.9$, $78$, $98$, $100$, and $95\%$ of draws (Table~\ref{tab:strbins}). A $60\%$ threshold would admit the $0.75$--$1.4$\,Gyr bin, whereas a $70\%$ threshold would exclude it. All bins from $1.4$\,Gyr onwards satisfy either criterion.

Second, we test whether the peak search intervals, $[1.0,1.7]$ and $[2.0,3.2]\,R_\oplus$, could artificially disfavour a valley shifted towards smaller radii at young ages. We compare a single Gaussian with a two-Gaussian mixture, both truncated to $1$--$4\,R_\oplus$ and convolved with each planet's radius uncertainty. The models are fitted to the observed counts without restricting their peak positions. 
We define $\Delta\mathrm{BIC}$ as the BIC of the single Gaussian minus that of the mixture, so that positive values favour two components. We also use a likelihood-ratio test, whose $p$-value is the fraction of $300$ samples simulated from the fitted single Gaussian with a likelihood ratio at least as large as observed.

Below $1.4$\,Gyr, the mixture is not preferred. For the $0.1$--$1$\,Gyr subset, $\Delta\mathrm{BIC}=-9.5$ and $p=0.51$. In the $0.1$--$0.75$\,Gyr bin, both fitted components lie near $2.0$--$2.1\,R_\oplus$, effectively describing one peak; in the $0.75$--$1.4$\,Gyr bin, $p=0.24$. 
From $1.4$\,Gyr onwards, both criteria favour the mixture in three of the four bins ($p\leq0.003$, the minimum resolvable with $300$ simulations), with the exception of the $2.05$--$2.7$\,Gyr bin ($p=0.07$).
The mixture is also favoured for the full sample ($\Delta\mathrm{BIC}=+27.6$) and the $1$--$2$\,Gyr subset, where it places the components at $1.2$ and $2.2\,R_\oplus$ ($p=0.003$). Across $300$ host-level bootstrap samples, the mixture is preferred by BIC in at most $19\%$ of replicates below $1.4$\,Gyr, compared with $35$--$83\%$ in the bins from $1.4$\,Gyr onwards.

Finally, we vary the KDE bandwidth over $0.030$, $0.0375$, $0.045$, $0.0525$, and $0.060$\,dex. For the full sample on observed counts, the valley position spans only $0.036\,R_\oplus$ on $R_p$ and $0.035\,R_\oplus$ on $R_p'$. In the bins from $1.4$\,Gyr onwards, among bandwidths for which at least two thirds of draws resolve the valley, the position changes by no more than $0.06\,R_\oplus$ on either radius scale, except for the $2.7$--$3.35$\,Gyr bin on $R_p$, where it decreases from $1.88$ to $1.75\,R_\oplus$ as the bandwidth increases. Stronger smoothing reduces the resolved fraction of the full sample from $0.94$ to $0.19$ across the tested bandwidths. Nevertheless, on the observed counts neither of the two youngest bins reaches the two-thirds threshold on $R_p$ at any tested bandwidth. Their lack of a consistently resolved valley is therefore not an artefact of the adopted smoothing scale.


\subsection{Robustness of the period--radius valley locus}
\label{app:svmrobust}

We assess the sensitivity of the reference valley locus (Eq.~\eqref{eq:slope}) to completeness weighting, mixture classification, and SVM regularisation.

The reference locus is fitted to observed counts. Weighting the SVM by $1/\bar S_i$ (Sect.~\ref{sec:compl}), while keeping the mixture labels fixed and rescaling the weights to unit mean so that $C_{\rm SVM}$ retains its reference value, increases the slope by $0.016$, corresponding to $0.6$ times its upper uncertainty. The weighted locus reclassifies only one of the $541$ planets and changes the counted SE fractions and the valley offset of Sect.~\ref{sec:ageshift} by at most $0.2\sigma$.

The reference mixture has full covariance. Adopting diagonal or shared covariance matrices instead gives slopes of $-0.091\pm0.017$ and $-0.084\pm0.020$, respectively.
The uncertainty quoted in Eq.~\eqref{eq:slope} combines in quadrature the bootstrap $16$th--$84$th percentile interval and half the range of slopes obtained with these three covariance models. Assigning class labels from radius alone, without accounting for orbital period, instead gives a shallower slope of $-0.038\pm0.012$.

Finally, varying the SVM soft-margin parameter gives $m=-0.116\pm0.025$ for $C_{\rm SVM}=1$ and $m=-0.091\pm0.023$ for $C_{\rm SVM}=100$, compared with $m=-0.096\pm0.022$ for the reference value $C_{\rm SVM}=10$. These uncertainties are the bootstrap $16$th--$84$th percentile half-widths and exclude the additional contribution from the mixture covariance model.


\section{The mixture likelihood}\label{app:compl}

\subsection{Modelling the detection and confirmation completeness}\label{app:compl:probit}
\paragraph{Detection.}
For planet $i$, we define $c_i$ as the log-radius at which the expected MES of Sect.~\ref{sec:compl} equals $8.28$, the value at which the recovery curve of \citet{Christiansen2017}, $\Gamma_{\rm CDF}(\mathrm{MES};\alpha,\theta)$ with shape $\alpha=30.87$ and scale $\theta=0.271$, reaches half of its ceiling. In the expected MES, $R_\star$ is the stellar radius of Sect.~\ref{sec:starpars}, the duration is the DR25 transit duration of the planet, and $\mathrm{CDPP}_{\rm dur}$ is interpolated linearly in duration between the CDPP values that the DR25 stellar table gives at fourteen durations from $1.5$ to $15$\,h, and outside that range is set to the value at the nearest tabulated duration, as in \citet{Burke2015}. The expected number of transits, $N_{\rm tr}$, is the observing baseline divided by the orbital period and multiplied by the duty cycle (the fraction of the baseline with valid observations). We model the probability that a planet with true log-radius $x'$ is detected and retained as a candidate as
\begin{equation}\label{eq:det}
p^{\rm det}_i(x')=C\,\Phi\!\left(\frac{x'-c_i-\delta_i}{s_D}\right),
\end{equation}
\begin{equation}\label{eq:delta}
\delta_i=d_0+d_b\,(b_i-0.45)+d_P\,(\log_{10}P_i-1.5)+d_L\,L_i.
\end{equation}
Here $C$ is the asymptotic detection efficiency (or ceiling) and $\Phi$ is the standard normal cumulative distribution.
In $\delta_i$, $b_i$ is the DR25 impact parameter of planet $i$ and $P_i$ its orbital period in days. $L_i$ is the CDPP slope of its host, taken from the DR25 stellar table, where it is the slope of a linear fit of $\log_{10}\mathrm{CDPP}$ against $\log_{10}$ of the transit duration over the six longest durations searched by the pipeline, from $7.5$ to $15$\,h \citep{BurkeCatanzarite2017,BurkeCatanzarite2017b}.
The detection probability reaches half its ceiling at $x'=c_i+\delta_i$, where $\delta_i$ corrects the threshold $c_i$ predicted by the expected MES for the dependence of the recovery on the planet and its host.
Around that point the detection probability goes from $16\%$ to $84\%$ of $C$ between $c_i+\delta_i-s_D$ and $c_i+\delta_i+s_D$. We measure $\delta_i$ and $s_D$ on the simulated transits that \citet{Christiansen2017} injected into the \textit{Kepler} pixel data and processed with the \textit{Kepler} DR25 pipeline, selected and fitted as described below. 
We determine $C$, $s_D$, $d_0$, $d_b$, $d_P$ and $d_L$ by maximising the Bernoulli likelihood of the $17\,627$ injections matched to the hosts and orbits of $\mathcal{D}_3$. An injection counts as a success when the pipeline recovers it and the Robovetter, which classifies detected signals as planet candidates or false positives \citep{Thompson2018}, retains it as a candidate. Its probability of success is given by Eq.~\eqref{eq:det}, with the injected signal in place of planet $i$. From the on-target transit injections, one per target, we keep those with at least three transits and a duration below $15$\,h, as in \citet{Christiansen2017}, with $P_{\rm orb}\le100$\,d and impact parameter $\le0.9$, as for the planets of $\mathcal{D}_3$, and on targets whose DR25 effective temperature, surface gravity and \textit{Kepler} magnitude lie within the ranges of the $404$ hosts of $\mathcal{D}_3$. 
With standard deviations from $100$ bootstrap resamplings of the targets, $d_0=-0.014\pm0.002$, $d_b=+0.081\pm0.002$, $d_P=-0.006\pm0.002$, $d_L=+0.039\pm0.005$ and $s_D=0.0393\pm0.0009$, all in dex of radius, and $C=0.940\pm0.005$. 
The fitted threshold correction $\delta_i$ increases with the impact parameter, which at fixed radius lowers the MES through limb darkening and the longer ingress and egress of transits far from the stellar centre. The fitted $s_D$ matches the width of $\Gamma_{\rm CDF}$, $0.039$\,dex once expressed in log-radius. The injections therefore shift the threshold of each planet by $\delta_i$ but leave the shape of the recovery curve unchanged. For each injection the threshold is computed as $c_i$ is for a planet, from the expected MES at the injected radius ratio and transit duration. 
The expected MES uses the geometric depth $(R_p/R_\star)^2$, the tabulated CDPP and a number of transits derived from the duty cycle, while the injected signals have their true limb-darkened depth, noise and number of transits. $\delta_i$ therefore also absorbs the effect of these approximations. Across $\mathcal{D}_3$, the radius at half the detection ceiling has a median of $0.95\,R_\oplus$, with 16th--84th percentiles of $0.65$--$1.37\,R_\oplus$.

\paragraph{Confirmation.}
The second selection stage is confirmation of a DR25 candidate, by which we mean that a candidate enters our catalogue and passes the planet-level criteria (c), (e) and (f) of Sect.~\ref{sec:sample}. We measure the \emph{confirmation rate} as the fraction of the $2603$ planet candidates of the DR25 KOI catalogue that are confirmed. These candidates have $P_{\rm orb}\le100$\,d, DR25 radii of $0.5$--$4\,R_\oplus$, and impact parameters $\le0.9$. They orbit $1965$ targets whose effective temperatures, surface gravities and \textit{Kepler} magnitudes $K_p$, from the DR25 stellar table \citep{Mathur2017}, lie within the ranges of the $\mathcal{D}_3$ hosts, the same ranges used to select the injections. Rotation periods and ages are not required for these stars; the age dependence is assessed separately below. The confirmation rate increases with measured MES, from $5\%$ at $\mathrm{MES}=7.1$--$10$ to $50\%$ at $10$--$12$ and $89\%$ above $20$. The confirmation of a candidate also rises with that of its siblings: at $\mathrm{MES}=10$--$12$, the rate is $77\%$ for candidates with a \emph{confirmed sibling} (another KOI around the same star listed as confirmed in the NEA), compared with $41\%$ for single candidates and $42\%$ for the $24$ candidates that share their star with other candidates, none of them confirmed.
Writing $\xi$ for the $\log_{10}$ of the MES reported by DR25, we model the probability that the candidate is confirmed as
\begin{equation}\label{eq:conf}
\eta(\xi,g)=\eta_{\max}[g]\,\Phi\!\left(\frac{\xi-\xi_0-\Delta_g\,g}{s_\eta}\right),
\end{equation}
where $\eta_{\max}[g]$ is the ceiling of the confirmation probability, one for each value of $g$,
$\xi_0$ the value of $\xi$ at which the probability reaches half of its ceiling for $g=0$,
$\Delta_g$ the shift of that value for $g=1$, and $s_\eta$ the analogue in $\xi$ of $s_D$ in Eq.~\eqref{eq:det}. 
The indicator $g$ is one if another KOI of the same star is listed as a confirmed planet in the NEA and zero otherwise, as $g_i$ in Eq.~\eqref{eq:wlogit}. 
We estimate $\xi_0$, $s_\eta$, $\Delta_g$, $\eta_{\max}[0]$ and $\eta_{\max}[1]$ by maximising the Bernoulli likelihood of the $2603$ candidates, in which a candidate counts as a success when it is confirmed. 
With standard deviations from $300$ bootstrap resamplings of the host stars, $\xi_0=1.056\pm0.006$, $s_\eta=0.057\pm0.005$, $\Delta_g=-0.066\pm0.010$, $\eta_{\max}[0]=0.843\pm0.014$ and $\eta_{\max}[1]=0.922\pm0.012$. We use the indicator $g$ because candidates with other DR25 candidates but no confirmed sibling are confirmed about as often as single ones at every MES, and less often than candidates with a confirmed sibling, and because the likelihood of Sect.~\ref{sec:hier} is conditional on the observed status of the other planets of the star. 
Because two confirmed planets of the same star necessarily qualify as each other's confirmed sibling, $g$ is not independent of confirmation.
As a control we replace $g$ with the multi-candidate indicator $g_{\rm mc}$, which identifies stars with multiple DR25 candidates. Unlike $g$, it does not depend on whether the other candidates are confirmed, but it describes the confirmations worse, by $2\Delta\ln\mathcal{L}=26.4$ with the same number of parameters.

\paragraph{Measured MES.}
Confirmation depends on the MES measured by the \textit{Kepler} DR25 pipeline, whereas the required selection function must depend on true planetary radius. We connect these quantities through
\begin{equation}\label{eq:mesmap}
\begin{split}
\hat\xi^{\rm exp}_i(x')={}&A_0+A_1\,(x'-c_i)+A_b\,(b_i-0.45)\\
&+A_P\,(\log_{10}P_i-1.5)+A_L\,L_i,
\end{split}
\end{equation}
\begin{equation}\label{eq:mesmeas}
\hat\xi_i(x')=\hat\xi^{\rm exp}_i(x')+z_{\rm inj}+z_0.
\end{equation}
Here $\hat\xi^{\rm exp}_i(x')$ is the $\log_{10}$ of the expected MES, as defined in the DR25 injection tests, of a transit of planet $i$ at true log-radius $x'$. We fit $A_0$, $A_1$, $A_b$, $A_P$ and $A_L$ by a linear regression of the $\log_{10}$ of the expected MES that \citet{Christiansen2017} tabulates for each of the $17\,627$ injections on the right-hand side of Eq.~\eqref{eq:mesmap} evaluated for that injection. We obtain $A_0=0.9702\pm0.0006$, $A_1=1.953\pm0.004$, $A_b=-0.160\pm0.001$, $A_P=-0.0046\pm0.0007$ and $A_L=-0.009\pm0.002$, with standard deviations from $100$ bootstrap resamplings of the targets and a residual scatter of $0.037$\,dex. 
The injection tests provide two MES values for each injected transit.
The expected MES, whose $\log_{10}$ is $\hat\xi^{\rm exp}_i(x')$, is computed from the depth of the signal and the noise of the star. The measured MES, instead, is the one that the transit search of the \textit{Kepler} DR25 pipeline actually obtains on the light curve, and it is the one that decides both the detection and the confirmation.
We denote by $\hat\xi_i(x')$ the mean $\log_{10}$ of the measured MES for a transit of planet $i$ at true log-radius $x'$, around which the measured $\xi$ scatters as a Gaussian with standard deviation $s_\xi$. 
The term $z_{\rm inj}=-0.059$\,dex brings the expected MES to the measured one. It is the mean of the $\log_{10}$ of the measured over the expected MES for the $5270$ of the $17\,627$ injections that the pipeline recovers with an expected MES of at least $10$, away from the detection threshold, and this ratio scatters by $0.046$\,dex. The standard deviation $s_\xi=0.059$\,dex adds in quadrature the residual scatter of the regression of Eq.~\eqref{eq:mesmap}, $0.037$\,dex, and that of the measured over the expected MES, $0.046$\,dex.
The zero point $z_0=-0.036$\,dex adapts the relation to the planets. It is the median, over the $389$ planets of $\mathcal{D}_3$ with a DR25 MES between $15$ and $60$, of the difference $u_i$ between the $\log_{10}$ of that MES and $\hat\xi^{\rm exp}_i+z_{\rm inj}$ evaluated at the measured radius of the planet.
The difference $u_i$ depends on the order in which the \textit{Kepler} DR25 pipeline detected the transit signals of the same star: its median is $-0.028$\,dex for the first signal and $-0.047$\,dex for the later ones, which the pipeline searches after removing cadences in and near previously detected transits \citep{Twicken2016}. The offset cannot be attributed to differences in transit depth, since the radius ratios of $\mathcal{D}_3$ agree closely with those of DR25 (median ratio $1.000$, 16th--84th percentile range $0.986$--$1.013$).
Near the detection threshold only the transits whose measured MES fluctuates upwards are detected, which biases the measured MES of the detected signals high. For the recovered injections, the log ratio of measured to expected MES rises from $-0.06$\,dex above an expected MES of $12$ to $-0.022$\,dex at $7.1$--$10$, and for the planets the median of $u_i$ likewise moves towards zero below $\mathrm{MES}=12$. Above $\mathrm{MES}=60$ the median of $u_i$ is $+0.005$\,dex, but these planets lie far above the confirmation threshold, where the confirmation probability is at its ceiling and $u_i$ does not affect it. We apply $z_0$ only to the confirmation probability, since confirmation is decided on the MES that DR25 tabulates for the candidates, the same MES on which $z_0$ is measured. The detection probability of Eq.~\eqref{eq:det} is calibrated directly on the injections and does not include $z_0$. Applying $z_0$ to it as well changes the age contrast $\bar f_{2\text{--}4}-\bar f_{<1}$ by $+0.002$.

\paragraph{Selection function.}

Because the same measured MES determines both detection (through the threshold of $7.1$ applied by the \textit{Kepler} DR25 pipeline) and confirmation, we evaluate the confirmation conditional on detection, and write the selection function $S_i(x')$, the probability that planet $i$ at true log-radius $x'$ is detected and confirmed, as
\begin{equation}\label{eq:sfull}
S_i(x')=\Phi\!\left(\frac{x'-c_i-\delta_i}{s_D}\right)
\frac{\mathcal{I}_i(x')}{\Phi\!\big([\hat\xi_i(x')-\log_{10}7.1]/s_\xi\big)},
\end{equation}
\begin{equation}\label{eq:confint}
\mathcal{I}_i(x')=\int_{\xi_{\rm cut}}^{\infty}\Phi\!\left(\frac{\xi-\xi_0-\Delta_g\,g_i-\mathcal{H}\,y_i}{s_\eta}\right)\mathcal{N}\!\big(\xi;\hat\xi_i(x'),s_\xi^2\big)\,\mathrm{d}\xi.
\end{equation}
The first factor of Eq.~\eqref{eq:sfull} is the detection probability $p^{\rm det}_i$ of Eq.~\eqref{eq:det} without its ceiling $C$. The ratio that follows is the confirmation probability conditional on detection. Its denominator is the probability that the measured MES exceeds the detection threshold. Its numerator $\mathcal{I}_i(x')$ averages the confirmation probability of Eq.~\eqref{eq:conf}, without its ceiling $\eta_{\max}[g_i]$, over the measured $\xi$ above $\xi_{\rm cut}$, with $\mathcal{N}(\xi;\hat\xi_i(x'),s_\xi^2)$ the Gaussian density of $\xi$ of Eq.~\eqref{eq:mesmeas}. 
The lower limit is $\xi_{\rm cut}=\log_{10}7.1$, or $\log_{10}15$ in the control restricted to $\mathrm{MES}\ge15$. In the threshold of the confirmation, $\mathcal{H}\,y_i$ is the age term described below. $S_i$ thus depends on planet $i$ through its host, $P_i$, $b_i$, its transit duration, $g_i$ and $t_i$.

We omit the constant $C$ and $\eta_{\max}[g_i]$ because a ceiling that does not depend on radius cancels in the normalisation of Eq.~\eqref{eq:znorm} (Sect.~\ref{sec:hier}). The paragraph on the detection ceiling below bounds the effect of a ceiling that rises with radius. The planet-level criteria (c), (e) and (f) enter through the confirmation, since a candidate counts as confirmed only if it passes them. 
The conditioning on detection uses the MES threshold alone and neglects the correlation between the measured MES and the Robovetter classification.
The Robovetter retains $84\%$ of the recovered injections with a measured MES of $7.1$--$8$ and $93$--$96\%$ of those at $8$--$15$, while only $5\%$ of the candidates with a MES of $7.1$--$10$ are confirmed, so the neglected correlation acts where the confirmation probability is already small. For the twelve planets of $\mathcal{D}_3$ that the DR25 vetting classifies as false positives (Sect.~\ref{sec:sample}), $S_i$ is computed as if the vetting had retained them as candidates.

\paragraph{Age term.}
Confirmation may also depend on stellar age, for example because magnetic activity is generally stronger in young stars. 
The $2603$ candidates orbit $1965$ stars. Of these, $1418$ host at least one confirmed planet, and Paper~I provides a gyrochronological age for $1258$ of them. For the remaining $547$ stars, which host only unconfirmed candidates, we derive ages with the procedure and rotation catalogues of Paper~I, obtaining an age for $381$ of them.
Of the $2603$ candidates, $695$, orbiting $496$ stars, have host ages that satisfy the criteria of Sect.~\ref{sec:sample}.
For these candidates, we add to the confirmation threshold of Eq.~\eqref{eq:conf} the term $\mathcal{H}\,y_i$, with
\begin{equation}
y_i=t_i^{-1/2}-t_{\rm ref}^{-1/2},
\end{equation}
where $t_i$ is the host age in Gyr and $t_{\rm ref}=2.59$\,Gyr is the median host age of these candidates. The $t^{-1/2}$ form approximates the decline of rotation and chromospheric activity with age \citep{Skumanich1972}.

We fit Eq.~\eqref{eq:conf} to all $2603$ candidates twice, using a Bernoulli likelihood in both cases. The first fit, without the age term, yields the values of $\xi_0$, $s_\eta$ and $\Delta_g$ quoted above. In the second, joint fit, we add $\mathcal{H}\,y_i$ to the confirmation threshold of the $695$ candidates whose host ages satisfy the host-level criteria of Sect.~\ref{sec:sample}. The remaining $1908$ candidates share a single, age-independent shift and therefore do not constrain $\mathcal{H}$. We estimate $\mathcal{H}$ from its likelihood profile.
For each value of $\mathcal{H}$ on a grid we refit the other parameters and record the maximum likelihood. Treated as a probability distribution of $\mathcal{H}$, this profile gives $\mathcal{H}=+0.033^{+0.036}_{-0.038}$\,dex\,Gyr$^{1/2}$, corresponding to a threshold $0.030$\,dex higher at $0.5$ than at $4$\,Gyr. With $g_{\rm mc}$ in place of $g$ in Eq.~\eqref{eq:conf} and in the joint fit, it is $\mathcal{H}=+0.050^{+0.037}_{-0.038}$\,dex\,Gyr$^{1/2}$. In Sect.~\ref{sec:fse}, we leave $\mathcal{H}$ free, using this likelihood profile as its prior, thereby propagating its uncertainty into the SE fractions.

In Eq.~\eqref{eq:confint} we take $\mathcal{H}$ from the joint fit and $\xi_0$, $s_\eta$ and $\Delta_g$ from the fit without the age term, because the availability of an age is not independent of confirmation. An age is available for $89\%$ of the stars that host a confirmed planet, often the candidate itself, but only for $70\%$ of the others. Candidates with a host age are therefore confirmed more often at a given MES, and in the joint fit their threshold is $\xi_0=1.036$, against $1.060$ for the others. This offset is absorbed by the shared shift of the candidates without an age, so it does not enter $\mathcal{H}$, which measures only how the threshold varies with age among the candidates with an age.
$\mathcal{H}$ would be biased only if the difference in age availability between the two groups of stars varied with age. Setting $\mathcal{H}=0$ raises the age contrast $\bar f_{2\text{--}4}-\bar f_{<1}$ by $0.013$ at the maximum a posteriori parameters of the three-interval form of $f_{\rm SE}^{\rm m}(t)$ (Sect.~\ref{sec:hier}), about a tenth of its uncertainty, so the age term has a negligible effect on the contrast.

We repeat the full fit of Sect.~\ref{sec:hier}, with $\mathcal{H}$ marginalised, on the $460$ planets with $\mathrm{MES}\ge15$, raising the lower limit $\xi_{\rm cut}$ of Eq.~\eqref{eq:confint} to $\log_{10}15$ (Sect.~\ref{sec:compl}). In this range the confirmation probability of Eq.~\eqref{eq:conf} is close to its ceiling and the age term has little effect. The age contrast $\bar f_{2\text{--}4}-\bar f_{<1}$ is then $+0.32^{+0.09}_{-0.11}$, against $+0.24^{+0.10}_{-0.10}$ on the full sample (Sect.~\ref{sec:resglobal}). 
At the maximum a posteriori parameters of the three-interval form of $f_{\rm SE}^{\rm m}(t)$, constant within the age intervals $0.1$--$1$, $1$--$2$ and $2$--$4$\,Gyr (Sect.~\ref{sec:hier}), replacing $y_i$ with a step function ($y_i=1$ below $1$\,Gyr and $0$ above) or with $\log_{10}(t_i/t_{\rm ref})$ changes the age contrast $\bar f_{2\text{--}4}-\bar f_{<1}$ by $-0.003$ and $+0.002$, respectively, and moving $t_{\rm ref}$ to $1.5$ or $4$\,Gyr changes it by $-0.002$ and $+0.001$.

\paragraph{Population completeness.}
The population completeness $\bar S$ of Sect.~\ref{sec:compl} is the probability that a planet of radius $R_p$ is selected, averaged over the host--orbit configurations of the planets $j$ of the sample or age subsample considered,
\begin{equation}\label{eq:sbar}
\bar S(R_p)=\frac{\sum_j \bar Z_j^{-1}\,S_j(x';P_j)}{\sum_j \bar Z_j^{-1}},
\qquad
x'=\log_{10}(R_p/R_\oplus),
\end{equation}
\begin{equation}\label{eq:sbari}
\bar S_i=\frac{\sum_j \bar Z_j^{-1}\,S_j(x_i;P_i)}{\sum_j \bar Z_j^{-1}},
\qquad
\bar Z_j=\bar w_j\,Z_{{\rm SE},j}+\big(1-\bar w_j\big)\,Z_{{\rm SN},j}.
\end{equation}
Weighting each configuration by $1/\bar Z_j$ corrects for the over-representation, among the detected planets, of the configurations that select planets easily. 
In Eq.~\eqref{eq:sbar}, $S_j(x';P_j)$ is the selection function of Eq.~\eqref{eq:sfull} for the host, impact parameter, confirmed-sibling status, age term and period $P_j$ of planet $j$. In Eq.~\eqref{eq:sbari}, $S_j(x_i;P_i)$ is the same function with the orbit of planet $j$ moved to the period $P_i$ of planet $i$, evaluated at the measured log-radius $x_i$ of planet $i$. Moving the orbit from $P_j$ to $P_i$ scales the transit duration as $(P_i/P_j)^{1/3}$ and the number of transits as $P_j/P_i$, and replaces $P_j$ with $P_i$ in Eqs.~\eqref{eq:delta} and \eqref{eq:mesmap}. $\bar S(R_p)$ thus describes the completeness of the sample as a function of radius and corrects the radius distributions, while $\bar S_i$ is the completeness of a planet like planet $i$ and gives its weight $1/\bar S_i$ in the weighted counts.

$\bar Z_j$ is the fraction of the fitted mixture that the configuration of planet $j$ selects, with $Z_{{\rm SE},j}$ and $Z_{{\rm SN},j}$ the fractions of the two components of Eq.~\eqref{eq:znorm} selected within the radius range of the sample. They are evaluated at the maximum a posteriori parameters of the three-interval form of $f_{\rm SE}^{\rm m}(t)$ (Sect.~\ref{sec:fse}), with $\mathcal{H}$ fixed at the median of its profile, $+0.033$\,dex\,Gyr$^{1/2}$. The weight $\bar w_j$ is that of Eq.~\eqref{eq:wlogit} for planet $j$, with $f_{\rm SE}^{\rm m}(t)$ held at a single value at all ages, $0.771$, at which the standardised weight $\bar f_{\rm SE}^{\rm m}$ of Eq.~\eqref{eq:wstd} equals $0.67$, the average of $\bar f_{<1}$, $\bar f_{1\text{--}2}$ and $\bar f_{2\text{--}4}$ weighted by the number of planets of $\mathcal{D}_3$ in each interval. The age trend measured by the fit therefore does not enter the weights with which the counts are compared across age. The ceilings $C$ and $\eta_{\max}[g_j]$ are omitted from $S_j$, so $\bar S$ is relative to a unit ceiling and tends to unity at large radii. Planets with $\bar S_i<0.25$ receive zero weight (Sect.~\ref{sec:compl}), which bounds the weights $1/\bar S_i$ at $4$.

\paragraph{Radius and age uncertainty implementation.}
For each planet, $x_i=\log_{10}(R_{p,i}/R_\oplus)$ and the radius error is approximated as symmetric Gaussian with standard deviation $\sigma_{x,i}$, obtained by averaging the reported upper and lower radius errors and dividing by $R_{p,i}\ln10$. The median asymmetry between the two error sides is $0.005$\,dex, smaller than the fitted intrinsic component widths ($0.155$\,dex for SEs and $0.081$\,dex for SNs). Repeating the maximum a posteriori fit with either error side changes the age contrast by at most $0.05\sigma$ and the component widths by less than $0.002$\,dex.

The conditional true-radius density entering Eq.~\eqref{eq:seldef} is
\begin{equation}\label{eq:postdef}
p(x'\mid x_i,{\rm k})\propto
\mathcal{N}(x';\mu_{\rm k},\sigma_{\rm k}^2)\,
\mathcal{N}(x_i;x',\sigma_{x,i}^2).
\end{equation}
We assume that selection at fixed true radius is independent of the uncertainty of the radius measurement. The probability that a planet of true log-radius $x'$ is measured within the analysis range is
\begin{equation}\label{eq:window}
W_i(x')=\Phi\!\left(\frac{\log_{10}4-x'}{\sigma_{x,i}}\right)
-\Phi\!\left(\frac{\log_{10}0.5-x'}{\sigma_{x,i}}\right).
\end{equation}

We represent the Paper~I age uncertainty by a two-sided Gaussian centred on the reported age, with lower and upper widths reproducing its quoted percentiles and equal probability on either side. We use $N_{\rm d}=200$ draws per planet. Draws are independent even for planets sharing a host; $293$ of the $581$ planets of $\mathcal{D}_3$ belong to $116$ multi-planet hosts. Using common age draws for planets around the same host changes the maximum a posteriori age contrast by $0.1\sigma$ and the $90\%$ upper bound on $t_{90}$ by $-0.1$\,Gyr, comparable to Monte Carlo scatter. Radius measurement errors are likewise treated as independent, although within-system correlations may be present.

The age term in the confirmation threshold, $\mathcal{H}\,y_i$, is evaluated at the Paper~I age estimate rather than at each sampled age. For fixed $\mathcal{H}$ and Gaussian component parameters, the component selection factors $\epsilon_{{\rm k},i}$ and selected fractions $Z_{{\rm k},i}$ are therefore constant across the age draws of a given planet; only the mixture weight $w_i(t_i^{(r)})$ changes $Z_i^{(r)}$. The age draws in Eq.~\eqref{eq:like} are equally weighted. Weighting them instead by their selected fractions changes the maximum a posteriori age contrast by $-0.01$ ($0.1\sigma$) and $t_{90}$ by less than $0.01$\,Gyr.

The period term is retained because allowing $\beta_P$ to vary improves the fit by $\Delta\mathrm{BIC}=40.7$ relative to $\beta_P=0$. 

\paragraph{Numerical evaluation.}
We evaluate the integrals $Z_{{\rm k},i}$ and $\epsilon_{{\rm k},i}$ of Sect.~\ref{sec:hier}, using Gauss--Legendre quadrature over $x'$ and the Gauss--Hermite quadrature over the true-radius posterior $p(x'\mid x_i,{\rm k})$, which is Gaussian with mean and variance
\begin{equation}\label{eq:truepost}
\bar x_{{\rm k},i}=\frac{\mu_{\rm k}\,\sigma_{x,i}^{2}+x_i\,\sigma_{\rm k}^{2}}{\sigma_{\rm k}^{2}+\sigma_{x,i}^{2}},
\qquad
\tilde\sigma_{{\rm k},i}^{2}=\frac{\sigma_{\rm k}^{2}\,\sigma_{x,i}^{2}}{\sigma_{\rm k}^{2}+\sigma_{x,i}^{2}},
\end{equation}
with $\mu_{\rm k}$ and $\sigma_{\rm k}$ the mean and width of component k and $\sigma_{x,i}$ the uncertainty of $x_i$. Since $\mathcal{H}$ is marginalised in the mixture fit of the planets (Sect.~\ref{sec:fse}), with the profile described above as its prior, we compute $S_i$ on a grid of values of $\mathcal{H}$ and interpolate linearly between them. Compared with the exact integrals for a selection given by a single normal cumulative distribution in log-radius, $\Phi[(x'-c_i)/s_D]$, for which they have a closed form, and with $S_i$ computed directly at values of $\mathcal{H}$ drawn from the posterior, the quadratures and the interpolation change $\ln\mathcal{L}$ by at most $2\times10^{-3}$. For comparison, in a Gaussian posterior a parameter moves by one standard deviation when $\ln\mathcal{L}$ changes by $0.5$. Numerical errors of this size therefore have no effect on the results.

\paragraph{Controls.}
We test whether confirmation depends on planetary class by adding a threshold shift $\Delta_{\rm SE}$ to Eq.~\eqref{eq:conf} for candidates whose DR25 radius lies below the valley locus of Eq.~\eqref{eq:slope} at their period (SEs), leaving the threshold unchanged for SNs. The argument of $\Phi$ in Eq.~\eqref{eq:conf} then becomes $(\xi-\xi_0-\Delta_g\,g-\Delta_{\rm SE}\,\mathcal{V}_{\rm SE})/s_\eta$, with $\mathcal{V}_{\rm SE}=1$ for the SEs and $\mathcal{V}_{\rm SE}=0$ for the SNs.
The fit gives $\Delta_{\rm SE}=-0.010\pm0.009$ in $\log_{10}\mathrm{MES}$, with the uncertainty from $200$ bootstrap resamplings of the host stars, and improves the likelihood by only $2\Delta\ln\mathcal{L}=1.2$ with one additional parameter. Thus, at a given MES, confirmation shows no significant dependence on planetary class. 
In each of the following controls, the three-interval form of $f_{\rm SE}^{\rm m}(t)$ is refitted to its maximum a posteriori with $\mathcal{H}$ fixed at the median of its profile, $+0.033$\,dex\,Gyr$^{1/2}$, and the age contrast $\bar f_{2\text{--}4}-\bar f_{<1}$ is compared with that of the same fit without the change, the reference fit of Sect.~\ref{sec:resglobal}. Excluding the twelve planets of $\mathcal{D}_3$ that the DR25 vetting classifies as false positives, although they are confirmed (Sect.~\ref{sec:sample}), changes it by $+0.036$, and using $g_{\rm mc}$ in place of $g$ changes it by $-0.013$. Drawing $d_0$, $d_b$, $d_P$, $d_L$, $s_D$, $\xi_0$, $s_\eta$ and $\Delta_g$ within their bootstrap errors and $z_0$ within the error of its median, $0.005$\,dex, independently in $40$ draws, moves it by at most $0.006$. In a separate set of $30$ draws, drawing the impact parameters within their DR25 errors moves it by at most $0.011$.

\paragraph{Detection ceiling.}
Equation~\eqref{eq:det} assumes a common, radius-independent detection ceiling $C$.
A positive $L_i$ means that the CDPP grows with the transit duration. For white noise the CDPP decreases with the transit duration, whereas astrophysical noise can make it grow \citep{BurkeCatanzarite2017b}, and targets with a positive $L_i$ have a suppressed recovery of transits \citep{BurkeCatanzarite2017}.
Hosts with $L_i>0.1$ account for $64\%$ of the planets of $\mathcal{D}_3$ in the $0.1$--$1$\,Gyr subset, compared with $11\%$ in the $2$--$4$\,Gyr subset. 
Around these hosts, the fraction of injections recovered and retained between $0.05$ and $0.2$\,dex above $c_i+\delta_i$ is $0.44$--$0.77$, compared with $0.92$--$0.96$ around hosts with $L_i\le0$. Yet all $20$ injections more than $0.2$\,dex above $c_i+\delta_i$ are recovered and retained, an outcome with probability $0.009$ under a constant ceiling. 
This indicates that the detection ceiling increases with radius for these hosts. As a test, for hosts with $L_i>0$ we replace $C$ in Eq.~\eqref{eq:det} with
\begin{equation}\label{eq:ceilrise}
C_i(x')=C_L+(C_0-C_L)\,\Phi\!\left(\frac{x'-c_i-\delta_i-x_{\rm r}}{s_{\rm r}}\right).
\end{equation}
Here $C_0$ is the ceiling of the hosts with $L_i\le0$, and $C_L$ the ceiling of the hosts with $L_i>0$ at small radii, with one value in each of three ranges of $L_i$, $0<L_i\le0.1$, $0.1<L_i\le0.2$ and $L_i>0.2$. As the radius increases, $C_i(x')$ goes from $C_L$ to $C_0$. It lies halfway between them at $x'=c_i+\delta_i+x_{\rm r}$, and goes from $16\%$ to $84\%$ of the change between $x'=c_i+\delta_i+x_{\rm r}-s_{\rm r}$ and $c_i+\delta_i+x_{\rm r}+s_{\rm r}$. We fit these parameters by maximum likelihood on the injections more than $0.05$\,dex above $c_i+\delta_i$. This injection fit gives $C_0=0.955\pm0.004$ and, in the three ranges of $L_i$, $C_L=0.944\pm0.026$, $0.889\pm0.037$ and $0.716\pm0.038$, with standard deviations from $100$ bootstrap resamplings of the targets. It places $s_{\rm r}$ at the smallest value it allows, $0.005$\,dex, far below $s_D$. The change from $C_L$ to $C_0$ is therefore a step function, with $C_i(x')=C_L$ if $x'<c_i+\delta_i+x_{\rm r}$ and $C_i(x')=C_0$ otherwise. The $95\%$ interval of $x_{\rm r}$ from the profile likelihood is $0.18$--$0.42$\,dex; it is wide because only $46$ of the injections around the hosts with $L_i>0$ lie more than $0.15$\,dex above $c_i+\delta_i$. Unlike the constant ceiling $C$, $C_i(x')$ does not cancel in the normalisation of Eq.~\eqref{eq:znorm}. Because the injections constrain $x_{\rm r}$ only loosely, we keep the constant ceiling $C$ in the mixture fit of Sect.~\ref{sec:hier}, which gives the results of Sect.~\ref{sec:resglobal}. To measure the effect of the rise of the detection ceiling of the hosts with $L_i>0$, we refit the three-interval form of $f_{\rm SE}^{\rm m}(t)$ to its maximum a posteriori with $\mathcal{H}$ fixed at the median of its profile, as in the controls of the previous paragraph, with $C_i(x')$ in place of $C$ and $x_{\rm r}=0.18$ or $0.42$\,dex, the two ends of its $95\%$ profile-likelihood interval. The SE fraction below $1$\,Gyr, where hosts with $L_i>0.1$ are common, then increases, and the age contrast $\bar f_{2\text{--}4}-\bar f_{<1}$, $+0.24^{+0.10}_{-0.10}$ in Sect.~\ref{sec:resglobal}, decreases by $0.014$ or $0.033$, while the probability that it is positive falls from
$0.995$ to $0.991$ or $0.986$. Thus, the constant ceiling raises the age contrast by at most $0.033$ and does not produce its positive sign.

\subsection{Priors and fitted parameters}\label{app:compl:posterior}
The parameters of the mixture fit of Sect.~\ref{sec:hier} are the four component parameters $\{\mu_{\rm SE},\sigma_{\rm SE},\mu_{\rm SN},\sigma_{\rm SN}\}$, $\beta_P$ and $\gamma$ of Eq.~\eqref{eq:wlogit}, the age term $\mathcal{H}$ of the confirmation threshold (Eq.~\eqref{eq:confint}), and the parameters of the functional form of $f_{\rm SE}^{\rm m}(t)$ in use, $(f_{<1},f_{1\text{--}2},f_{2\text{--}4})$ or $(f_0,f_\infty,\tau_{\rm E})$. We sample them with \texttt{emcee} \citep{ForemanMackey2013}, with $40$ walkers over $16\,000$ steps for the three-interval form and two independent chains of $24$ walkers over $10\,000$ steps for the exponential form, discarding the first third of every chain as burn-in. Priors are flat within $\mu_{\rm SE}\in\log_{10}[0.8,1.9]$, $\mu_{\rm SN}\in\log_{10}[1.9,3.5]$ (the shared boundary at $1.9\,R_\oplus$ enforces the component ordering), $\sigma_{\rm SE}\in[0.02,0.30]$ and $\sigma_{\rm SN}\in[0.03,0.40]$\,dex, $\beta_P\in[-8,8]$ per dex, $\gamma\in[-8,8]$, $f_{<1},f_{1\text{--}2},f_{2\text{--}4},f_0,f_\infty\in[0.02,0.98]$ and $\log_{10}(\tau_{\rm E}/\mathrm{Gyr})\in[-1.5,1.5]$, with two exceptions. The prior on $\mathcal{H}$ is its likelihood profile on the $2603$ DR25 candidates (Appendix~\ref{app:compl:probit}), within $[-0.14,0.20]$\,dex\,Gyr$^{1/2}$. In the exponential form, the prior on $f_\infty$ is a Gaussian truncated to $[0.02,0.98]$, centred on the median of $f_{2\text{--}4}$ from the three-interval form with half its $16$--$84\%$ range as width.

Fitted on $\mathcal{D}_3$ ($N=581$), the three-interval form places the SE component at $\mu_{\rm SE}=0.042^{+0.018}_{-0.022}$ in $\log_{10}(R_p/R_\oplus)$, that is $1.10^{+0.05}_{-0.05}\,R_\oplus$, with width $\sigma_{\rm SE}=0.155^{+0.019}_{-0.015}$\,dex, and the SN component at $\mu_{\rm SN}=0.387^{+0.008}_{-0.009}$, that is $2.44^{+0.05}_{-0.05}\,R_\oplus$, with width $\sigma_{\rm SN}=0.081^{+0.008}_{-0.007}$\,dex, with $\beta_P=-2.55^{+0.43}_{-0.47}$ per dex and $\gamma=-0.57\pm0.31$. The component parameters lie inside their prior bounds, with unimodal marginals. The posterior of $\mathcal{H}$, $+0.016^{+0.034}_{-0.035}$\,dex\,Gyr$^{1/2}$, keeps the width of its prior, $+0.033^{+0.036}_{-0.038}$\,dex\,Gyr$^{1/2}$, so the planets of $\mathcal{D}_3$ add little information on $\mathcal{H}$ to that of the $2603$ candidates, and its uncertainty accounts for $1.4\%$ of the variance of the age contrast $\bar f_{2\text{--}4}-\bar f_{<1}$, a negligible contribution.

The exponential form, with the plateau propagated from the three-interval form, gives $f_0=0.38^{+0.30}_{-0.25}$, $f_\infty=0.795^{+0.040}_{-0.041}$, and $\log_{10}(\tau_{\rm E}/\mathrm{Gyr})=-0.30^{+0.38}_{-0.49}$, with the component parameters within $0.12\sigma$ of those of the three-interval form. The fitted plateau $f_\infty$ refers to a planet at $P_{\rm ref}$ without a confirmed sibling (Eq.~\eqref{eq:wlogit}); averaging the corresponding mixture weights over the periods and confirmed-sibling status of the planets of $\mathcal{D}_3$ (Eq.~\eqref{eq:wstd}) gives the standardised plateau $\bar f_\infty=0.69\pm0.04$.
The posteriors of $f_0$ and $\tau_{\rm E}$ extend to the bounds of their priors, so the data set no lower limit on $t_{90}$, only an upper bound.
Over the $88\%$ of the posterior draws in which $\bar f_{\rm SE}^{\rm m}$ starts below $0.9\,\bar f_\infty$, $t_{90}$ is below $1.9$\,Gyr with $90\%$ probability and below $2.6$\,Gyr with $95\%$ probability.

To test the effect of the Gaussian prior on $f_\infty$, we refit the exponential form with a flat prior on $f_\infty$ within $[0.02,0.98]$, sampled with $40$ walkers over $16\,000$ steps. The fit gives $f_\infty=0.82\pm0.06$, and over the $88\%$ of the posterior draws in which $\bar f_{\rm SE}^{\rm m}$ starts below $0.9\,\bar f_\infty$, $t^{f}_{90}$, the analogue of $t_{90}$ for this test fit, is below $3.8$\,Gyr with $90\%$ probability.

\subsection{Tests on mock catalogues}\label{app:compl:injrec}
We test the mixture fit of Sect.~\ref{sec:hier} on mock catalogues that reproduce $\mathcal{D}_3$ planet by planet. Each of the $581$ planets keeps its orbit, host, confirmed-sibling status and radius uncertainty, and the age estimate of its host is taken as its true age. We draw its class with the weight of Eq.~\eqref{eq:wlogit} under a chosen form of $f_{\rm SE}^{\rm m}(t)$, and its true log-radius from the Gaussian component of that class. The draw is accepted with probability $S_i$ (Eq.~\eqref{eq:sfull}) if the radius, measured with the planet's own error, falls inside $0.5$--$4\,R_\oplus$, and is repeated otherwise, so that every mock catalogue holds $581$ planets that passed the selection. The age of each planet is then displaced from the true one by a random draw from the two-sided uncertainty of its host age (Sect.~\ref{sec:starpars}). 
The component parameters, $\beta_P$, and $\gamma$ are set at the maximum a posteriori of the exponential form fitted to $\mathcal{D}_3$ with a flat prior on $f_\infty$ and $\mathcal{H}$ fixed at the median of its profile, $+0.033$\,dex\,Gyr$^{1/2}$. The same fit gives $f_0=0.02$, $f_\infty=0.81$ and $\log_{10}(\tau_{\rm E}/\mathrm{Gyr})=-0.35$. Each mock catalogue is refitted as in Sect.~\ref{sec:hier} at a reduced cost, with $24$ walkers over $3000$ steps, $20$ age draws per planet and $\mathcal{H}$ fixed, with the three-interval form and then with the exponential form of $f_{\rm SE}^{\rm m}(t)$, whose plateau is propagated from the three-interval form fitted to the same mock catalogue.

We first test whether the fit recovers a known rise of the SE fraction. We draw $24$ mock catalogues with $f_{\rm SE}^{\rm m}(t)$ equal to the exponential form with $f_0=0.02$, $f_\infty=0.81$ and $\log_{10}(\tau_{\rm E}/\mathrm{Gyr})=-0.35$ obtained above, for which the age contrast $\bar f_{2\text{--}4}-\bar f_{<1}$ is $+0.156$ and $t_{90}=1.08$\,Gyr. The age contrast fitted on the mock catalogues has a median of $+0.18$, and its $16$--$84\%$ interval contains $+0.156$ in $18$ of the $24$, against $16$ expected. 
The $90\%$ upper bound on $t_{90}$ lies above $1.08$\,Gyr in $21$ of the $24$, with a median of $2.4$\,Gyr. We then test a slower rise, with $\tau_{\rm E}$ three times longer and $t_{90}=3.2$\,Gyr.
The bound moves up with it, to a median of $4.6$\,Gyr, lies above $3.2$\,Gyr in $12$ of $12$ mock catalogues, and is never as low as the $1.9$\,Gyr measured on $\mathcal{D}_3$ (Sect.~\ref{sec:resglobal}). The measured bound therefore follows from the data, not from the prior on $f_\infty$, and disfavours a rise as slow as $t_{90}=3.2$\,Gyr.

Finally, we test whether the fit produces a rise where there is none. In $50$ mock catalogues with $f_{\rm SE}^{\rm m}(t)$ held constant with age at $0.81$, fitted with the three-interval form only and with $2000$ steps, the age contrast scatters around zero, with a mean of $+0.02\pm0.01$. Only one of the mock catalogues reaches an age contrast $\bar f_{2\text{--}4}-\bar f_{<1}$ at least as large and as significant as the one measured on $\mathcal{D}_3$ (Sect.~\ref{sec:resglobal}), equal to $+0.24$ and positive in $99.5\%$ of the posterior. The fit therefore shows no significant bias in the age contrast when the SE weight does not change
with age.

\subsection{The mixture components across age}\label{app:compl:widths}
The model of Sect.~\ref{sec:hier} holds the means and widths of the two components fixed with age. We let $\mu_{\rm k}$ (${\rm k}={\rm SE},{\rm SN}$) vary linearly with the age estimate $t_i$ of each host, as $\mu_{\rm k}+\kappa^{t}_{\rm k}(\log_{10}t_i-\langle\log_{10}t\rangle)$ with $\langle\log_{10}t\rangle$ the median over $\mathcal{D}_3$, or $\ln\sigma_{\rm k}$ in the same way, and refit the three-interval form to its maximum a posteriori as in the controls of Appendix~\ref{app:compl:probit}. The likelihood improves by $2\Delta\ln\mathcal{L}=2.2$ for the means and $0.4$ for the widths, that is $\Delta\mathrm{BIC}=10.5$ and $12.4$ in favour of fixed components ($22.1$ with both free).

With the two planetary classes split at the fixed boundary $R_p=1.71\,R_\oplus$, the locus of Eq.~\eqref{eq:slope} at the $10$\,d reference period, as in Sect.~\ref{sec:resmass}, the median radii of the two classes also show no significant change with age. We compute them with the Monte Carlo of Sect.~\ref{sec:ageshift}, with each planet's class reassigned on every draw. On $\tilde{\mathcal{D}}_3$, where $195$ of the $541$ planets lie below the class boundary, and with each planet weighted by $1/\bar S_i$ (Sect.~\ref{sec:compl}), the SE median radius is $1.32^{+0.05}_{-0.05}$, $1.31^{+0.05}_{-0.04}$ and $1.38^{+0.03}_{-0.03}\,R_\oplus$ in the $0.1$--$1.5$, $1.5$--$2.5$ and $2.5$--$4$\,Gyr age bins, and the unbinned slope is $\mathrm{d}\log R_{\rm SE}/\mathrm{d}\log t=+0.018^{+0.022}_{-0.022}$ ($+0.016^{+0.020}_{-0.020}$ on the observed counts). 
The SN median radius is $2.27^{+0.08}_{-0.07}$, $2.35^{+0.07}_{-0.06}$ and $2.40^{+0.05}_{-0.05}\,R_\oplus$ in the same three bins, and $\mathrm{d}\log R_{\rm SN}/\mathrm{d}\log t=+0.015^{+0.019}_{-0.017}$ ($+0.016^{+0.019}_{-0.017}$).
Classified instead at each planet's own period, along the locus of Eq.~\eqref{eq:slope}, the two slopes are $+0.003^{+0.022}_{-0.022}$ and $+0.006^{+0.019}_{-0.020}$ ($+0.002$ and $+0.011$ on the observed counts), within the errors of the values above.

In the model of Sect.~\ref{sec:hier} the components are symmetric and keep their means and shapes at all ages. Thus, the valley of the mixture, the minimum $x_{\rm min}$ of its radius distribution, can move only through the change of the SE weight with age, a higher SE weight moving the minimum to larger radii. We test whether this, or the age dependence of the means fitted above, can reproduce the shift of the locus at $10$\,d measured in Sect.~\ref{sec:ageshift}, $0.07$\,dex ($0.24^{+0.07}_{-0.27}\,R_\oplus$) lower in the $1$--$2$ than in the $2$--$4$\,Gyr subset. 
In the exponential form of $f_{\rm SE}^{\rm m}(t)$, the standardised SE weight averaged over the planets of each subset rises by only $0.015^{+0.036}_{-0.015}$ from the $1$--$2$ to the $2$--$4$\,Gyr subset, since the rise is mostly complete before $2$\,Gyr (Sect.~\ref{sec:resglobal}), and $x_{\rm min}$ rises by $0.004^{+0.009}_{-0.004}$\,dex, a twentieth of the measured shift. With the means varying with age at the slopes obtained from the fit
above, $x_{\rm min}$, which depends mostly on $\mu_{\rm SN}$ ($\partial x_{\rm min}/\partial\mu_{\rm SN}=0.82$ against $0.18$ for $\mu_{\rm SE}$), rises by $0.013^{+0.009}_{-0.009}$\,dex, a fifth of the measured shift of Sect.~\ref{sec:ageshift}, and the whole shift would require $\kappa^{t}_{\rm SN}=0.26$, against the fitted $+0.035\pm0.027$. If the shift is real, it therefore comes from a change in the shape of a component, such as the change in the asymmetry of the radius distribution of the SEs that Sect.~\ref{sec:ageshift} measures and Sect.~\ref{sec:disc} interprets, which the mixture does not model.


\subsection{Quantile skewness of the super-Earth distribution}
\label{app:skewtest}

For planets with $R_p<R_{\rm gap}(10\,\mathrm{d})$, let $\ell_{16}$, $\ell_{50}$ and $\ell_{84}$ denote the 16th, 50th and 84th percentiles of $\log_{10}R_p$. We define the quantile skewness as
\begin{equation}
\mathcal{A}_{\rm skew} =
\frac{\ell_{84}+\ell_{16}-2\ell_{50}}
{\ell_{84}-\ell_{16}}.
\end{equation}
Positive values indicate a distribution with a peak at smaller radii and a tail towards the class boundary; negative values indicate the reverse. The statistic vanishes for a symmetric distribution.

To test whether the mixture model of Sect.~\ref{sec:fse} can reproduce the observed difference in skewness, we generated $2000$ mock catalogues from posterior draws of the three-interval model of $f_{\rm SE}^{\rm m}(t)$, incorporating the selection function of Sect.~\ref{sec:compl} and the measured stellar ages. Since the two components are symmetric and retain fixed shapes across age, skewness can arise only from component overlap and selection.

For each catalogue, we compute the weighted difference between the $1$--$2$ and $2$--$4$\,Gyr subsets. At the measured ages, the observed difference is $+0.28$ ($+0.39$ unweighted). Mock catalogues produce an absolute difference at least this large in $19\%$ ($5\%$ unweighted) of cases. Thus, the weighted skewness difference does not provide conclusive evidence for an intrinsic evolution of the component shapes at this sample size.

\section{Checks for systematic effects}\label{app:robust}

\subsection{Selection of hosts with a gyrochronological age}\label{app:resgate}
A gyrochronological age exists only where a rotation period is detected, and the detection of a period favours large, long-lived spots, and so young, active hosts \citep{WalkowiczBasri2013,McQuillan2014}.
The \emph{parent sample} of $\tilde{\mathcal{D}}_3$ holds $1794$ planets around $1324$ hosts. It comprises the planets of $\tilde{\mathcal{D}}_1$ that are DR25 KOIs around hosts with the DR25 detection metrics, as for $\mathcal{D}_3$ (Sect.~\ref{sec:sample}).
Only $10$ of these hosts are excluded because no rotation period is detected for them in any of the catalogues of Sect.~\ref{sec:db}. 
All further exclusions come from criterion~(g) of Sect.~\ref{sec:sample}, which requires a non-limit gyrochronological age within $0.1$--$4$\,Gyr and with a reliability tier $\le2$. In particular, of the $1314$ hosts with a rotation period, the age model of Paper~I returns no age for $111$ and only an age limit for another $195$, most of them beyond $4$\,Gyr. Of the $1008$ hosts with a non-limit age, $330$ have a reliability tier $>2$, and $297$ of the remaining $678$ are older than $4$\,Gyr. The $541$ planets of $\tilde{\mathcal{D}}_3$ orbit the remaining $381$ hosts.

Every requirement of criterion~(g) acts on the host, and a selection on the host alone cancels in the normalisation of Eq.~\eqref{eq:znorm}. Criterion~(g) can therefore bias the age trend of the mixture only if, at equal host properties, the hosts of SEs and those of SNs pass one of its requirements at different rates, or if the host mix changes with age and the SE fraction depends on the host. We test the first scenario on the parent sample defined above, requirement by requirement, and the second on $\tilde{\mathcal{D}}_3$ and, with the mixture, on $\mathcal{D}_3$.

For each requirement of criterion~(g), we model with a logistic regression the probability that a planet reaching the requirement passes it. The variables of the regression, all entering linearly, are the class label $Y_i$ ($Y_i=1$ below the valley locus of Eq.~\eqref{eq:slope}), the host colour, the \textit{Kepler} magnitude, the logarithms of the stellar radius and of the orbital period, and the number of planets of the host in the parent sample. For each planet, we predict from the fit its probability of passing with $Y_i=1$ and with $Y_i=0$.
The class term, defined as the mean difference between these two probabilities over the planets, measures whether the hosts of SEs and SNs pass the requirement at different rates at fixed host properties.

Unless otherwise stated, uncertainties in this subsection are estimated from $2000$ host-level bootstrap resamplings, with host ages redrawn from their two-sided Gaussian $\sigma_{\rm tot}$ distributions (Sect.~\ref{sec:starpars}) for age-dependent quantities. Comparisons before and after changes to the sample or host mix use the same bootstrap resamples and age draws, preserving their covariance.

The class term is consistent with zero for the requirements that the age model returns an age, $+0.010^{+0.013}_{-0.014}$, and that the age is not a limit, $+0.030^{+0.021}_{-0.022}$. The requirement of an age within $0.1$--$4$\,Gyr selects on age itself and cannot bias the SE fraction at a given age.

The requirement of a reliability tier $\le2$ for the age is the only requirement that, at equal host properties, the hosts of SEs pass less often than those of SNs, by $0.077^{+0.026}_{-0.027}$. This requirement removes $170$ planets with an age within $0.1$--$4$\,Gyr. Admitting them raises $\Delta f_{\rm SE}^{\rm c}$, the difference between the counted SE fractions $f_{\rm SE}^{\rm c}$ (Sect.~\ref{sec:ageshift}) of the $2$--$4$ and $0.1$--$1$\,Gyr bins, by $0.046^{+0.015}_{-0.041}$, while the difference between the $1$--$2$ and $0.1$--$1$\,Gyr bins changes by only $-0.004^{+0.037}_{-0.027}$. 
The extended sample of Appendix~\ref{app:altsamples:extage}, which admits ages with a reliability tier $>2$, also shows the rise of the counted SE fraction with age. The selection of the hosts with an age therefore does not produce the rise of the SE fraction with age.

The host mix also changes across age. The K and late-G hosts, $(G_{\rm BP}-G_{\rm RP})_0>0.95$, hold $75\%$ of the planets at $0.1$--$1$\,Gyr and $74\%$ at $1$--$2$\,Gyr, against $61\%$ at $2$--$4$\,Gyr. These hosts also have a higher SE fraction, $0.41\pm0.03$ against $0.33\pm0.03$, so the host mix raises the SE fraction of the younger bins and works against the rise. Fitted with the likelihood of Sect.~\ref{sec:hier} on the $377$ planets of the K and late-G hosts of $\mathcal{D}_3$ alone, with $\mathcal{H}$ held at the median of its profile, the mixture gives an age contrast $\bar f_{2\text{--}4}-\bar f_{<1}=+0.30^{+0.10}_{-0.11}$, against $+0.24^{+0.09}_{-0.09}$ for the full sample fitted in the same way, with posterior intervals. 
To standardise the host mix of the three age bins, $0.1$--$1$, $1$--$2$ and $2$--$4$\,Gyr, to that of the whole of $\tilde{\mathcal{D}}_3$, we fit a logistic regression of the class label $Y_i$ on the age bin and, linearly, on one of three sets of host properties: the colour, the colour and the logarithm of the stellar radius, or the logarithm of the host mass and [Fe/H]. For each age bin, we then assign all the planets of $\tilde{\mathcal{D}}_3$ to that bin and average their predicted probabilities of being a SE. This gives the SE fraction of the bin at the host mix of $\tilde{\mathcal{D}}_3$. 
For the three sets of host properties, the standardisation moves $\Delta f_{\rm SE}^{\rm c}$ by $+0.002^{+0.006}_{-0.004}$, $+0.002^{+0.008}_{-0.006}$ and $+0.011^{+0.012}_{-0.009}$, respectively. The host mix therefore does not produce the rise of the SE fraction with age.

\subsection{Host colour and the offset of the valley locus across age}\label{app:colourmatch}

The hosts of the $2$--$4$\,Gyr subset ($N=371$) are bluer than those of the $1$--$2$\,Gyr subset ($N=115$), with a median $(G_{\rm BP}-G_{\rm RP})_0$ lower by $0.054$\,mag. As the SN sizes depend on host colour (Sect.~\ref{sec:resmass}), part of the offset of the valley locus of Sect.~\ref{sec:ageshift}, $0.24^{+0.07}_{-0.27}\,R_\oplus$ at $10$\,d between the two subsets, could follow from that difference. With $\kappa^{\rm col}_{\rm SN}=-0.12\pm0.02$ per mag, the median radius of the SNs of the $2$--$4$\,Gyr subset is larger by about $0.04\,R_\oplus$ than it would be at the colour of the $1$--$2$\,Gyr subset, while the SE sizes show no significant dependence on colour. If the valley locus moved with the SNs, the colour difference would raise it in the $2$--$4$\,Gyr subset by at most $0.03\,R_\oplus$, a negligible part of the offset.

To test this directly, we build a sample of $2$--$4$\,Gyr planets with the colour distribution of the $1$--$2$\,Gyr subset. We divide the colour range into eight bins, each holding the same number of planets of the $1$--$2$\,Gyr subset, and draw from each bin, with replacement, three times as many $2$--$4$\,Gyr planets as there are $1$--$2$\,Gyr planets, so that the matched sample has about the size of the full $2$--$4$\,Gyr subset. 
The median colours of the $1$--$2$\,Gyr subset and of this colour-matched $2$--$4$\,Gyr sample then differ by less than $0.005$\,mag. We measure the offset of the valley locus at $10$\,d twice, between the $1$--$2$\,Gyr subset and the full $2$--$4$\,Gyr subset and between the $1$--$2$\,Gyr subset and the colour-matched $2$--$4$\,Gyr sample. 
As in Sect.~\ref{sec:prplane}, we fit the valley locus on the observed counts of $\tilde{\mathcal{D}}_3$, without completeness weights, and re-derive the SE and SN labels on which it is fitted in each of $1000$ bootstrap resamplings, here of the planets rather than of the hosts. 
As medians over the resamplings, the valley locus of the $1$--$2$\,Gyr subset lies below that of the full $2$--$4$\,Gyr subset by $0.21^{+0.10}_{-0.22}\,R_\oplus$ and below that of the colour-matched $2$--$4$\,Gyr sample by $0.22^{+0.11}_{-0.23}\,R_\oplus$. The difference between the two offsets, which is the part due to the colour difference, is $+0.01^{+0.07}_{-0.08}\,R_\oplus$. The colour difference between the age subsets therefore does not produce the offset of the valley locus.

\subsection{Unresolved stellar companions}\label{app:dilution}

An unresolved stellar companion of the host dilutes the transit depth and leads to an underestimate of the planet radius. Imaging of the host gives a radius correction factor, ${\rm RCF}\ge1$, that removes this bias for the detected companions if the planet orbits the primary star \citep{Furlan2017}. 
We do not apply this correction to the radii adopted in this paper, and here we test its effect. Of the $541$ planets of $\tilde{\mathcal{D}}_3$, $172$ orbit a host around which \citet{Furlan2017} resolve a companion within $4$\arcsec\ and report an RCF. Another $360$ orbit hosts that they observed without reporting a companion, and for these we take ${\rm RCF}=1$. The remaining $9$ orbit hosts that they did not observe and are excluded here. Companions below the detection limits of the imaging remain uncorrected. 
The RCF is small for most of the $172$ planets, with a median of $1.011$.
With the RCF applied, $8$ planets cross the valley locus of Eq.~\eqref{eq:slope} from the SE to the SN side, 
lowering the completeness-weighted counted SE fraction $f_{\rm SE}^{{\rm c},w}$ (Sect.~\ref{sec:ageshift}) by $0.010^{+0.006}_{-0.005}$ and shifting the valley position by $-0.012^{+0.012}_{-0.024}\,R_\oplus$. The uncertainties in this subsection are estimated from $1000$ host-level bootstrap resamplings, propagating the reported RCF errors. The valley-position shift is evaluated over the $694$ resamplings in which both the corrected and uncorrected valleys are resolved.

The planets with hosts of $0.1$--$2$\,Gyr ($N=167$) and of $2$--$4$\,Gyr ($N=365$) have a resolved companion in $27\%$ and $35\%$ of cases, respectively, with a median RCF among these of $1.021$ against $1.009$. Applying the correction lowers $f_{\rm SE}^{{\rm c},w}$ of both subsets, that of the $0.1$--$2$\,Gyr planets by $0.013^{+0.017}_{-0.013}$ more than that of the $2$--$4$\,Gyr planets, or by $0.041^{+0.050}_{-0.039}$ more when the sample is split at $1$\,Gyr. The correction therefore cannot produce the rise of the SE fraction with age of Sects.~\ref{sec:ageshift} and~\ref{sec:resglobal} and would, if anything, enlarge it.

Planet radii underestimated preferentially around young hosts would also lower the valley locus of the $1$--$2$\,Gyr subset, a way in which unresolved companions could produce the offset of the valley locus between the $1$--$2$ and $2$--$4$\,Gyr subsets (Sect.~\ref{sec:ageshift}). 
We therefore refit the valley locus on the $1$--$2$ and $2$--$4$\,Gyr subsets ($N=113$ and $365$) with the corrected radii. To isolate the effect of the radii, we use the same host resamplings and the same SE and SN labels as with the uncorrected radii. Correcting the radii changes the offset of the valley locus at $10$\,d by $+0.00\pm0.02\,R_\oplus$, and the valley locus of the $1$--$2$\,Gyr subset lies below that of the $2$--$4$\,Gyr subset in $74\%$ of the resamplings with and without the radius correction. 
With the SE and SN labels re-derived on the corrected radii, the offset changes by $+0.01^{+0.05}_{-0.05}\,R_\oplus$. Unresolved companions therefore do not produce the offset of the valley locus.

\subsection{Survey composition of the age-resolved sample}
\label{app:surveymix}

The multi-survey sample $\mathcal{D}_2$ contains planets discovered by \textit{Kepler}, K2, and TESS, whose selection functions are not directly comparable. Over $[0.5,4]\,R_\oplus$, the observed SE fractions without completeness correction are $0.42$ for \textit{Kepler} ($N=586$), $0.40$ for K2 ($N=82$), and $0.17$ for TESS ($N=76$).

The relative contributions of these surveys also vary with stellar age. TESS contributes $23$ of the $97$ planets younger than $1$\,Gyr ($24\%$), but only $53$ of the $649$ planets aged $1$--$4$\,Gyr ($8\%$). Consequently, age-dependent changes in the observed SE fraction of $\mathcal{D}_2$ may partly reflect differences in survey composition.

Unlike \textit{Kepler} DR25, K2 and TESS detections originate from heterogeneous searches without a common per-star detection model. An equivalent completeness correction (Sect.~\ref{sec:compl}) cannot therefore be applied to the entire multi-survey sample. This motivates the use of $\mathcal{D}_3$ for the reference population analysis.

\section{Implementation and checks of the host-property analysis}
\label{app:hostdemog}

The model of Sect.~\ref{sec:results} is defined by Eq.~\eqref{eq:mumass}; the details of uncertainty propagation, metallicity adjustment, and direct within-class regressions follow. Unless derived from the mixture model, uncertainties in this appendix are estimated from $1000$ host-level bootstrap resamples, with all regressions refitted in each replicate.

\paragraph{Mixture-model implementation and host uncertainties.}
We test for an additional dependence of the component means on age by including $\kappa^{t}_{\rm k}(\log_{10}t_i-\langle\log_{10}t\rangle)$ in Eq.~\eqref{eq:mumass}, where the ages are sampled as in Sect.~\ref{sec:hier} and $\langle\log_{10}t\rangle$ is the median over $\mathcal{D}_3$. The improvement is modest ($2\Delta\ln\mathcal{L}=3.6$ for two additional parameters), with $\Delta\mathrm{BIC}=9.1$ favouring the model without age-dependent means. Including these terms changes $\kappa^{M}_{\rm SN}$ and $\kappa^{\rm Fe}_{\rm SN}$ by less than half their uncertainties. We therefore retain age-independent component means, consistent with Sect.~\ref{sec:hier} and Appendix~\ref{app:compl:widths}.

The host-property mixture fits use the $581$ planets of $\mathcal{D}_3$, the three-interval parametrisation of $f_{\rm SE}^{\rm m}(t)$, and component widths independent of host properties. We fix the age-dependent selection parameter $\mathcal{H}$ to $+0.033$\,dex\,Gyr$^{1/2}$, the median of its likelihood profile from the DR25 planet candidates (Appendix~\ref{app:compl:probit}).

We adopt uniform priors on the host-property slopes, with $\kappa^{M}_{\rm k}\in[-1.5,1.5]$, $\kappa^{\rm Fe}_{\rm k}\in[-0.5,0.5]$, $\kappa^{P}_{\rm k}\in[-0.3,0.3]$, $\kappa^{t}_{\rm k}\in[-0.5,0.5]$, $\beta_M\in[-10,10]$, and $\beta_{\rm Fe}\in[-20,20]$, with slopes expressed per dex. Fits without metallicity terms instead use $\kappa^{M}_{\rm k}\in[-2,2]$ and $\beta_M\in[-8,8]$. For the colour fits of Sect.~\ref{sec:resmass}, the corresponding prior ranges are $[-0.6,0.6]$ per mag for the component means and $[-8,8]$ per mag for the SE weight.

We propagate uncertainties in host properties by averaging each host likelihood, obtained as the product of its planetary likelihoods $\mathcal{L}_i$ (Eq.~\eqref{eq:like}), over $32$ joint draws of $\log_{10}M_\star$, [Fe/H], and $\log_{10}R_\star$. Drawing host mass and metallicity directly around their measured values would introduce regression dilution: measurement errors broaden the observed distribution of the predictors, biasing individual slopes towards zero and potentially transferring part of the dependence between correlated predictors.

To account for this effect, we model the intrinsic distribution of $\log_{10}M_\star$ and [Fe/H] across the $404$ hosts as a bivariate Gaussian. Its mean is estimated from the observed distribution, while its covariance is obtained by subtracting the mean measurement-error variances from the observed variances, assuming independent errors in mass and metallicity. 
Asymmetric uncertainties are approximated by their mean. The inferred intrinsic variance accounts for $81\%$ of the observed variance in $\log_{10}M_\star$ and $58\%$ in [Fe/H]. We draw the true values of each host from their distribution conditional on the measured values, which shrinks them towards the population mean, with less shrinkage for more precise measurements.
Mass estimates from stellar models may have errors correlated with those of [Fe/H]. Assuming a correlation of $+0.5$ changes the four slopes $\kappa^{M}_{\rm k}$ and $\kappa^{\rm Fe}_{\rm k}$ by at most $0.05$ at the maximum a posteriori.

We also propagate stellar-radius uncertainties because $R_p$ scales with $R_\star$. Correlated errors in stellar mass and radius could otherwise produce a spurious relation between planet size and host mass. We draw $\log_{10}R_\star$ from its asymmetric uncertainties, assuming a correlation of $0.9$ with the error in $\log_{10}M_\star$, equal to the observed correlation between these quantities across $\mathcal{D}_3$.
Setting this error correlation to zero changes the four slopes $\kappa^{M}_{\rm k}$ and $\kappa^{\rm Fe}_{\rm k}$ by less than $0.05$ at the maximum a posteriori.
Each draw updates the planet radius, the selection function $S_i(x')$ through the expected MES and its threshold $c_i$, and the radius-window selection of Eq.~\eqref{eq:window}. To avoid counting the stellar-radius uncertainty twice, we reduce $\sigma_{x,i}$ in Eq.~\eqref{eq:like} to the uncertainty in $R_p/R_\star$, subtracting the stellar-radius contribution in quadrature and imposing a floor of $0.01$\,dex.

\paragraph{Completeness-weighted metallicity comparison.}
We use $\tilde{\mathcal{D}}_3$, comprising $541$ planets around $381$ hosts, and control for de-reddened host colour, stellar mass, \textit{Kepler} magnitude $K_p$, gyrochronological age, and orbital period. To account for selection, each planet receives a weight $v_i=1/\bar S_i$, where $\bar S_i$ is the population completeness of Sect.~\ref{sec:compl}, evaluated at its radius and period within its metallicity subset. We exclude $13$ planets for which $\bar S_i<0.25$ in either subset, ensuring that both cover a comparable radius range.

Splitting the remaining sample at $[\mathrm{Fe/H}]=0$ gives $198$ metal-poor and $330$ metal-rich planets. The median [Fe/H] uncertainty is $0.08$\,dex, and $40\%$ of planets have metallicities within one measurement uncertainty of the solar value. Misclassification between the two subsets, if independent of planetary class and the control variables, tends to dilute the estimated metallicity effect. A direct comparison of the SE fractions of the two subsets would be confounded because metal-poor hosts are generally fainter, less massive, and redder, and their planets have longer orbital periods (host-level Kolmogorov--Smirnov tests give $p=2\times10^{-5}$ for brightness, $4\times10^{-5}$ for mass, and $0.018$ for colour; the planet-level period comparison gives $p=0.0049$, whereas the host-age distributions do not differ significantly, with $p=0.12$). We therefore estimate the metallicity effect at fixed control variables.

Let $\zeta_i=1$ identify a metal-rich host and $\zeta_i=0$ a metal-poor host. We classify planets as SEs ($Y_i=1$) or SNs ($Y_i=0$) using the period-dependent valley locus of Eq.~\eqref{eq:slope}, as in Sect.~\ref{sec:ageshift}. The vector $X_i$ contains five standardised control variables, in the order $(G_{\rm BP}-G_{\rm RP})_{0,i}$, $M_{\star,i}$, $\log_{10}P_i$, $K_{p,i}$, and $\log_{10}t_i$. Including $K_p$ accounts for residual brightness-dependent detection effects, since the two metallicity subsets differ in mean magnitude by $0.49$\,mag.

We first estimate the \emph{propensity score}, $e(X)=P(\zeta=1\mid X)$, the probability that a host belongs to the metal-rich subset given its control variables. For each planet, we obtain $\hat e_i$ from a weighted logistic regression,
\begin{equation}\label{eq:propensity}
\operatorname{logit}\hat e_i
=\hat\omega_0+\sum_{j=1}^{5}\hat\omega_jX_{j,i},
\end{equation}
whose coefficients maximise
$\sum_i v_i[\zeta_i\ln\hat e_i+(1-\zeta_i)\ln(1-\hat e_i)]$ without regularisation.

We define the population-average metallicity level effect as
\begin{equation}\label{eq:deltafe}
\delta_{\rm Fe}
=\frac{1}{\sum_i v_i}\sum_i v_i
\left[\psi(1,X_i)-\psi(0,X_i)\right],
\end{equation}
where $\psi(z,X)=P(Y=1\mid\zeta=z,X)$ is the probability of belonging to the SE class at metallicity label $z$ and fixed control variables. Thus, $\delta_{\rm Fe}$ measures the difference in SE fraction between the two metallicity conditions, averaged over the completeness-weighted population. For every planet, both $\psi(1,X_i)$ and $\psi(0,X_i)$ are evaluated regardless of its observed metallicity label.

We estimate $\psi$ through a second weighted logistic regression of $Y$ on $\zeta$ and the same five control variables, with no interaction terms or regularisation. Its coefficients maximise
$\sum_i v_i[Y_i\ln\hat\psi(\zeta_i,X_i)+(1-Y_i)\ln(1-\hat\psi(\zeta_i,X_i))]$.

We then estimate $\delta_{\rm Fe}$ with the augmented inverse-probability weighting (AIPW) estimator \citep{Robins1994,LuncefordDavidian2004},
\begin{equation}\label{eq:aipw}
\hat\delta_{\rm Fe}
=\frac{1}{\sum_i v_i}\sum_i v_i\Bigg[
\hat\psi(1,X_i)-\hat\psi(0,X_i)
+\left(\frac{\zeta_i}{\hat e_i}
-\frac{1-\zeta_i}{1-\hat e_i}\right)\hat r_i
\Bigg],
\end{equation}
where $\hat r_i=Y_i-\hat\psi(\zeta_i,X_i)$ is the residual of the outcome model. The augmentation term corrects the modelled SE probabilities using residuals weighted by the inverse propensity of each planet's observed metallicity subset. Given the completeness weights, and assuming no unmeasured confounders and overlapping propensity scores, the estimator is consistent if either the propensity model or the outcome model is correctly specified \citep{LuncefordDavidian2004}.

After weighting by $v_i/\hat e_i$ and $v_i/(1-\hat e_i)$, the effective sample sizes are approximately $255$ metal-rich and $42$ metal-poor planets. The uncertainty is therefore dominated by the metal-poor subset. We obtain
$\hat\delta_{\rm Fe}=+0.03^{+0.12}_{-0.11}$, consistent with zero ($0.3\sigma$).

The result is robust to several analysis choices. Shifting the classification boundary by $\pm0.05$ in $\log_{10}R_p$, omitting the completeness weights, removing either $K_p$ or colour from the controls, or splitting at the median metallicity ($[\mathrm{Fe/H}]=+0.03$) yields $\hat\delta_{\rm Fe}$ between $-0.02$ and $+0.10$, always within the quoted uncertainty. For the median-metallicity split, the completeness weights and exclusions are recomputed. Treating [Fe/H] instead as a continuous predictor in the outcome model, without splitting the sample, gives a completeness-weighted mean derivative of the SE probability of $-0.06^{+0.15}_{-0.15}$ per dex, corresponding to $-0.01^{+0.04}_{-0.04}$ across the $0.24$\,dex difference in mean metallicity between the two metallicity subsets.

\paragraph{Colour fits and direct regressions.}
We also replace host mass in Eqs.~\eqref{eq:mumass} and \eqref{eq:wlogit} with the de-reddened \textit{Gaia} colour $(G_{\rm BP}-G_{\rm RP})_0$, centred on its median over $\mathcal{D}_3$, while setting $\kappa^{\rm Fe}_{\rm k}=\beta_{\rm Fe}=\kappa^{P}_{\rm k}=0$. The resulting slopes are
$\kappa^{\rm col}_{\rm SN}=-0.12\pm0.02$ and
$\kappa^{\rm col}_{\rm SE}=+0.05\pm0.08$ per mag. Colour uncertainties are neglected because their median ($0.003$\,mag) is much smaller than the dispersion across hosts ($0.25$\,mag). Thus, SNs are systematically smaller around redder, generally lower-mass hosts, whereas SE sizes show no significant colour dependence.

As a check that does not use the mixture model, we fit the relation between planet radius and host properties directly within the observed SE and SN classes. We divide planets at $R_p=1.71\,R_\oplus$, corresponding to the valley locus at the reference period of $10$\,d (Eq.~\eqref{eq:slope}), and regress $\log_{10}R_p$ against $\log_{10}M_\star$ within each class. The least-squares fits use $\tilde{\mathcal{D}}_3$, with completeness weights $1/\bar S$ evaluated within each colour half. The slopes, with host-level bootstrap uncertainties, are $-0.06\pm0.07$ for SEs and $+0.35\pm0.06$ for SNs per dex of host mass (Fig.~\ref{fig:massdecomp}). The corresponding direct-regression slopes against colour are $+0.02\pm0.02$ per mag for SEs and $-0.08\pm0.02$ per mag for SNs. Recomputing $\bar S$ in five equal-count bins of host mass or colour, and excluding planets with $\bar S<0.25$ in any bin, changes all slopes by less than half their uncertainties.

The direct regressions confirm that SN radius increases with host mass, whereas SE radius does not show a significant trend. Because these regressions omit metallicity, the appropriate mixture comparison is a fit with $\kappa^{\rm Fe}_{\rm k}=\beta_{\rm Fe}=0$. In that case, the SN mass slope increases to $\kappa^{M}_{\rm SN}=+0.58\pm0.11$ ($2\Delta\ln\mathcal{L}=25.2$ relative to zero), absorbing part of the metallicity dependence correlated with host mass.

The smaller slope from direct regression is expected for two reasons. First, classification at a fixed observed radius introduces a truncation effect: at lower host masses, intrinsically smaller SNs are more likely to fall below $1.71\,R_\oplus$ and be classified as SEs, increasing the mean radius of the remaining SNs and flattening the relation. Second, unlike the mixture model, the regression does not correct for errors in host mass, which dilute the fitted slope. For the metallicity dependence, adding [Fe/H] and $\log_{10}P$ to the regression, with the weights $v_i$ of the metallicity comparison, gives a SN metallicity slope of $+0.04\pm0.04$ per dex, compared with $\kappa^{\rm Fe}_{\rm SN}=+0.15\pm0.08$ from the mixture. Mock catalogues generated with $\kappa^{\rm Fe}_{\rm SN}=0.15$ yield a mean regression slope of $+0.06$ (standard deviation $0.03$) over $500$ realisations, confirming the expected attenuation.

\section{Alternative samples}\label{app:altsamples}
\subsection{Extended age coverage}\label{app:altsamples:extage}
The extended sample $\mathcal{E}$ is built on the same \textit{Kepler} DR25 base as $\mathcal{D}_3$, under the same selection criteria, with three restrictions relaxed. 
The tier requirements of criteria~(c) and~(g) are widened to $\le$4, and the age range is extended beyond $4$\,Gyr. The gyrochronological ages given as limits, which criterion~(g) excludes, enter wherever their bound alone decides the age bin: upper limits at or below $0.75$\,Gyr in the $0$--$0.75$\,Gyr bin and lower limits above $4$\,Gyr in the bin beyond $4$\,Gyr. We define $\tilde{\mathcal{E}}$ as $\mathcal{E}$ restricted to radii of $1$--$4\,R_\oplus$, as $\tilde{\mathcal{D}}_3$ is for $\mathcal{D}_3$ (Sect.~\ref{sec:global}). 
The sample $\tilde{\mathcal{E}}$ holds $1656$ planets around $1201$ hosts: $1416$ with non-limit ages, $31$ with upper and $209$ with lower limits. They are divided into the age bins of Fig.~\ref{fig:d3bins}, with the first extended to $0$--$0.75$\,Gyr, plus a bin beyond $4$\,Gyr.
The limits are not perturbed in the Monte Carlo, since their bound fixes their bin. 
As in Sect.~\ref{sec:ageshift}, the counts are weighted by $1/\bar S_i$, with $\bar S$ averaged over the configurations of $\tilde{\mathcal{E}}$ in the age bin of each planet, and the observed counts are given alongside. 
The locus refitted on $\tilde{\mathcal{E}}$ with the SVM of Sect.~\ref{sec:prplane}, $m=-0.055\pm0.010$ with $R_{\rm gap}\simeq 1.79\,R_\oplus$ at $10$\,d, sets the classifier and $R_p'$ throughout this subsection. It is flatter than the locus of $\tilde{\mathcal{D}}_3$, and $\tilde{\mathcal{E}}$ holds $62\%$ F and G hosts against $35\%$ in $\tilde{\mathcal{D}}_3$. The F and G hosts of $\tilde{\mathcal{E}}$ place the valley higher than its K and late-G hosts, at $1.85^{+0.05}_{-0.02}$ against $1.70^{+0.03}_{-0.02}\,R_\oplus$ at $10$\,d, with a locus flatter in $90\%$ of the paired resamples (slopes $-0.064\pm0.013$ and $-0.098\pm0.022$).

\begin{figure*}[htbp]\centering
\includegraphics[width=\textwidth]{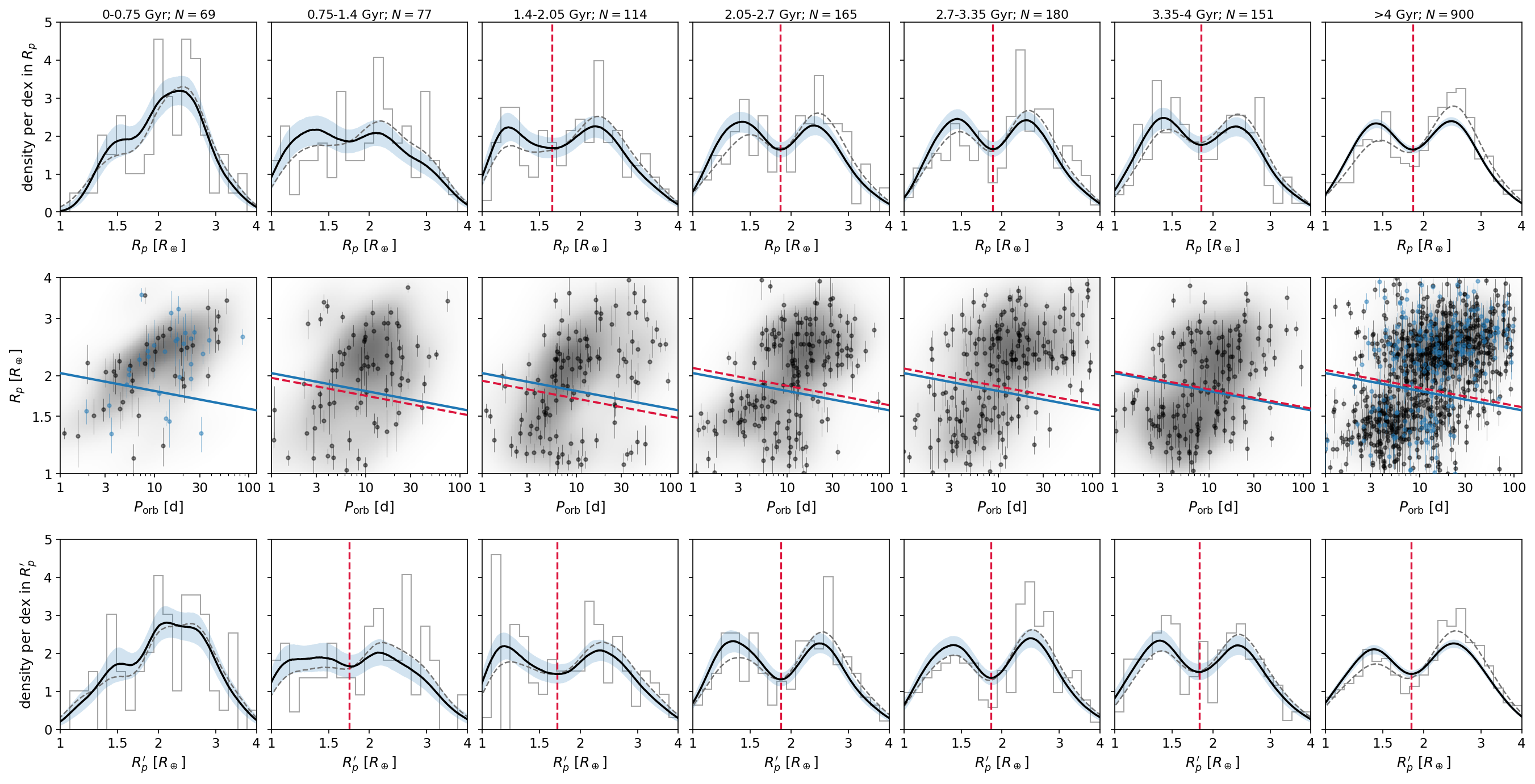}
\caption{Seven age bins of the extended sample $\tilde{\mathcal{E}}$ (columns).
\textit{Top:} distribution of $R_p$. On the observed counts: grey outline, histogram; grey dashed curve, median of the fixed-bandwidth KDE over Monte Carlo draws that resample the planets and perturb the non-limit ages, while the age limits are kept fixed.
With each planet weighted by $1/\bar S_i$, with $\bar S$ averaged over the configurations of $\tilde{\mathcal{E}}$ in its age bin: black curve and blue band, median and 16th--84th percentile interval of the same KDE; crimson dashed line, the marginalised position of the valley minimum, drawn where the weighted counts resolve the valley in at least two thirds of the draws. 
\textit{Middle:} period--radius plane on the observed counts. Background: two-dimensional density. Points: the planets with their period and radius uncertainties, in black where the age is not a limit and in blue where it is a limit (upper limits in the $0$--$0.75$\,Gyr bin, lower limits beyond $4$\,Gyr). $N$ counts both the planets with non-limit ages and those with limits. Blue solid line: the locus refitted on $\tilde{\mathcal{E}}$, identical in every panel. Crimson dashed line: the valley minimum of each bin at that fixed slope, its position read on $R_p'$ (Eq.~\eqref{eq:straight}, with the $\tilde{\mathcal{E}}$ locus) with the weights $1/\bar S_i$, drawn where the weighted counts resolve the valley on $R_p'$ in at least two thirds of the draws. 
\textit{Bottom:} distribution of $R_p'$ (Eq.~\eqref{eq:straight}, with the $\tilde{\mathcal{E}}$ locus), drawn as in the top row.}
\label{fig:e7bin}
\end{figure*}

In the five bins from $0.75$ to $4$\,Gyr, $\tilde{\mathcal{E}}$ reproduces the morphology of $\tilde{\mathcal{D}}_3$ (Fig.~\ref{fig:e7bin}), with the valley resolved on $R_p'$ in at least $71\%$ of the draws. The $0$--$0.75$\,Gyr bin, raised from $29$ to $69$ planets by the relaxed tiers and the upper limits, resolves it in $44\%$ of the draws, and the bin beyond $4$\,Gyr, with $900$ planets, in all of them. The weighted fraction $f_{\rm SE}^{{\rm c},w}$ is $0.32^{+0.06}_{-0.06}$, $0.53^{+0.04}_{-0.04}$ and $0.50^{+0.03}_{-0.03}$ in the $0$--$1$, $1$--$2$ and $2$--$4$\,Gyr bins, respectively. It rises by $+0.18^{+0.06}_{-0.07}$ from the first to the last bin ($+0.14^{+0.05}_{-0.06}$ on the observed counts).

Beyond $4$\,Gyr, $f_{\rm SE}^{{\rm c},w}$ is $0.44^{+0.02}_{-0.02}$ ($0.37^{+0.02}_{-0.02}$ on the observed counts), $0.05^{+0.03}_{-0.03}$ below the $2$--$4$\,Gyr bin. Across that edge the sample changes in composition. Beyond $4$\,Gyr, $75\%$ of the planets orbit F and G hosts and $31\%$ orbit beyond $20$\,d, where SEs make up a smaller fraction of the planets of $\tilde{\mathcal{E}}$, against $51\%$ and $24\%$ in the $2$--$4$\,Gyr bin. 
To separate the effect of age from that of composition, we standardise the weighted SE fractions $f_{\rm SE}^{{\rm c},w}$ of the age bins. We fit a logistic regression of the SE label on the age bin, host colour, orbital period, age reliability tier and \textit{Kepler} magnitude, weighted by $1/\bar S_i$, and use it to predict the SE fraction of each age bin for the composition of the whole of $\tilde{\mathcal{E}}$.
The difference in $f_{\rm SE}^{{\rm c},w}$ between the bin beyond $4$\,Gyr and the $2$--$4$\,Gyr bin then becomes $-0.014^{+0.028}_{-0.029}$, or $+0.004^{+0.028}_{-0.029}$ with the lower limits left out. 
The rise of $f_{\rm SE}^{{\rm c},w}$ from the $0$--$1$ to the $2$--$4$\,Gyr bin remains $+0.18\pm0.06$ ($3.1\sigma$), and is $+0.15\pm0.06$ with stellar radius and [Fe/H] added to the regression.

We fit the likelihood of Sect.~\ref{sec:hier} to the $773$ planets of $\mathcal{E}$ with non-limit ages of $0.1$--$4$\,Gyr and radii of $0.5$--$4\,R_\oplus$, with the selection function and priors used on $\mathcal{D}_3$ (Appendix~\ref{app:compl:posterior}). 
In the three-interval form of $f_{\rm SE}^{\rm m}(t)$, the age contrast, $\bar f_{2\text{--}4}-\bar f_{<1}$, is $+0.26^{+0.09}_{-0.09}$, positive in $99.8\%$ of the posterior. 
In the exponential form, the Gaussian prior on $f_\infty$ is taken from $f_{2\text{--}4}$ of the three-interval form fitted to the same planets (Sect.~\ref{sec:resglobal}). In the $89\%$ of the posterior draws in which $\bar f_{\rm SE}^{\rm m}$ starts below $0.9\,\bar f_\infty$, $t_{90}$ is below $1.7$\,Gyr with $90\%$ probability and below $2.2$\,Gyr with $95\%$ probability.

\subsection{Stellar and planetary radii from \texorpdfstring{\citet{Berger2026}}{Berger et al. (2026)}}\label{app:altsamples:homog}
The stellar and planetary quantities of this paper are drawn from several literature sources and selected one parameter at a time (Sect.~\ref{sec:planetpars} and Paper~I). We repeat the analysis taking the stellar masses and radii and the planet radii and orbital periods from a single input catalogue, the \textit{Gaia}--\textit{Kepler}--TESS-host catalogue of \citet[][hereafter B26]{Berger2026}. 
The rotation periods, the spectroscopic metallicities and the de-reddened \textit{Gaia} colours remain those adopted in the rest of the paper. The gyrochronological ages are refitted with the masses of B26 under the same per-star likelihood and the same random seeds. 
We denote by $\mathcal{K}$ the sample of the planets for which B26 provide all four of these quantities and whose hosts have a rotation period and a \textit{Gaia} colour. Applying the criteria that define $\tilde{\mathcal{D}}_3$ (Sect.~\ref{sec:sample}) to these values gives the sample $\tilde{\mathcal{K}}_3$, which holds $514$ planets around $358$ hosts, against $541$ planets around $381$ hosts in $\tilde{\mathcal{D}}_3$.

The ages change little with the masses of B26. On the $343$ stars that $\tilde{\mathcal{K}}_3$ and $\tilde{\mathcal{D}}_3$ share, the ages change by a median of $0.003$\,Gyr, and by less than $1\%$ for $86\%$ of the stars, since the ages are set mostly by rotation period and colour, which are the same in $\tilde{\mathcal{K}}_3$ and $\tilde{\mathcal{D}}_3$. 
On $\tilde{\mathcal{K}}_3$, the locus fitted on the observed counts is $m=-0.101\pm0.019$ with $R_{\rm gap}\simeq1.70\,R_\oplus$ at $10$\,d, against $-0.096\pm0.022$ and $1.71\,R_\oplus$ on $\tilde{\mathcal{D}}_3$, and sets the classifier and $R_p'$ for $\tilde{\mathcal{K}}_3$. 
With the counts weighted by $1/\bar S_i$ at the radii of $\tilde{\mathcal{K}}_3$, the valley on $R_p'$ sits at $1.72^{+0.05}_{-0.05}\,R_\oplus$ with depth $0.42^{+0.06}_{-0.07}$, resolved in more than $99\%$ of the draws. The weighted fraction $f_{\rm SE}^{{\rm c},w}$ is $0.39^{+0.08}_{-0.08}$, $0.51^{+0.06}_{-0.05}$ and $0.43^{+0.03}_{-0.03}$ in the $0.1$--$1$, $1$--$2$ and $2$--$4$\,Gyr bins, respectively.
The difference in $f_{\rm SE}^{{\rm c},w}$ between the $2$--$4$ and $0.1$--$1$\,Gyr bins is $+0.03^{+0.09}_{-0.08}$, and between the $1$--$2$ and $0.1$--$1$\,Gyr bins $+0.12^{+0.10}_{-0.09}$. Compared with $\tilde{\mathcal{D}}_3$ on the same host resamples, both differences are lower in $\tilde{\mathcal{K}}_3$ by $0.03\pm0.04$, and the fraction of each bin changes by at most $0.03$. 
After standardisation on host colour, orbital period and \textit{Kepler} magnitude, as in Appendix~\ref{app:altsamples:extage}, $f_{\rm SE}^{{\rm c},w}$ becomes $0.30\pm0.07$, $0.47^{+0.04}_{-0.05}$ and $0.45\pm0.03$ in the $0.1$--$1$, $1$--$2$ and $2$--$4$\,Gyr bins, respectively. Its difference between the $2$--$4$ and $0.1$--$1$\,Gyr bins becomes $+0.15^{+0.07}_{-0.08}$, as on $\tilde{\mathcal{D}}_3$, where the same standardisation gives $+0.15\pm0.07$. 
After the standardisation, the $2$--$4$\,Gyr bin lies $0.02\pm0.05$ below the $1$--$2$\,Gyr bin, against $0.08\pm0.06$ before. Most of this change comes from the orbital period differences across the bins, since standardising on the period alone already gives a difference of $+0.12^{+0.08}_{-0.07}$ between the $2$--$4$ and $0.1$--$1$\,Gyr bins. 

We fit the likelihood of Sect.~\ref{sec:hier} to the $553$ planets of $\mathcal{K}$ that pass the criteria of $\mathcal{D}_3$, with the selection function and priors used on $\mathcal{D}_3$ (Appendix~\ref{app:compl:posterior}) and the detection threshold recalibrated on the radii of $\mathcal{K}$. 
In the three-interval form of $f_{\rm SE}^{\rm m}(t)$, the age contrast, $\bar f_{2\text{--}4}-\bar f_{<1}$, is $+0.22^{+0.10}_{-0.10}$, positive in $98.9\%$ of the posterior. 
In the exponential form, the Gaussian prior on $f_\infty$ is taken from $f_{2\text{--}4}$ of the three-interval form fitted to the same planets. In the $83\%$ of the posterior draws in which $\bar f_{\rm SE}^{\rm m}$ starts below $0.9\,\bar f_\infty$, $t_{90}$ is below $2.0$\,Gyr with $90\%$ probability and below $2.9$\,Gyr with $95\%$ probability.

The radii adopted in this paper and those of B26 differ in the source of the planet-to-star radius ratio (Appendix~\ref{app:planetparams:radius} against the per-mission tables used by B26), and only in part in the stellar radius. This is because the adopted stellar radius is derived from the estimates of the most reliable method available (Paper~I), and the estimate of B26 enters it for $365$ of the $381$ hosts of $\tilde{\mathcal{D}}_3$ and is the only one used for $114$. 
We compare the two sets of radii on the $2073$ planets that pass the criteria of $\mathcal{D}_1$ with both the adopted values and those of B26, and have a radius of $1$--$4\,R_\oplus$ in both. The comparison is unweighted, since both sets of radii describe the same planets. For each planet we divide the radius of B26 by the radius adopted in this paper. 
Across the $2073$ planets this ratio has a median of $1.00$, and it lies between $0.91$ and $1.09$ for $90\%$ of them. The valley of the full sample lies at $1.77^{+0.03}_{-0.01}\,R_\oplus$ with both sets of radii, with a depth of $0.30^{+0.03}_{-0.02}$ with the adopted radii and $0.31^{+0.03}_{-0.02}$ with those of B26. Across age, we repeat the comparison on the $483$ of these planets that belong to $\tilde{\mathcal{D}}_3$, in the six age bins of Sect.~\ref{sec:ageshift}. 
Both sets of radii resolve the valley in at least two thirds of the draws in the $1.4$--$2.05$, $2.7$--$3.35$ and $3.35$--$4$\,Gyr bins.
In the $1.4$--$2.05$\,Gyr bin the minimum lies at $1.49\pm0.07\,R_\oplus$ with the adopted radii and $1.45^{+0.06}_{-0.05}\,R_\oplus$ with those of B26, in the $2.7$--$3.35$\,Gyr bin at $1.81\pm0.06$ and $1.81\pm0.04\,R_\oplus$, and in the $3.35$--$4$\,Gyr bin at $1.77\pm0.10$ and $1.82\pm0.09\,R_\oplus$. 
In the $2.05$--$2.7$\,Gyr bin the adopted radii place the minimum at $1.77\pm0.08\,R_\oplus$, and those of B26 at $1.81^{+0.08}_{-0.11}\,R_\oplus$, but the latter resolve the valley in only $59\%$ of the draws. 
With either set of radii, the minimum therefore lies at smaller radii in the $1.4$--$2.05$\,Gyr bin than in the later bins.

\end{document}